\documentclass[]{article}

\usepackage{graphicx}
\usepackage{epstopdf, epsfig}

\usepackage{color}
\usepackage{natbib}
\usepackage[shortlabels]{enumitem}
\usepackage[margin=5pt,skip=0pt]{subcaption}
\usepackage{placeins}
\usepackage{bibentry}
\usepackage{amsmath}
\usepackage{amssymb} 
\usepackage{cancel}
\usepackage{xcolor}
\usepackage{bm}         
\usepackage{tabularx}
\usepackage{txfonts}		
\usepackage{mathrsfs}
\usepackage{booktabs}
\usepackage[capitalise,nameinlink,noabbrev]{cleveref}
\crefname{equation}{}{}
\crefname{enumi}{}{} 

\graphicspath{{imgs/}}

\newcolumntype{x}[1]{>{\centering\arraybackslash\hspace{0pt}}m{#1}}		
\newcommand\capspace{\vspace{0.2cm}}  

\title{Global stability of rigid-body-motion fluid-structure-interaction problems}
\author{P. S. Negi, A. Hanifi and D. S. Henningson}

\begin{document}

\maketitle

\begin{abstract}
	A rigorous derivation and validation for linear fluid-structure-interaction (FSI) equations for a rigid-body-motion problem is performed in an Eulerian framework. We show that the ``added-stiffness" terms arising in the formulation of \cite{fanion00} vanish at the FSI interface in a first-order approximation. Several numerical tests with rigid-body motion are performed to show the validity of the derived formulation by comparing the time evolution between the linear and non-linear equations when the base flow is perturbed by identical small-amplitude perturbations. In all cases both the growth rate and angular frequency of the instability matches within $0.1\%$ accuracy. The derived formulation is used to investigate the phenomenon of symmetry breaking for a rotating cylinder with an attached splitter-plate. The results show that the onset of symmetry breaking can be explained by the existence of a zero-frequency linearly unstable mode of the coupled fluid-structure-interaction system. Finally, the structural sensitivity of the least stable eigenvalue is studied for an oscillating cylinder, which is found to change significantly when the fluid and structural frequencies are close to resonance.
\end{abstract}

\section{Introduction}
Fluid-structure interaction (FSI) studies span a vast and diverse range of applications -- from natural phenomenon such as the fluttering of flags \citep{shelley11}, phonation \citep{heil11}, blood flow in arteries \citep{freund14}, path of rising bubbles \citep{ern12} \textit{etc.}, to the more engineering applications of aircraft stability \citep{dowell01}, vortex induced vibrations \citep{williamson04}, compliant surfaces \citep{riley88,kumaran03} \textit{etc}. The phenomena that emerge out of an FSI problem often exhibit highly non-linear, dynamically rich and complex behavior with different flow regimes such as fluttering and tumbling of plates falling under gravity \citep{mittal04}, or the unsteady path of rising bubbles \citep{guillaume01,ern12}. Despite its non-linear nature, the initial transitions from one state to another may often be governed by a linear instability mechanism. The simplified case of a linear instability of a parallel boundary layer over a compliant surface was first derived and investigated by \cite{benjamin59,benjamin60} and \cite{landahl62}. For boundary layers over Krammer-type compliant surfaces, the linear hydrodynamic instability study was performed by \cite{carpenter85,carpenter86} using this formulation and subsequently several studies have used the same formulation for studying flow instabilities over flexible surfaces, for example \cite{carpenter90,davies97} \textit{etc}. 

When the parallel flow assumption is no longer valid a global approach needs to be taken. From the perspective of global instabilities in fluid-structure-interaction problems, the first such study was reported by \cite{cossu00} for the instability of a spring mounted cylinder. The authors considered a rigid 2D cylinder that was free to oscillate in the cross-stream direction, subject to the action of a spring-mass-damper system. The linearized equations were solved in a non-inertial frame of reference attached to the cylinder. For low (solid to fluid) mass density ratios, the authors report that the critical Reynolds number for vortex shedding drops to $Re_{c}\approx23$. \cite{mittal05} also report a similarly low critical $Re$ for a cylinder oscillating in both streamwise and transverse directions. A non-inertial reference is also used by \cite{navrose16} to study the lock-in phenomenon of cylinders oscillating in cross-stream through the linear stability analysis. More recently, Magnaudet and co-workers have approached the problem of rising and falling bodies through the perspective of linear instability. \cite{fabre11}  developed a quasi-static model to describe the stability of heavy bodies falling in a viscous fluid. \cite{assemat12} perform a linear study of thin and thick two-dimensional plates falling under gravity. The problem is again formulated in a non-inertial frame of reference of the moving body undergoing both translation and rotation. The authors showed a quantitative agreement between the quasi-static model of \cite{fabre11} and the linear stability results for high mass-ratio cases, although the agreement systematically deteriorated as the mass ratio was reduced. \cite{tchoufag14} performed similar studies for three-dimensional disks and thin cylinders, \cite{tchoufag14a} apply the linear stability analysis to spheroidal bubbles and \cite{lozano16} for oblate bubbles. 

In general the linear FSI investigations have largely followed one of two methodologies. Either through the parallel flow approach or through frames of reference attached to the rigid body in motion. There are a few exceptions to his. \cite{lesoinne00} formulated the linear FSI problem in the Arbitrary-Eulerian-Lagrangian (ALE) form of the inviscid Navier--Stokes where they treated the grid velocity as a pseudo-variable. \cite{fanion00} have derived a more general formulation starting from the ALE equations in the weak form,  which has been used by \cite{fernandez03i,fernandez03ii}. The authors derive a formulation independent of the fluid grid-velocity but both the boundary conditions and the fluid-stresses at the FSI interface are modified. The velocity continuity boundary condition transforms to a transpiration boundary condition while additional stress terms at the interface are obtained comprising of higher order derivatives of the base flow. These additional stresses have been termed as ``added stiffness" terms \citep{fernandez03i,fernandez03ii}. \cite{goza18} have used an immersed boundary method for the global stability analysis of inverted flag-flapping. In a very recent development, \cite{pfister19} also derive the the linear FSI formulation with the ALE framework. In their formulation the fluid and structural quantities are solved on the moving material points. The standard Navier--Stokes and divergence equations are therefore modified to account for the motion of the material points.

In the current work we follow the methodology of \cite{fanion00} and \cite{fernandez03i} which, in spirit is similar to the methodology used by \cite{benjamin60} and \cite{landahl62} for compliant wall cases. However, unlike \cite{fanion00} we proceed with the linearization of the equations in their strong form. The final expressions for the linearized equations are all evaluated on a stationary grid (Eulerian formulation). The resulting expressions for the linearized FSI problem are very similar to the ones obtainted by \cite{fanion00} however some crucial differences arise for the stresses at the interface. The ``added-stiffness" terms arising in the work of \cite{fanion00} are shown to vanish at first-order. While we invoke no special assumptions for the linearization of the fluid equations (other than small-amplitude perturbations), the current capabilities of our numerical solver limits the class of FSI problems that can be validated. Therefore we confine the focus of the current work to FSI problems undergoing rigid-body motion and defer a more general formulation and validation to future work. The derived formulation for rigid-body linear FSI is numerically validated by comparing linear and non-linear evolution of different cases which are started from the same base flow state, perturbed by identical small amplitude disturbances. It is shown that the linear and non-linear simulations evolve nearly identically through several orders of magnitude of growth of the perturbations. The non-linear cases eventually reach a saturated state while the linear simulations continue with the exponential growth. The derived equations are then used to investigate the instability of an oscillating cylinder at subcritical Reynolds numbers, for an ellipse in both pure translation and pure rotation, and for the case of symmetry breaking in a cylinder with an attached splitter-plate. Finally in section~\ref{fsi_struc_sensitivity} we investigate the structural sensitivity of the least stable eigenvalue in the coupled FSI problem of an oscillating cylinder at $Re=50$, with varying structural parameters.

The remainder of the paper is organized as follows. In section~\ref{fsi_linearization} we describe the problem in a general setting and derive the linearized equations for a fluid-structure-interaction problem for rigid-body motion. Numerical validation and results for the derived formulation are presented in section~\ref{fsi_instability_results}. In section~\ref{fsi_struc_sensitivity} we introduce the adjoint problem for the linear FSI system of a cylinder oscillating in crossflow and show the changes in structural sensitivity of the unstable eigenvalue. Section~\ref{conclusion} concludes the paper.

\section{Linearization of FSI}
\label{fsi_linearization}
\subsection{General problem description}
\label{problem_description}
\begin{figure}
	\centering
	\includegraphics[width=0.9\linewidth]{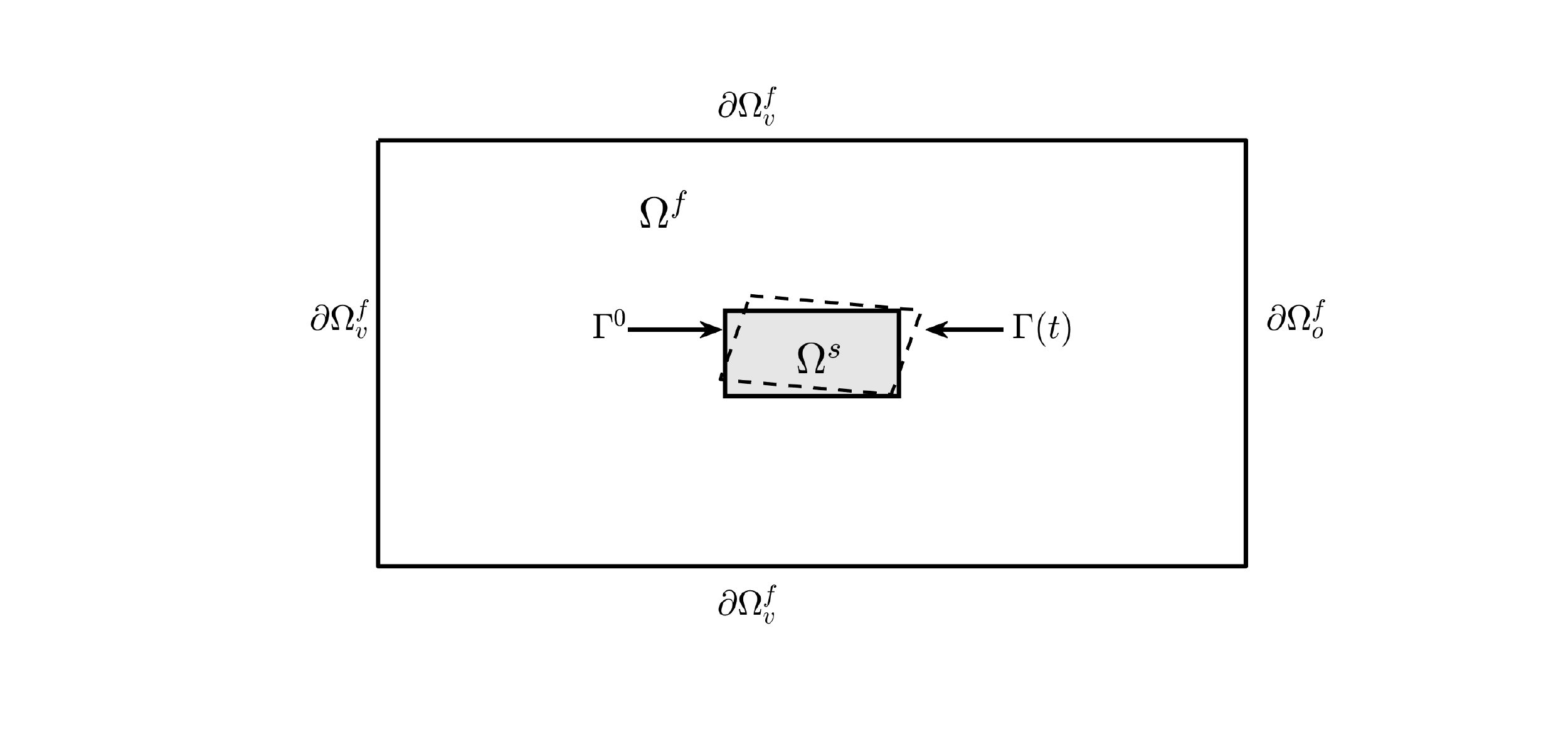}
	\caption{Domain for a general fluid-structure interaction problem. The equilibrium position of the interface is marked by $\Gamma^{0}$ and the perturbed position is marked by $\Gamma(t)$}
	\label{fig:fsi_domain}
\end{figure}
We primarily follow the index notation with the implied Einstein summation. Bold face notation is used to represent vectors where necessary. Consider a fluid-structure-interaction problem as illustrated in figure~\ref{fig:fsi_domain}. The bounded region in white marked by $\Omega^{f}$ represents the fluid part of the domain and the gray region marked by $\Omega^{s}$ represents the structural domain. The combined fluid and structural regions are bounded by the boundaries represented by $\partial\Omega^{f}_{v}$ and $\partial\Omega^{f}_{o}$. Here $\partial\Omega^{f}_{v}$ represents the far-field boundary conditions which define the problem and $\partial\Omega^{f}_{o}$ represents the open boundary condition applied for computations performed on a finite sized domain. The structural domain is bounded by the time varying fluid-structure interaction interface $\Gamma(t)$ on which the fluid forces act. The Navier--Stokes equations, along with the incompressibility constraint govern the evolution of the fluid in $\Omega^{f}$. For a problem with moving interfaces, the Navier--Stokes is usually formulated in the Arbitrary-Lagrangian-Eulerian (ALE) formulation \citep{ho90,ho91} defined on moving material points as
\begin{subequations}
	\begin{eqnarray}
	\left.\frac{\partial U_{i}}{\partial t}\right\vert_{\mathbf{W^{g}}} + (U_{j} -W^{g}_{j})\frac{\partial U_{i}}{\partial x_{j}} = -\frac{\partial P}{\partial x_{i}}  + \frac{1}{Re}\frac{\partial^{2} U_{i}}{\partial x_{j}\partial x_{j}}\label{NS},\\
	\frac{\partial U^{0}_{i}}{\partial x_{i}} = 0 \label{incompressibility}.
	\end{eqnarray}
\end{subequations}
Here $\mathbf{U}$ and $P$ represents the fluid velocity and pressure respectively, while $\mathbf{W^{g}}$ represents the velocity of the material points. We refer to the time derivative in this formulation $(\partial U_{i}/\partial t\vert_{\mathbf{W^{g}}})$ as the ALE time derivative evaluated when following a material point with velocity $\mathbf{W^{g}}$. Note that $\mathbf{W^{g}}$ is not uniquely defined. The only restriction on material point velocity is for the points on the interface $\Gamma(t)$, where the fluid velocity is equal to the velocity of the interface points. Typically $\mathbf{W^{g}}$ is defined by some function which smoothly extends the velocity at the interface to the rest of the fluid domain.

The equations governing the structural motion depend on the type of modeling or degrees of freedom of the structure being considered. Confining the discussion to rigid-body motion problems we represent the structural equations in a general form as
\begin{eqnarray}
	\mathcal{M}\frac{\partial^{2} \eta_{i}}{\partial t^{2}} + \mathcal{D}\frac{\partial \eta_{i}}{\partial t} + \mathcal{K} \eta_{i} = \mathcal{F}_{i} \label{struc}.
\end{eqnarray}
Here $\mathcal{M}$ is a generalized inertia, $\mathcal{D}$ is a generalized damping and $\mathcal{K}$ is a generalized stiffness. $\mathcal{F}_{i}$ represents the fluid forces acting on $\Gamma(t)$. The equation represents a typical linear spring-mass-damper system for rigid-body motion. The structural degrees of freedom are represented by $\eta_{i}$, whose definition depends on the type of modeling used for the structure. For example, for a cylinder free to oscillate in one direction subject to a spring-damper action (as in \cite{cossu00}), $\eta_{i}$ represents the position of the center of mass in that direction, $\mathcal{M}$ is the cylinder mass, $\mathcal{D}$ is the damping coefficient and $\mathcal{K}$ is the spring stiffness constant. $\mathcal{F}_{i}$ is the integral of the fluid forces acting on the cylinder in the $i^{th}$ direction. 

The fluid and structural domains are coupled at the FSI interface through a no-slip and no penetration condition. Defining $\mathbf{x}^{\Gamma}$ as the (time-dependent) position of points on the interface $\Gamma(t)$ and $\mathbf{v}^{\Gamma}$ as their instantaneous velocity, we may write the boundary condition as
\begin{eqnarray}
 	W^{g}_{i} = U_{i} = \frac{d x_{i}^{\Gamma}}{dt}= v^{\Gamma}_{i}, & \text{on $\Gamma(t)$}\label{noslip}.
\end{eqnarray}
The coupled problem is now defined by equations~\ref{NS} and \ref{incompressibility} in the fluid domain, equation \ref{struc} in the structural domain, coupled with the no-slip condition at the interface (equation ~\ref{noslip}). The problem is completed by the application of appropriate Dirichlet boundary conditions on $\partial \Omega^{f}_{v}$ and open boundary conditions on $\partial \Omega^{f}_{o}$. We assume that the outer domain boundaries always remain stationary.

\subsection{Steady-state}
For the problem described in Section~\ref{problem_description}, consider a full system state given by $(\mathbf{U}^{0},P^{0},\boldsymbol{{\eta}}^{0})$ (referred to as the base flow state), defined on material points $\mathbf{x}^{0}$, which satisfies the time-independent incompressible Navier--Stokes and structural equations (along with the respective boundary conditions). \textit{i.e.}
\begin{subequations}
	\begin{eqnarray}
	   U^{0}_{j}\frac{\partial U^{0}_{i}}{\partial x_{j}} +\frac{\partial P^{0}}{\partial x_{i}}  - \frac{1}{Re}\frac{\partial^{2} U^{0}_{i}}{\partial x_{j}\partial x_{j}} = 0\label{NS0}, \\
		\frac{\partial U^{0}_{i}}{\partial x_{i}} = 0 \label{incompressibility0}, \\
	    \mathcal{K} \eta^{0}_{i} - \mathcal{F}^{0}_{i} = 0\label{struc0}, \\
	    U_{i} = v^{\Gamma}_{i} = 0; & \text{on $\Gamma^{0}$}.
	\end{eqnarray}
\end{subequations}
Where $\mathcal{F}^{0}_{i}$ are the fluid forces acting on the structure at equilibrium and $\Gamma^{0}$ is the equilibrium position of the FSI interface.

\subsection{Linearization of the structure}
We begin with the linearization of the structural equations. In defining the generalized structural equation~\ref{struc} we make an implicit assumption that the structural equations are inherently linear, which we consider appropriate for the class of rigid-body motion problems that are being considered. In principle non-linear spring-mass-damper systems may also be considered but for the moment we restrict the discussion to linear spring-mass-damper systems.
Given an initial stationary solution $\eta^{0}_{i}$ of the structural equation, we may introduce a perturbed state given by the superposition of the stationary state and small amplitude perturbations, \textit{i.e.} $\eta_{i} = \eta^{0}_{i} + \eta'_{i}$. Similarly, we decompose the fluid forces acting on the structure as $\mathcal{F}_{i} = \mathcal{F}^{0}_{i} + \mathcal{F}'_{i}$, where $\mathcal{F}'_{i}$ are the fluid forces acting on structure due to the (yet unknown) linearized fluid perturbations. Introducing the decomposition into equation \ref{struc} we obtain the equations for the structural perturbations
\begin{eqnarray}
	\mathcal{M}\frac{\partial^{2} \eta'_{i}}{\partial t^{2}} + \mathcal{D}\frac{\partial\eta'_{i}}{\partial t} + \mathcal{K} \eta^{0}_{i} + \mathcal{K} \eta'_{i} - \mathcal{F}^{0}_{i} - \mathcal{F}'_{i}  = 0.  \label{struc_linear}
\end{eqnarray}
Equation~\ref{struc_linear} will be completed if we can evaluate the expression for the term $\mathcal{F}'_{i}$ arising due to the linearized fluid perturbations. The two systems (fluid and structure) are coupled through the term $\mathcal{F}'_{i}$ and the velocity boundary conditions at the interface.  

\subsection{Taylor expansion based linearization of the fluid equations}
\label{taylor_linearization}
To begin with, we define a linear invertible mapping between the material points in the reference (equilibrium) configuration and the perturbed points as 
\begin{subequations}
	\begin{eqnarray}
	\mathbf{x} &=& [\mathcal{I} + \mathcal{R'}]\mathbf{x}^{0} \label{mapping_decomposition}.
	\end{eqnarray}
\end{subequations}
Here $\mathcal{I}$ is the identity matrix and $\mathcal{R}'$ can be considered to be a ``perturbation matrix" for the material points. Using such a decomposition allows us to conveniently express the inverse mapping
\begin{eqnarray}
\frac{\partial \mathbf{x}^{0}}{\partial \mathbf{x}} = \left(\frac{\partial \mathbf{x}}{\partial \mathbf{x}^{0}}\right)^{-1} = \left[\mathcal{I} + \mathcal{R'}\right]^{-1} \label{mapping_inverse} \nonumber
\end{eqnarray}
For a small deformation, we may assume $||\mathcal{R'}||\ll1$ and the matrix inverse may then be written as
\begin{eqnarray}
\left[\mathcal{I} + \mathcal{R'}\right]^{-1} = \mathcal{I} - \mathcal{R'} + \mathcal{R'}^{2} - \mathcal{R'}^{3} \ldots \nonumber
\end{eqnarray}
Retaining only the first order term of the expansion, one obtains the approximate inverse as
\begin{eqnarray}
	\frac{\partial \mathbf{x}^{0}}{\partial \mathbf{x}} = \left(\frac{\partial \mathbf{x}}{\partial \mathbf{x}^{0}}\right)^{-1} \approx \left[\mathcal{I} - \mathcal{R'}\right] \label{geometric_linearization}
\end{eqnarray}
We refer to equation~\ref{geometric_linearization} as the \emph{geometric linearization} of the problem. This allows us to conveniently reformulate all derivative terms evaluated on the perturbed grid in terms of the derivatives in the reference configuration. In addition, since $\mathcal{R}'$ is a perturbation matrix, this allows us to identify higher order terms arising due to this geometric non-linearity and later discard them when we retain only the first order terms. In what follows, all quantities of the form $\partial/\partial x_{j}^{0}$ represent the evaluation of derivatives in the reference configuration.

Next we define a perturbation field for the fluid components $(\mathbf{u}',p')$. The perturbation field is also defined on the same equilibrium grid $\mathbf{x}^{0}$ on which the base flow $(\mathbf{U}^{0},P^{0})$ has been defined. The total velocity and pressure fields on the perturbed locations are then evaluated using a superposition of the first order Taylor expansions for the base flow and the perturbation field, \textit{i.e}

\begin{subequations}
	\begin{eqnarray}
		U_{i}(x,y,z) &=& U^{0}_{i} + \left(\frac{\partial U^{0}_{i}}{\partial x^{0}_{j}}\Delta x_{j} \right)  + u'_{i} + \left(\frac{\partial u'_{i}}{\partial x^{0}_{j}}\Delta x_{j} \right), \label{taylor_vel}\\
		P(x,y,z) &=& P^{0} + \left(\frac{\partial P^{0}}{\partial x^{0}_{j}}\Delta x_{j} \right)  + p' + \left(\frac{\partial p'}{\partial x^{0}_{j}}\Delta x_{j} \right) \label{taylor_pr}.
	\end{eqnarray} 
\end{subequations}
The last term in equations~\ref{taylor_vel} and \ref{taylor_pr} is dropped since it represents a second order quantity arising due to the interaction of fluid and geometric perturbation terms. This results in the expressions for the total velocity and pressure as 
\begin{subequations}
	\begin{align}
		u^{\xi}_{i}(x,y,z) &= \left(\frac{\partial U^{0}_{i}}{\partial x^{0}_{j}}\Delta x_{j} \right), &	p^{\xi}(x,y,z) &= \left(\frac{\partial P^{0}}{\partial x^{0}_{j}}\Delta x_{j} \right), \nonumber \\	
		U_{i}(x,y,z) &= U^{0}_{i} + u^{\xi}_{i}  + u'_{i} \label{vel_taylor_expansion}, \\
		P(x,y,z) &= P^{0} + p^{\xi}  + p' \label{pr_taylor_expansion}.
	\end{align}
\end{subequations}
Where, equations~\ref{vel_taylor_expansion} and \ref{pr_taylor_expansion} are derived from equations~\ref{taylor_vel} and \ref{taylor_pr} respectively after dropping the second-order terms, and using the compact notations $u^{\xi}_{i}$ and $p^{\xi}$ to represent the first order quantities of the Taylor expansion of the base flow velocity and pressure respectively. Physically, this means that for a point moving in space, we ignore changes in the perturbation field experienced due to the motion of the point, however we include the first-order changes in the base flow field. The expressions in equation~\ref{vel_taylor_expansion} amount to a triple decomposition of the total velocity field, where the three terms $U^{0}_{i},u^{\xi}_{i}$ and $u'_{i}$ signify the base flow field, the perturbation of the base field observed by a point due to its motion, and the perturbation velocity field respectively. Additionally, associated with the motion of the material points, a velocity of the material points can also be defined such that 
\begin{eqnarray}
	w_{i} = \frac{\partial x_{i}}{\partial t} = \frac{\partial \Delta x_{i}}{\partial t}.
	\label{material_vel}
\end{eqnarray}
Taking the time derivative of equation~\ref{vel_taylor_expansion} one obtains the ALE time derivative of the total velocity along the trajectory of the material point motion
\begin{eqnarray}
	\left.\frac{\partial U_{i}}{\partial t}\right\vert_{\mathbf{w}} &=& \frac{\partial U^{0}_{i}}{\partial t} + \frac{\partial }{\partial t}\left(\frac{\partial U^{0}_{i}}{\partial x^{0}_{j}}\Delta x_{j} \right) +\frac{\partial u'_{i}}{\partial t}, \nonumber \\
	\implies\left.\frac{\partial U_{i}}{\partial t}\right\vert_{\mathbf{w}} &=& \left(w_{j}\frac{\partial U^{0}_{i}}{\partial x^{0}_{j}}\right)  +\frac{\partial u'_{i}}{\partial t}. \label{ale_derivative1}
\end{eqnarray}

Next we substitute the triple decomposition of the velocity and pressure fields into the governing equations for the fluid motion and perform the geometric linearization so that all derivative quantities are consistently evaluated based on the reference configuration. Starting with the divergence-free constraint at the perturbed configuration leads to the following set of expressions
\begin{eqnarray}
\frac{\partial U_{i}}{\partial x_{i}} &=&\left\lbrace
\begin{split}
\underbrace{\frac {\partial U^{0}_{i}}{\partial x^{0}_{i}}}_{I}
+ \underbrace{\Delta x_{j}\frac{\partial }{\partial x^{0}_{j}}\left(\frac{\partial U^{0}_{i}}{\partial x^{0}_{i}}\right)}_{II}
+\underbrace{\frac{\partial u'_{i}}{\partial x^{0}_{i}}}_{III} \\
-\underbrace{\frac{\partial u'_{i}}{\partial x^{0}_{k}}\mathcal{R'}_{ki} }_{IV}
-\underbrace{\frac{\partial }{\partial x^{0}_{k}}\left(\frac{\partial U^{0}_{i}}{\partial x^{0}_{j}}\right)\Delta x_{j}\mathcal{R'}_{ki}}_{V}
-\underbrace{\frac{\partial U^{0}_{i}}{\partial x^{0}_{j}}\frac{\partial \Delta x_{j}}{\partial x^{0}_{k}}\mathcal{R'}_{ki}}_{VI}.
\end{split}\right. \label{incompressibility_taylor_expansion}
\end{eqnarray}
Term $I$ vanishes since it represents the divergence-free constraint of the base flow in the reference configuration. Similarly, term $II$ contains the same divergence-free constraint inside the brackets and hence also vanishes. Terms $IV,V$ and $VI$ all represent terms that are second order in the perturbations quantities $\mathbf{u}',\mathbf{\Delta x}$ and $\mathcal{R'}$. Therefore, to a first-order approximation these quantities can be dropped. This leaves the final divergence constraint as 
\begin{eqnarray}
	\frac{\partial U_{i}}{\partial x_{i}} = \frac{\partial u'_{i}}{\partial x^{0}_{i}} = 0. \label{incompressibility_final}
\end{eqnarray}
Note that the derivative is evaluated in the reference configuration (but satisfies the divergence-free constraint on the perturbed locations up to a first order approximation).

In a similar manner, we may introduce the triple decomposition of the velocity and pressure field into the ALE form of the Navier--Stokes evaluated at the perturbed locations as
\begin{eqnarray}
	\begin{split}
	\left(U^{0}_{j}\frac{\partial U^{0}_{i}}{\partial x_{j}} + 
	U^{0}_{j}\frac{\partial u^{\xi}_{i}}{\partial x_{j}} + 
	u^{\xi}_{j}\frac{\partial U^{0}_{i}}{\partial x_{j}}\right) +
	\left( U^{0}_{j}\frac{\partial u'_{i}}{\partial x_{j}} + 
	u'_{j}\frac{\partial U^{0}_{i}}{\partial x_{j}}\right) 
	+\left(	\left.\frac{\partial U_{i}}{\partial t}\right\vert_{\mathbf{w}}	- w_{j}\frac{\partial U_{i}}{\partial x_{j}}	\right)
	\\=
	\left(-\frac{\partial P^{0}}{\partial x_{i}} + \frac{1}{Re}\frac{\partial^{2}U^{0}_{i}}{\partial x_{j}\partial x_{j}}\right) 
	+ \left(-\frac{\partial p^{\xi}}{\partial x_{i}} + \frac{1}{Re}\frac{\partial^{2}u^{\xi}_{i}}{\partial x_{j}\partial x_{j}}\right)
	+\left(-\frac{\partial p'}{\partial x_{i}} + \frac{1}{Re}\frac{\partial^{2}u'_{i}}{\partial x_{j}\partial x_{j}}\right).
	\end{split} \label{NS2}
\end{eqnarray}
The non-linear transport terms have already been dropped. We now substitute the ALE time derivative from equation~\ref{ale_derivative1} and introduce the geometric linearization to consistently evaluate derivatives based on the original configuration. A full expansion of all the terms can be found in Appendix~\ref{appA_fsi}. We write the final form of the Navier--Stokes obtained after the expansion and simplification of the terms and dropping all terms higher than first order:
\begin{eqnarray}
\begin{split}
\left[U^{0}_{j}\frac{\partial U^{0}_{i}}{\partial x^{0}_{j}}   +
\frac{\partial P^{0}}{\partial x^{0}_{i}}	-
\frac{1}{Re}\frac{\partial^{2} U^{0}_{i}}{\partial x^{0}_{j}\partial x^{0}_{j}}
\right] +
\Delta x_{l}\frac{\partial }{\partial x^{0}_{l}}\left[U^{0}_{j}\frac{\partial U^{0}_{i}}{\partial x^{0}_{j}}   +
\frac{\partial P^{0}}{\partial x^{0}_{i}}	-
\frac{1}{Re}\frac{\partial^{2} U^{0}_{i}}{\partial x^{0}_{j}\partial x^{0}_{j}}
\right] \\
+
\left[\frac{\partial u'_{i}}{\partial t} +
U^{0}_{j}\frac{\partial u'_{i}}{\partial x^{0}_{j}}   +
u'_{j}\frac{\partial U^{0}_{i}}{\partial x^{0}_{j}} +
\frac{\partial p'}{\partial x^{0}_{i}}	-
\frac{1}{Re}\frac{\partial^{2} u'_{i}}{\partial x^{0}_{j}\partial x^{0}_{j}}
\right] = 0.
\end{split} \label{NS3}
\end{eqnarray}
The terms in first square bracket may be identified as the steady state equation for the base flow in the reference configuration. The terms inside the second square bracket are also the steady state equations in the reference configuration. In fact, the set of terms from the first two brackets together represent the \emph{first-order Taylor expansion of the steady-state equations for the base flow at the perturbed points}. Thus all the terms in the first two brackets vanish. On hindsight the final set of expressions obtained seems rather obvious. In fact one may observe the first three terms in the expansion of the divergence-free constraint and realize that they amount to a similar expression. Terms $I$ and $II$ in equation~\ref{incompressibility_taylor_expansion} together form the first-order Taylor expansion of the base flow divergence-free constraint at the perturbed locations. Since the solution of the steady-state equations and the base field divergence is identically zero everywhere, their Taylor expansions also vanish at the perturbed locations. Thus we are left with the final linear  equations for the perturbed quantities which is also independent of the arbitrarily defined velocity of the grid motion $\mathbf{w}$:
\begin{eqnarray}
\begin{split}
\frac{\partial u'_{i}}{\partial t}
+U^{0}_{j}\frac{\partial u'_{i}}{\partial x^{0}_{j}}
+u'_{j}\frac{\partial U^{0}_{i}}{\partial x^{0}_{j}}
+\frac{\partial p'}{\partial x^{0}_{i}}
-\frac{1}{Re}\frac{\partial^{2} u'_{i}}{\partial x^{0}_{j}\partial x^{0}_{j}}
= 0.
\end{split} \label{NS4}
\end{eqnarray}

\subsection{Linearized boundary conditions}
Before proceeding to evaluate the total linearized forces on the perturbed points, we consider the global balance of fluxes for the base flow at the equilibrium and the perturbed positions. This allows us to evaluate the forces arising due to the variation of the boundary in a steady base flow. We use the conservative form of the equations, starting with the base flow state in the equilibrium configuration and evaluate the integral over the whole domain
\begin{eqnarray}
	\frac{\partial}{\partial x^{0}_{j}} (U^{0}_{i}U^{0}_{j} - \sigma^{0}_{ij}) &=& 0, \nonumber \\
	\implies \int_{\Omega^{0}}\frac{\partial}{\partial x^{0}_{j}} (U^{0}_{i}U^{0}_{j} - \sigma^{0}_{ij})d\Omega &=& 0, \nonumber \\
		\left.
		\begin{split}
			\implies \int_{\partial \Omega^{f}_{v}} (U^{0}_{i}U^{0}_{j} - \sigma^{0}_{ij})n^{0}_{j}dS + \int_{\partial \Omega^{f}_{o}} (U^{0}_{i}U^{0}_{j}	- \sigma^{0}_{ij})n^{0}_{j}dS \\
			- \int_{\Gamma^{0}}\sigma^{0}_{ij}n^{0}_{j}dS 
		\end{split}\right\rbrace &=& 0.  \label{base_mom_flux_balance}
\end{eqnarray}
Here $n^{0}_{j}dS$ represents the the local surface vector in the equilibrium configuration. For compact notation we have denoted the stresses due the fluid at equilibrium as $\sigma^{0}_{ij}$, defined as
 \begin{eqnarray}
	\sigma^{0}_{ij} = -P^{0}\delta_{ij} +\frac{1}{Re}\left(\frac{\partial U^{0}_{i}}{\partial x^{0}_{j}}  + \frac{\partial U^{0}_{j}}{\partial x^{0}_{i}} \right)  \label{base_forces}.
\end{eqnarray}
The three integrals in equation~\ref{base_mom_flux_balance} represent the momentum flux through the far-field ($\partial\Omega^{f}_{v}$), outflow ($\partial\Omega^{f}_{o}$) and FSI interface ($\Gamma^{0}$) respectively. In a similar manner the integral over the perturbed domain may be performed and without loss of generality we may assume that the outer boundaries $\partial\Omega^{f}_{v}$ and $\partial\Omega^{f}_{o}$ remain fixed, leading to an expression for the momentum flux at the perturbed surfaces as
\begin{eqnarray}
	\int_{\Omega}\frac{\partial}{\partial x_{j}} (U^{0}_{i}U^{0}_{j} + u^{\xi}_{i}U^{0}_{j} + U^{0}_{i}u^{\xi}_{j} - \sigma^{0}_{ij} - \sigma^{\xi}_{ij}) d\Omega &=& 0, \nonumber \\
	\left. 
	\begin{split}
		\implies \int_{\partial \Omega^{f}_{v}} (U^{0}_{i}U^{0}_{j} - \sigma^{0}_{ij})n^{0}_{j}dS + \int_{\partial \Omega^{f}_{o}} (U^{0}_{i}U^{0}_{j}	- \sigma^{0}_{ij})n^{0}_{j}dS \\
		- \int_{\Gamma^{0}}\sigma^{0}_{ij}n^{0}_{j}dS  - \int_{\Gamma^{0}}\sigma^{0}_{ij}n'_{j}dS - \int_{\Gamma^{0}}\sigma^{\xi}_{ij}n^{0}_{j}dS
	\end{split}
	\right\rbrace &=& 0. \label{basep_mom_flux_balance}
\end{eqnarray}
Where $n'_{j}dS$ represents the change in the surface vector due to the motion of the boundary and we have dropped the second order terms for convection $u^{\xi}_{i}u^{\xi}_{j}$ and the surface force $\sigma^{\xi}_{ij}n'_{j}$.  Equation~\ref{basep_mom_flux_balance} represents a first-order approximation of the fluxes at the perturbed boundaries. The expression $\sigma^{\xi}_{ij}$ is simply the first-order term of the Taylor expansion of the base flow-stresses, defined as
 \begin{eqnarray}
	\sigma^{\xi}_{ij} = \Delta x_{k}\frac{\partial }{\partial x^{0}_{k}}\left[-P^{0}\delta_{ij} +\frac{1}{Re}\left(\frac{\partial U^{0}_{i}}{\partial x^{0}_{j}}  + \frac{\partial U^{0}_{j}}{\partial x^{0}_{i}} \right)\right]  \label{basep_forces}.
\end{eqnarray}
Subtracting equation~\ref{base_mom_flux_balance} from \ref{basep_mom_flux_balance} one obtains
\begin{eqnarray}
	\int_{\Gamma^{0}}\sigma^{0}_{ij}n'_{j}dS + \int_{\Gamma^{0}}\sigma^{\xi}_{ij}n^{0}_{j}dS &=& 0. \label{basep_zero_force}
\end{eqnarray}
The first term represents the variation of the base flow forces due to the change in surface normal and the second term represents the variation due to the change in boundary position. To a first-order approximation these terms balance to zero. Note that no assumption has been made on the type of deformation at the boundary and the condition holds for any arbitrary deformation of the boundary. We note that these are precisely the terms that arise in \cite{fanion00,fernandez03i} and \cite{fernandez03ii} that have been termed as ``added-stiffness". To a first-order approximation they simply sum up to zero and play no role in the linear dynamics. One may now evaluate the total linearized forces arising from equation~\ref{NS3} on the perturbed FSI interface and obtain the linearized boundary conditions for the stress balance (equation~\ref{struc_linear}) as
\begin{eqnarray}
	\left.
	 \begin{split}
	 	\mathcal{M}\frac{\partial^{2} \eta'_{i}}{\partial t^{2}} + \mathcal{D}\frac{\partial\eta'_{i}}{\partial t} + \mathcal{K} (\eta^{0}_{i} + \eta'_{i}) +\int_{\Gamma^{0}}\sigma^{0}_{ij}n^{0}_{j}dS \\  
		 + \int_{\Gamma^{0}}\sigma^{0}_{ij}n'_{j}dS
		 + \int_{\Gamma^{0}}\sigma^{\xi}_{ij}n^{0}_{j}dS + \int_{\Gamma^{0}}\sigma'_{ij}n^{0}_{j}dS
	 \end{split}\right\rbrace = 0.  \nonumber \\
	\implies \mathcal{M}\frac{\partial^{2} \eta'_{i}}{\partial t^{2}} + \mathcal{D}\frac{\partial\eta'_{i}}{\partial t} + \mathcal{K}\eta'_{i} +  \int_{\Gamma^{0}}\sigma'_{ij}n^{0}_{j}dS = 0, \label{stress_continuity_lin}
\end{eqnarray}
where $\sigma'_{ij}$ is the fluid stress due to the perturbation field $(\mathbf{u}',p')$ defined as
 \begin{eqnarray}
	\sigma'_{ij} = -p'\delta_{ij} +\frac{1}{Re}\left(\frac{\partial u'_{i}}{\partial x^{0}_{j}}  + \frac{\partial u'_{j}}{\partial x^{0}_{i}} \right)  \label{pert_forces}.
\end{eqnarray}

The velocity continuity boundary conditions on the moving interface $\Gamma(t)$ simply become
\begin{eqnarray}
	U_{i} = v^{\Gamma}_{i}& & on\ \Gamma(t) \nonumber \\
	\implies \Delta x^{\Gamma}_{k}\frac{\partial U^{0}_{i}}{\partial x_{k}} + u'_{i} = v^{\Gamma}_{i}& &  \label{fsi_bc}
\end{eqnarray}
Thus the perturbation velocities must account for the Taylor expansion term of the base flow at the perturbed boundary.  \cite{fanion00,fernandez03i,fernandez03ii} refer to this as the transpiration boundary condition.
  
 The linear FSI problem for the perturbations is turned into a standard linear perturbation problem with the exception of the modified boundary conditions (transpiration instead of no-slip). The additional forces at the perturbed boundary are simply due to the perturbation field $(\mathbf{u}',p')$ and no ``added stiffness" terms arise in a linear approximation. One may immediately notice that the final form is a generalization of the form derived by \cite{benjamin59,benjamin60} and \cite{landahl62} for parallel flow cases. We make a note that while in the current work the structural equations comprise of rigid-body motion equations, the linearization of the fluid equations did not invoke any such assumption. For a more general FSI problem, one would only need to focus on the linearization of the structural equations and the fluid linearization remains unchanged. In the next section several examples are considered for rigid-body motion in symmetric and asymmetric configurations to show the validity of the current approach.

\section{Linear Instability Results}
\label{fsi_instability_results}
\subsection{Numerical method}
Numerical tests are performed to validate the linear equations for FSI derived in the previous section. The cases considered are an oscillating cylinder with a spring-damper action, oscillating and rotating ellipse initially held at an angle to the flow, and rotating cylinder-splitter body.  All computations were performed using a high-order spectral-element method (SEM) code \citep{nek5000}. The code uses $n^{th}$ order Lagrange interpolants at Gauss-Lobatto-Legendre (GLL) points for the representation of the velocity and $(n-2)^{th}$ order interpolants at Gauss-Legendre (GL) points for the representation of the pressure in a $\varmathbb{P}_{n}-\varmathbb{P}_{n-2}$ formulation \citep{maday89}. A third-order backward difference is used for the time integration of the equations. The viscous terms are evaluated implicitly while extrapolation is used for the non-linear terms. Over-integration is used for a consistent evaluation of the non-linear terms and a relaxation term based on the high-pass filtered velocity field (HPF-RT) is used to stabilize the method \citep{negi17}. The stabilization method is based on the ADM-RT method used for the large-eddy simulation (LES) of transitional flows \citep{schlatter04,schlatter06} and has been validated with channel flows and flow over wings in \cite{negi18}. Moving boundaries are treated using the ALE formulation \citep{ho90,ho91} and the fluid and structural equations are coupled using the Green's function decomposition approach whereby the geometrical non-linearity is evaluated explicitly via extrapolation, while the added-mass effects are treated implicitly \citep{fischer17}. In addition we have also implemented a fully implicit fixed-point nonlinear iteration method \citep{kuttler08} for coupling of the fluid and structural equations. Both methods give nearly identical results in all tested cases. All quantities are non-dimensionalized using the fluid density $\rho$, free-stream velocity $U_{\infty}$, and the cylinder/ellipse diameter $D$. 

\subsection{Oscillating cylinder at sub-critical Reynolds Numbers}
\begin{figure}
	\centering
	\includegraphics[width=0.60\textwidth]{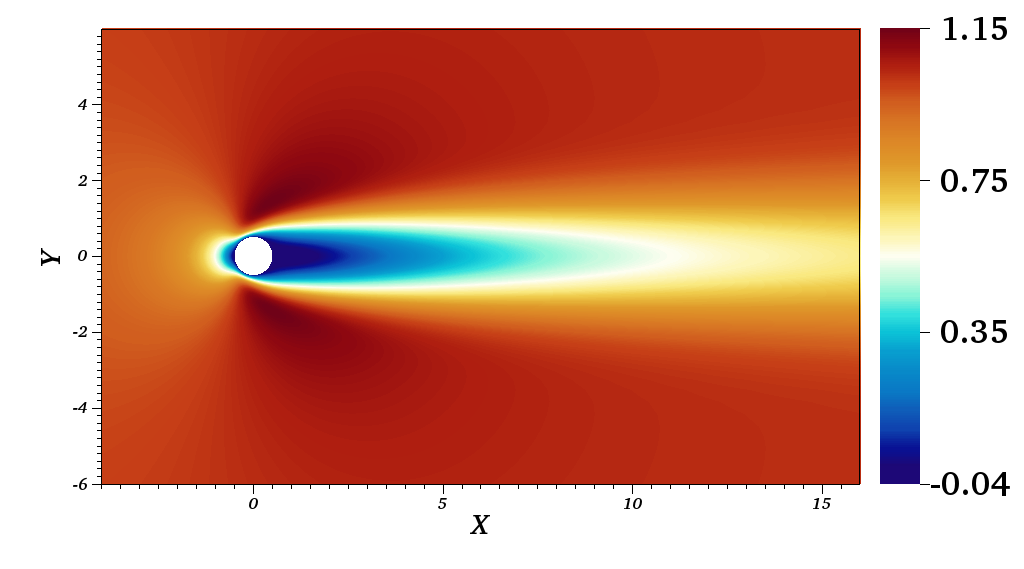}
	\caption{Streamwise velocity of the base flow state for a cylinder in cross-flow at $Re=23.512$}
	\label{fig:re23_base}
	\vspace{-0.5cm}
\end{figure}

For the first case we investigate a 2D circular cylinder free to oscillate in the cross-stream direction subject to the action of a spring-damper system. The Reynolds number of the flow based on the cylinder diameter is $Re=23.512$. The inlet of the computational domain is $25$ diameters upstream of the cylinder while the outflow boundary is $60$ diameters downstream of the cylinder. The lateral boundaries are 50 diameters away on either side. A uniform inflow Dirichlet boundary condition is applied on the inflow and the lateral boundaries while the stress-free boundary condition is applied on the outflow boundary. The computational domain is discretized using $2284$ spectral-elements which are further discretized into $10\times10$ GLL points for a $9^{th}$ order polynomial representation for the velocity. This amounts to a total of $228400$ degrees of freedom in the domain. The base flow for all cases was calculated by keeping structure fixed at its initial position. At the $Re=23.512$ the flow with a fixed cylinder is linearly stable. The convergence to steady state was accelerated by using BoostConv \citep{citro17}. Figure~\ref{fig:re23_base} shows the calculated base flow state (streamwise velocity). 

We denote $\eta$ as the vertical position of the cylinder and the structural equation is modeled using a spring-mass-damper system for the vertical displacement of the cylinder. Following the earlier notation we denote the fluid domain as $\Omega^{f}$, the structural domain as $\Omega^{s}$ and the FSI interface as $\Gamma$. In all our cases, the structural equation contains a second order derivative in time. The order of the time derivative is reduced by introducing a variable substitution for the velocity $\varphi$ of the structure.
The combined non-linear equations for the FSI system can be written as
\begin{subequations}
	\begin{align}
	\left.\frac{\partial U_{i}}{\partial t}\right\vert_{\mathbf{W^{g}}} + (U_{j} -W^{g}_{j})\frac{\partial U_{i}}{\partial x_{j}} +\frac{\partial P}{\partial x_{i}}  - \frac{1}{Re}\frac{\partial^{2} U_{i}}{\partial x_{j}\partial x_{j}}   =& 0, & \text{in } \Omega^{f},\label{NS_cossu}\\
	\frac{\partial U_{i}}{\partial x_{i}} =  &0,  &\text{in } \Omega^{f}, \label{incompressibility_cossu} \\
	\mathcal{M}\frac{d \varphi}{d t} + \mathcal{D}\varphi + \mathcal{K}\eta - \mathcal{F}_{2} =& 0, &\text{for } \Omega^{s} \label{struc_cossu}. \\
		\frac{\partial \eta}{\partial t}  - \varphi =& 0, &\text{for } \Omega^{s}, \label{struc_cossu2} \\
	U_{1} = & 0, & \text{on } \Gamma, \label{fsi_bc_cossu1} \\
	U_{2} -\varphi = & 0, & \text{on } \Gamma, \label{fsi_bc_cossu2} \\
	\mathcal{F}_{i} - \oint_{\Gamma} \left[p\delta_{ij} -\frac{1}{Re}\left(\frac{\partial u_{i}}{\partial x_{j}}  + \frac{\partial u_{j}}{\partial x_{i}} \right)   \right]n_{j}\partial\Omega  = &0, & \text{fluid forces on } \Gamma.
	 \label{fluid_forces_cossu}
	\end{align}
\end{subequations}
The parameters for the structural system are specified in table~\ref{tab:struc_param}. The mass $\mathcal{M}$ corresponds to a density ratio of $7$ (solid to fluid), and the damping constant $\mathcal{D}$ is set to $0.754\%$ of the critical damping.  $\omega_{n} = \sqrt{\mathcal{K}/\mathcal{M}}$, represents the undamped natural frequency of the spring-mass system, $\lambda_{s}=-\omega_{n}(\zeta \pm i\sqrt{1-\zeta^{2}})$, represents the damped structural eigenvalue of the spring-mass-damper system, with $\zeta=\mathcal{D}/(2\sqrt{\mathcal{K}\mathcal{M} })$ being the damping ratio.
\renewcommand{\arraystretch}{1.5}
\begin{table}
	\centering
	\begin{tabular}{x{1.5cm} x{1.5cm} x{2cm} x{1.5cm} x{3cm}}
		\toprule
		$\mathcal{M}$ & $\mathcal{K}$ & $\mathcal{D}$ & $\omega_{n}$ &  $\lambda_{s}$ \\
		$5.4977$	   	     & $3.4802$ 	   &  $6.597\times 10^{-2}$ 	& $0.7956$ & $-0.006\pm0.7956i$\\
		\bottomrule
	\end{tabular}
	\capspace
	\caption{Structural parameters for a 2D cylinder oscillating in the cross-stream direction.} \label{tab:struc_param}
\end{table}
Following the Taylor expansion based triple decomposition of the total fluid velocity and pressure, the linearized system of equations governing the fluid and structural perturbations $(\mathbf{u}',p',\eta',\varphi')$ can be written as
\begin{subequations}
	\label{cylinder_lin_NS}
	\begin{align}
	\frac{\partial u'_{i}}{\partial t} + U^{0}_{j}\frac{\partial u'_{i}}{\partial x^{0}_{j}} + u'_{j}\frac{\partial U^{0}_{i}}{\partial x^{0}_{j}} +\frac{\partial p'}{\partial x^{0}_{i}}  - \frac{1}{Re}\frac{\partial^{2} u'_{i}}{\partial x^{0}_{j}\partial x^{0}_{j}}   =& 0, & \text{in } \Omega^{f} \label{NS_cossu_lin},\\
	\frac{\partial u'_{i}}{\partial x^{0}_{i}} =  &0,  &\text{in } \Omega^{f}, \label{incompressibility_cossu_lin} \\
	u'_{i} =& 0, &\text{on } \partial\Omega_{v} \label{dir_dirichlet}, \\
	\sigma'_{ij}n^{0}_{j} =& 0, &\text{on } \partial\Omega_{o} \label{dir_outflow}, \\
	\mathcal{M}\frac{d \varphi'}{d t} + \mathcal{D}\varphi' + \mathcal{K}\eta' - \mathcal{F}'_{2} =& 0, &\text{for } \Omega^{s} \label{struc_cossu_lin}, \\
	\frac{\partial \eta'}{\partial t}  - \varphi' =& 0, &\text{for } \Omega^{s} \label{struc_cossu_lin2}, \\
	u'_{1} + \eta'\frac{\partial U^{0}_{1}}{\partial x^{0}_{2}} = & 0, & \text{on } \Gamma, \label{fsi_bc_cossu_lin1} \\
	u'_{2} + \eta'\frac{\partial U^{0}_{2}}{\partial x^{0}_{2}} - \varphi' = & 0, & \text{on } \Gamma, \label{fsi_bc_cossu_lin2} \\
	\mathcal{F}'_{i} - \oint_{\Gamma^{0}}\left[p'\delta_{ij} -\frac{1}{Re}\left(\frac{\partial u'_{i}}{\partial x^{0}_{j}}  + \frac{\partial u'_{j}}{\partial x^{0}_{i}} \right)\right]n^{0}_{j}\partial\Omega  = &0, & \text{fluid forces on } \Gamma. \label{fluid_forces_cossu_lin}
	\end{align}
\end{subequations}
Note that even though the cylinder is free to move only in the vertical direction, the streamwise perturbation velocity $u'_{1}$ is non-zero at the cylinder surface due to the Taylor expansion term of the base flow. Additionally, when evaluating the fluid forces in equation~\ref{fluid_forces_cossu_lin}, the direction of the normal $n^{0}_{j}$, is based on the equilibrium position of the FSI interface.

Before calculating the spectra of linearized problem we compare the evolution of the linear and non-linear flow cases when the base flow is perturbed by identical small-amplitude perturbations. If the perturbation amplitude is small then non-linear effects will be negligible and one can expect the linear and non-linear evolution to be the same. The solutions would eventually diverge due to amplitude saturation in the non-linear case. Performing this comparison has another advantage in this particular case. Given the low Reynolds number of the case, we expect the dynamics of be governed by a single least damped global mode. Therefore by tracking the growth in perturbation amplitude through the linear regime, one can determine both the frequency and the growth rate of the unstable mode from the non-linear simulations. This allows us to validate \emph{both the frequency and the growth rate} obtained from the linearized equations. 

\begin{figure}
	\centering
	\begin{subfigure}{0.48\textwidth}
		\includegraphics[width=1.0\textwidth]{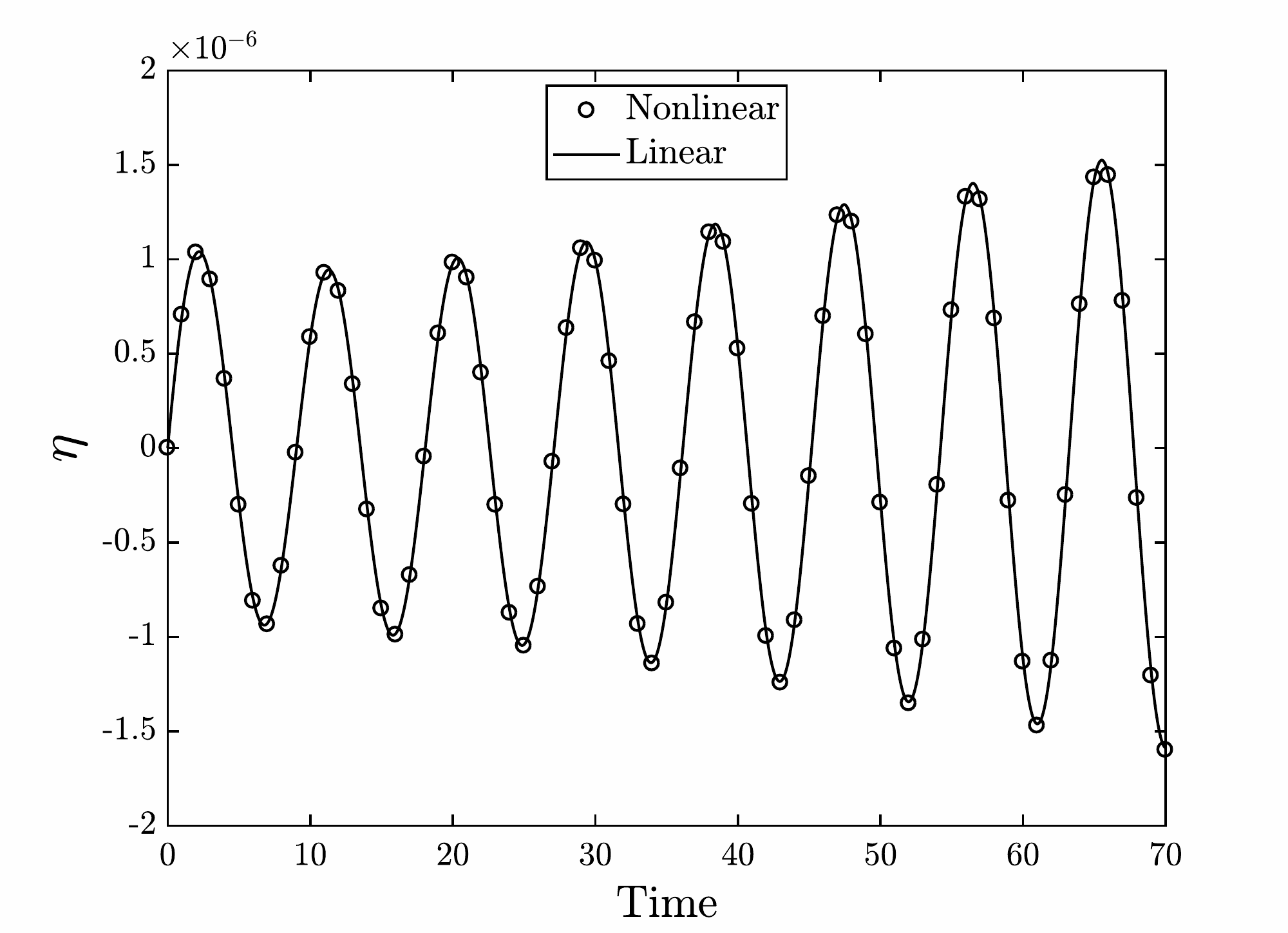}
		\caption{}
		\label{fig:re23_evol}		
	\end{subfigure}
	\begin{subfigure}{0.48\textwidth}
		\includegraphics[width=1.0\textwidth]{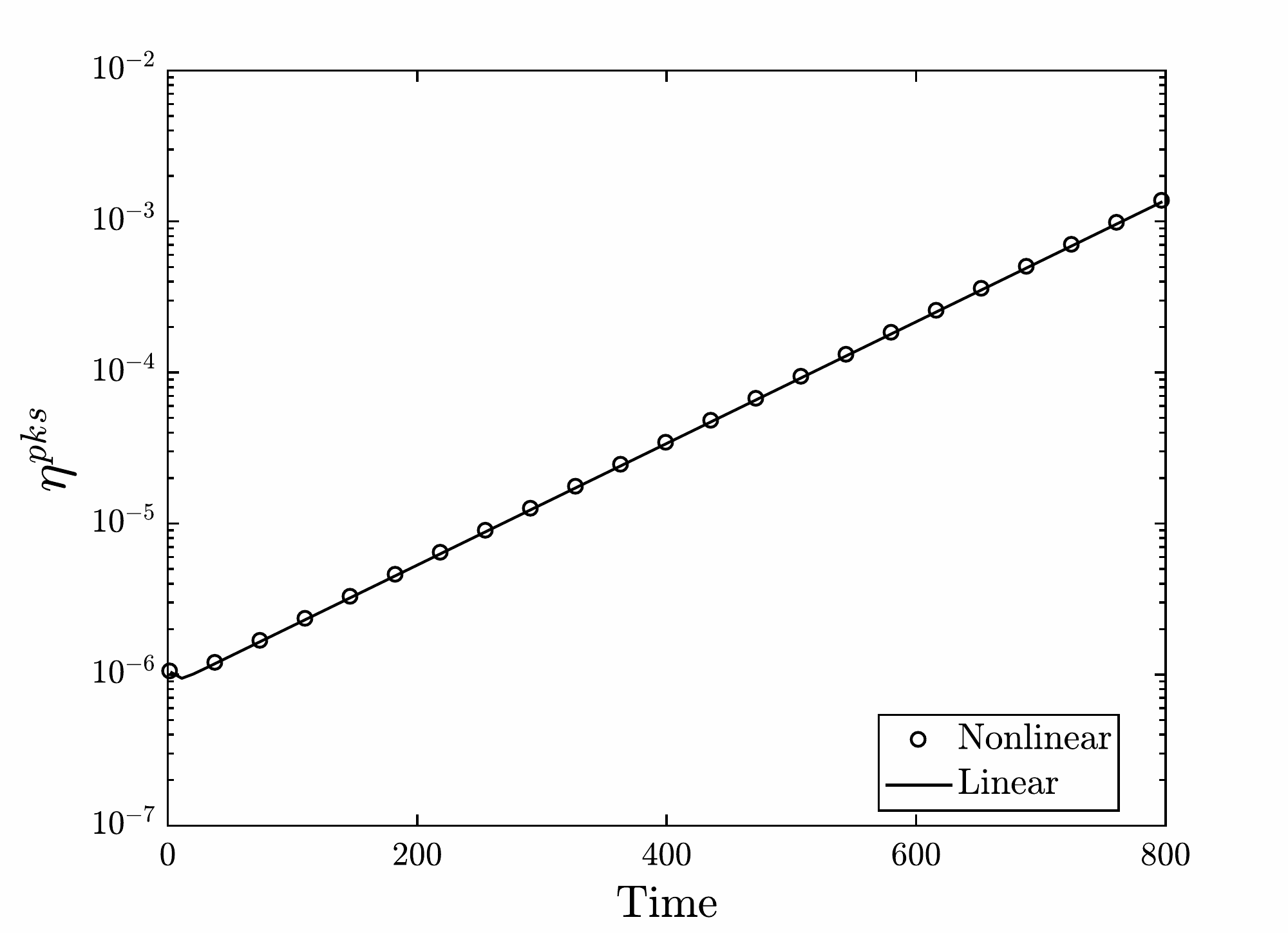}
		\caption{}
		\label{fig:re23_pks}
    \end{subfigure}
	\caption{Comparison of linear and non-linear evolution of cylinder position $\eta$ for identical small amplitude disturbance. Figure (a) shows the time evolution of $\eta$ in the first few cycles of the oscillation. Figure (b) shows the evolution of peak amplitudes of the oscillation in a semi-log scale.}
\end{figure}

The base flow is disturbed by pseudo-random perturbations of order $O(10^{-6})$ and the flow evolution is tracked. Figure~\ref{fig:re23_evol} shows the variation of $\eta$ with time during the initial stages of the evolution. Both the linear and non-linear results fall on top of each other providing the first evidence of the validity of the derived linear equations. Note that the non-linear simulations have been performed using the ALE framework including the mesh movement, while there is no mesh motion in the linear equations which have been derived to be independent of the grid velocity $\mathbf{W^{g}}$. The time evolution of $\eta$ clearly indicates a single growing mode in the flow. By tracking the peak amplitudes of the oscillations, denoted as $\eta^{pks}$, one can determine the the growth rate and the frequency of the unstable mode.  Figure~\ref{fig:re23_pks} shows the time evolution of $\eta^{pks}$ in a semi-log plot. After an initial transient phase both the growth rate and angular frequency of the disturbances stabilize to a constant value and the $\eta^{pks}$ plot traces a straight line in the semi-log plot, signifying exponential growth in time.

\begin{figure}
	\centering
	\includegraphics[width=0.60\textwidth]{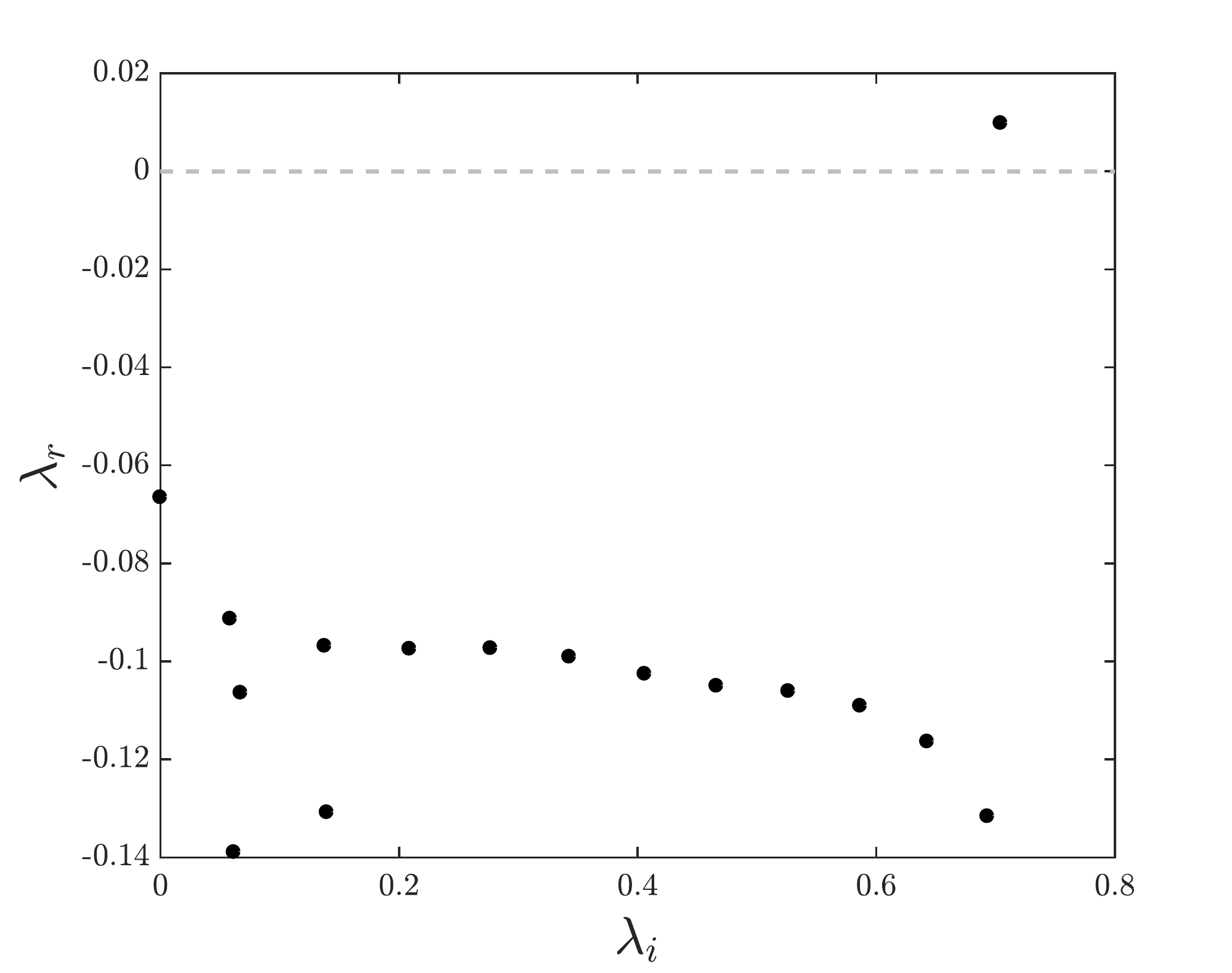}
	\caption{One-sided eigenspectra for a cylinder in cross-flow at $Re=23.512$. }
	\label{fig:re23_spectra}
\end{figure}

\begin{figure}
	\centering
	\includegraphics[width=0.49\textwidth]{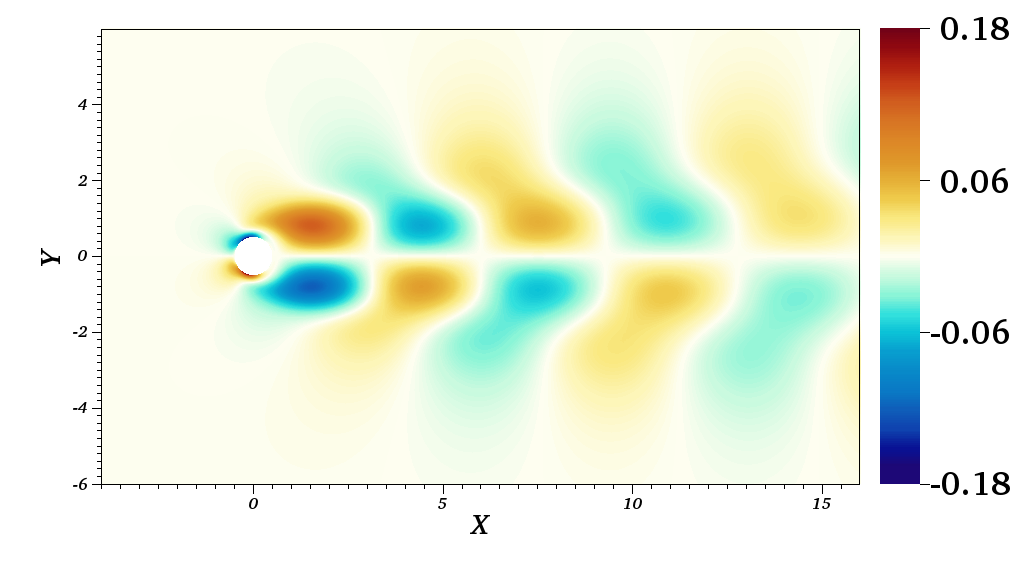}
	\includegraphics[width=0.49\textwidth]{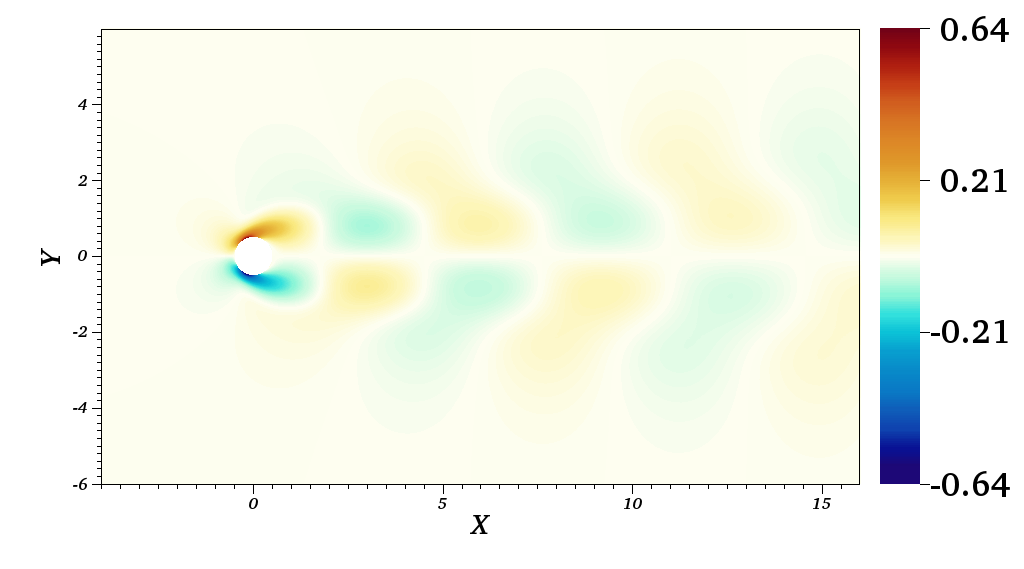}\hfill
	\caption{(Left) Real and (right) Imaginary parts of the streamwise velocity component of the eigenvector corresponding to the unstable eigenvalue for $Re=23.512$.}
	\label{fig:re23_eigenvector}
\end{figure}

Finally, we introduce the ansatz $(\mathbf{u}',p',\eta',\varphi') = (\hat{\mathbf{u}},\hat{p},\hat{\eta},\hat{\varphi})e^{\lambda t}$ to reduce the linear equations into an eigenvalue problem for the angular frequency $\lambda$. The system is solved by estimating the eigenvalues of the time-stepping operator. The method was first used by \cite{eriksson85} and has been used in several previous works by \cite{barkley02,bagheri09,rahkola17} \textit{etc.} The eigenvalue estimation is done using the implicitly restarted Arnoldi method \citep{sorensen92} implemented in the open source software package ARPACK \citep{lehoucq98}.  Figure~\ref{fig:re23_spectra} shows the one-sided eigenspectra obtained with a single unstable mode. $\lambda_{r}$ represents the real part of the eigenvalue (growth rate) and $\lambda_{i}$ represents the imaginary part of the eigenvalue (angular frequency). 
In table~\ref{tab:re23_unstable} we report the comparison of estimates obtained through the non-linear simulations, linear simulations and the Arnoldi method. For both the linear and non-linear simulations the initial transient period of 10 oscillation cycles was discarded  and the estimates are evaluated from the time period when both the growth rate and frequency have stabilized. All three methods have a very good agreement with each other, with the relative difference in the growth rate being less than $0.1\%$. The streamwise velocity component of the eigenvector for the unstable frequency is shown in figure~\ref{fig:re23_eigenvector}.

\begin{table}
	\centering
	\begin{tabular}{x{2.0cm}x{2.8cm}x{2.8cm}x{2.8cm}x{2.8cm} }		
		Case  						& Non-linear &  Linear & Arnoldi \\
		\toprule
		Oscillation ($Re=23.512$)   	    &  $9.86\times10^{-3}\pm0.704i$ & $9.85\times10^{-3}\pm0.704i$ & $9.86\times10^{-3} \pm 0.704i$\\
		\bottomrule
	\end{tabular}
	\capspace
	\caption{Unstable eigenvalue estimates obtained from different methods for an oscillating cylinder at $Re=23.512$.} \label{tab:re23_unstable}
\end{table}

Thus we find the flow is unstable for an oscillating cylinder at nearly half the critical Reynolds number for the stationary case. \cite{cossu00} also found instability at the same Reynolds number for a cylinder oscillating in the cross-flow direction, albeit for slightly different structural parameters. 

\begin{figure}
	\centering
	\includegraphics[width=0.60\textwidth]{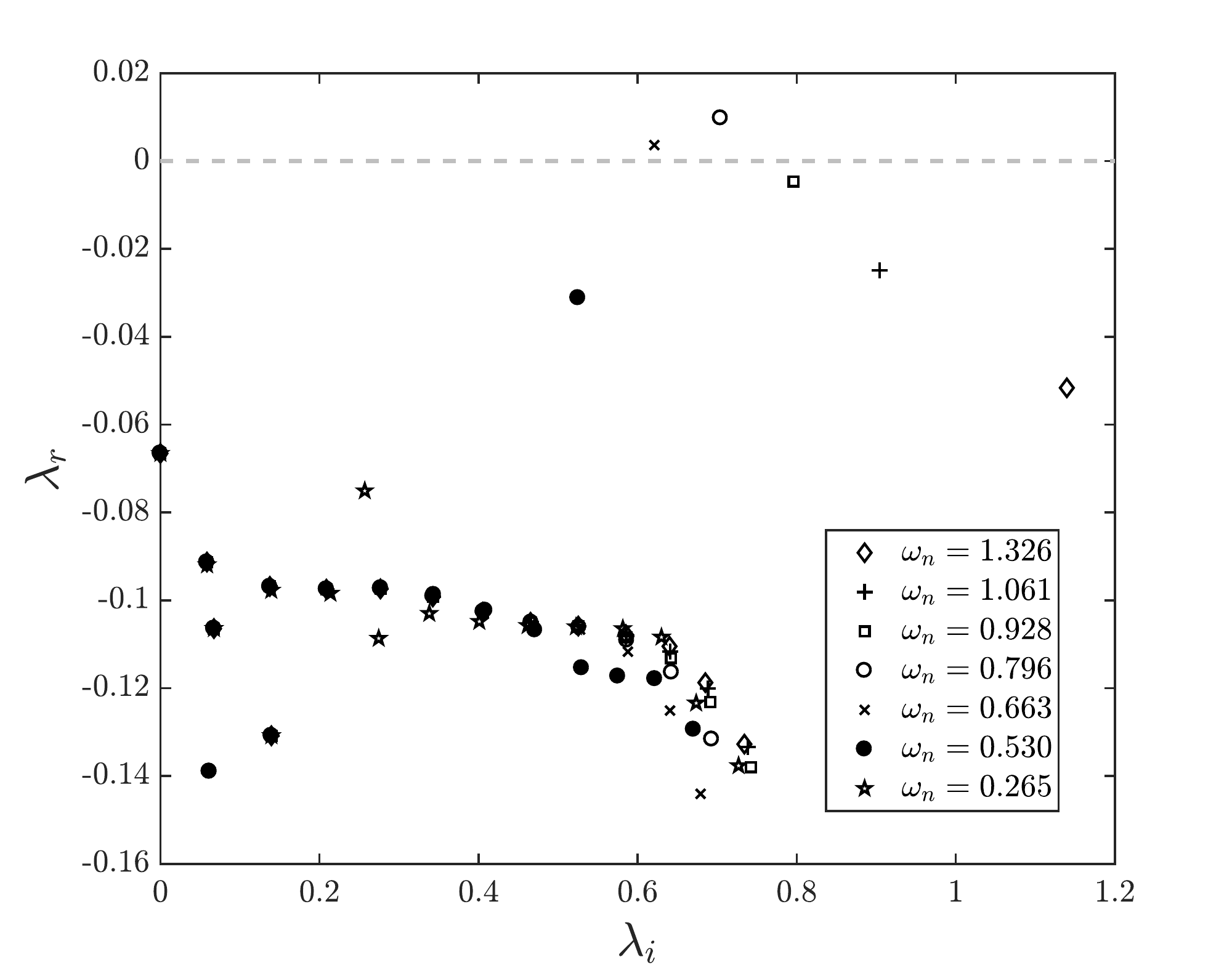}
	\caption{One-sided eigenspectra for a cylinder in cross-flow at $Re=23.512$ for varying undamped frequency, $\omega_{n}$, of the spring-mass system.}
	\label{fig:re23_spectra_k}
\end{figure}

The undamped natural frequency of the structural system $\omega_{n}$ is varied parametrically to investigate the changes in instability while keeping the density ratio constant. The damping constant is also varied such that the damping is always equal to $0.754\%$ of the critical damping of the spring-mass-damper system. The one-sided spectra of the various cases is shown in figure~\ref{fig:re23_spectra_k}. The instability of the system arises for a narrow range of $\omega_{n}$ with the peak of the instability centered around $\omega_{n}\approx0.796$. The system  rapidly becomes stable again as the structural frequency is varied away from the peak value. In the range where the system is unstable, the unstable frequency of the combined FSI system is close to the undamped frequency $\omega_{n}$. Much of the low frequency spectra remains unaffected by the variation of $\omega_{n}$. We note that some of the results are in contrast with some of the findings of \cite{cossu00} who performed global stability of the oscillating cylinder at the same Reynolds number and density ratio. In their study, the authors find a low frequency unstable mode with eigenvalue $\lambda = 1.371\pm 0.194i$ for a structural eigenvalue of $\lambda_{s}=-0.01 \pm 1.326i$. The flow case marked with diamonds in figure~\ref{fig:re23_spectra_k} corresponds to the same case investigated by \cite{cossu00}. While we find the system to be unstable at the sub-critical Reynolds number of $Re=23.512$, we do not find the instability for the same structural parameters. Unlike \cite{cossu00}, we also do not find the existence of a low frequency unstable mode within the investigated structural parameters. In all our investigated cases, the effect of the structure on the spectrum remains confined close to the angular frequency of the spring-mass-damper system. (Note that the non-dimensionalization of length in \cite{cossu00} is with respect to the radius of the cylinder while it is with respect to the diameter in the current study. Hence the respective angular frequencies are doubled in the current study). 

\begin{figure}
	\centering
	\begin{subfigure}{0.48\textwidth}
		\includegraphics[width=1.0\textwidth]{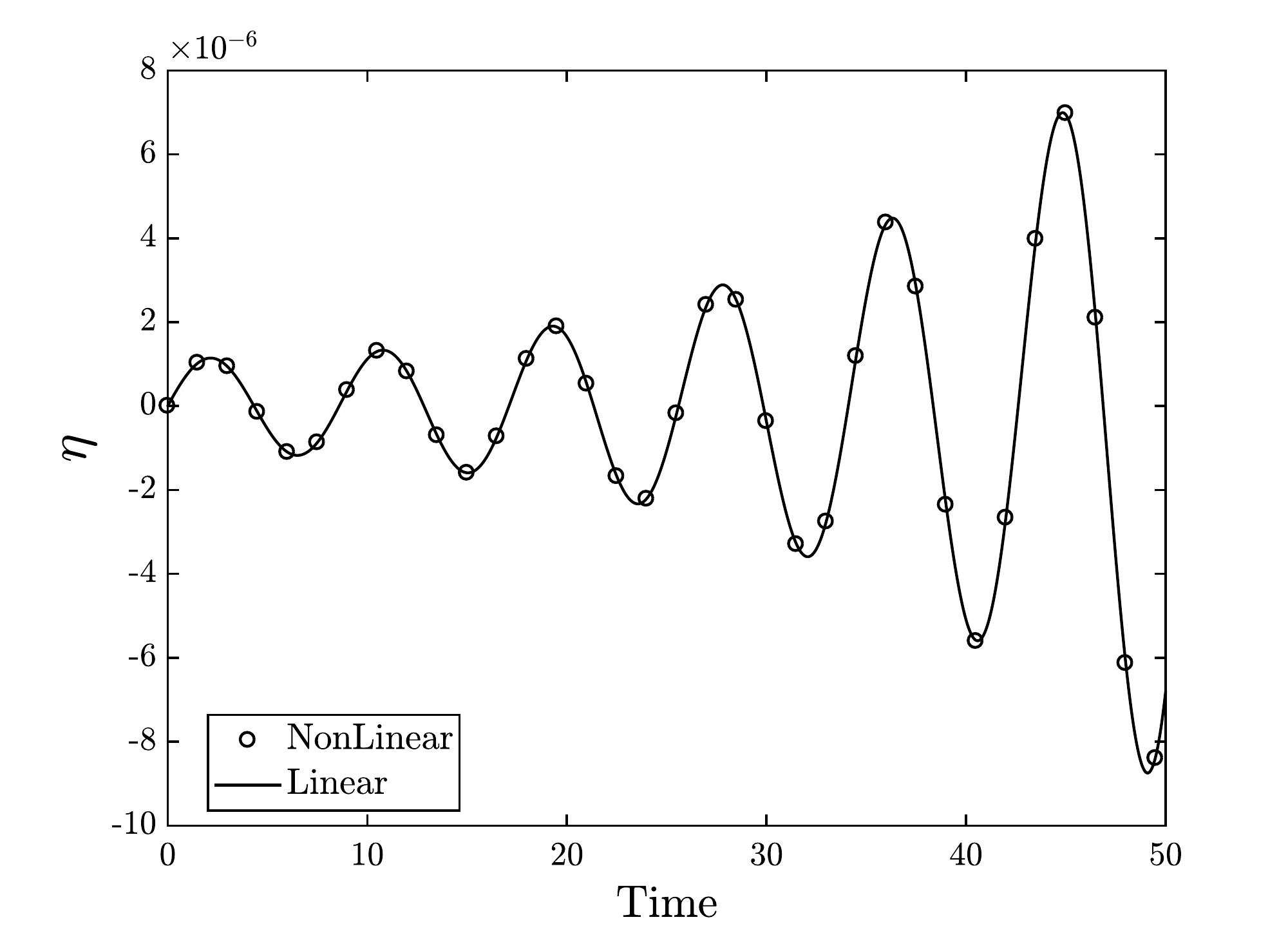}
		\caption{}
		\label{fig:navrose40_evol}		
	\end{subfigure}
	\begin{subfigure}{0.48\textwidth}
		\includegraphics[width=1.0\textwidth]{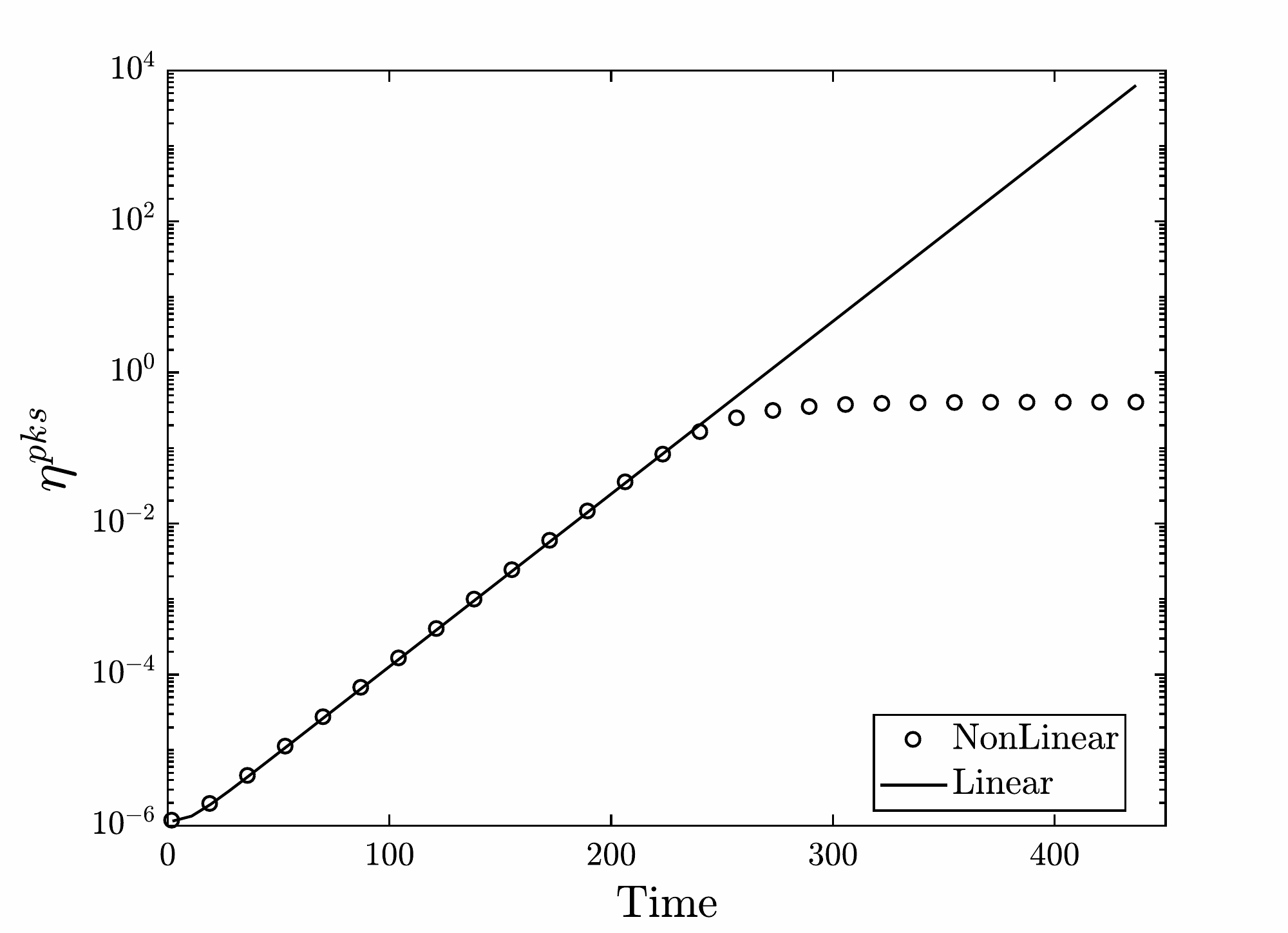}
		\caption{}
		\label{fig:navrose40_pks}
	\end{subfigure}
	\caption{Comparison of linear and non-linear evolution of cylinder position $\eta$ at $Re=40$ and $U^{*}=8$. (a) Time evolution of $\eta$ in the first few cycles of the oscillation. (b) Evolution of peak amplitudes in a semi-log scale.}
\end{figure}

We consider another sub-critical case at $Re=40$ which has been investigated in \cite{navrose16}. In this case the authors consider an oscillating cylinder in cross-flow, with a mass ratio of 10 and without any structural damping ($\mathcal{D}=0$) . The results are reported for the variation of the reduced velocity $U^{*}$ which is defined as $U^{*}=U_{\infty}/(f_{n}D)$, where $f_{n}=\omega_{n}/(2\pi)$ is the natural frequency of the spring-mass system. We investigate the case with $U^{*}=8$, again comparing evolution of small-amplitude perturbations in the linear and the non-linear cases. The initial cycles and the evolution of peak amplitude is shown in figure~\ref{fig:navrose40_evol} and \ref{fig:navrose40_pks}. The estimated eigenvalues from the linear, non-linear and Arnoldi method are reported in table~\ref{tab:navrose40_unstable}.  The estimates match to within $0.1\%$ accuracy. For the non-linear simulations the amplitude of oscillations saturates at $y/D=0.4$. This compares well with the saturation amplitude reported in \cite{navrose16} for this case.

\begin{table}
	\centering
	\begin{tabular}{x{2.0cm}x{2.8cm}x{2.8cm}x{2.8cm}x{2.8cm} }
		Case  						& Non-linear &  Linear & Arnoldi \\
		\toprule
		Oscillation ($Re=40$)   	    &  $5.26\times10^{-2}\pm0.738i$ & $5.26\times10^{-2}\pm0.738i$ & $5.26\times10^{-2} \pm 0.738i$ \\
		\bottomrule
	\end{tabular}
	\capspace
	\caption{Unstable eigenvalue estimates obtained from different methods for an oscillating cylinder at $Re=40$ and $U^{*}=8$.} \label{tab:navrose40_unstable}
\end{table}

\begin{figure}
	\centering
	\begin{subfigure}{0.48\textwidth}
		\includegraphics[width=1.0\textwidth]{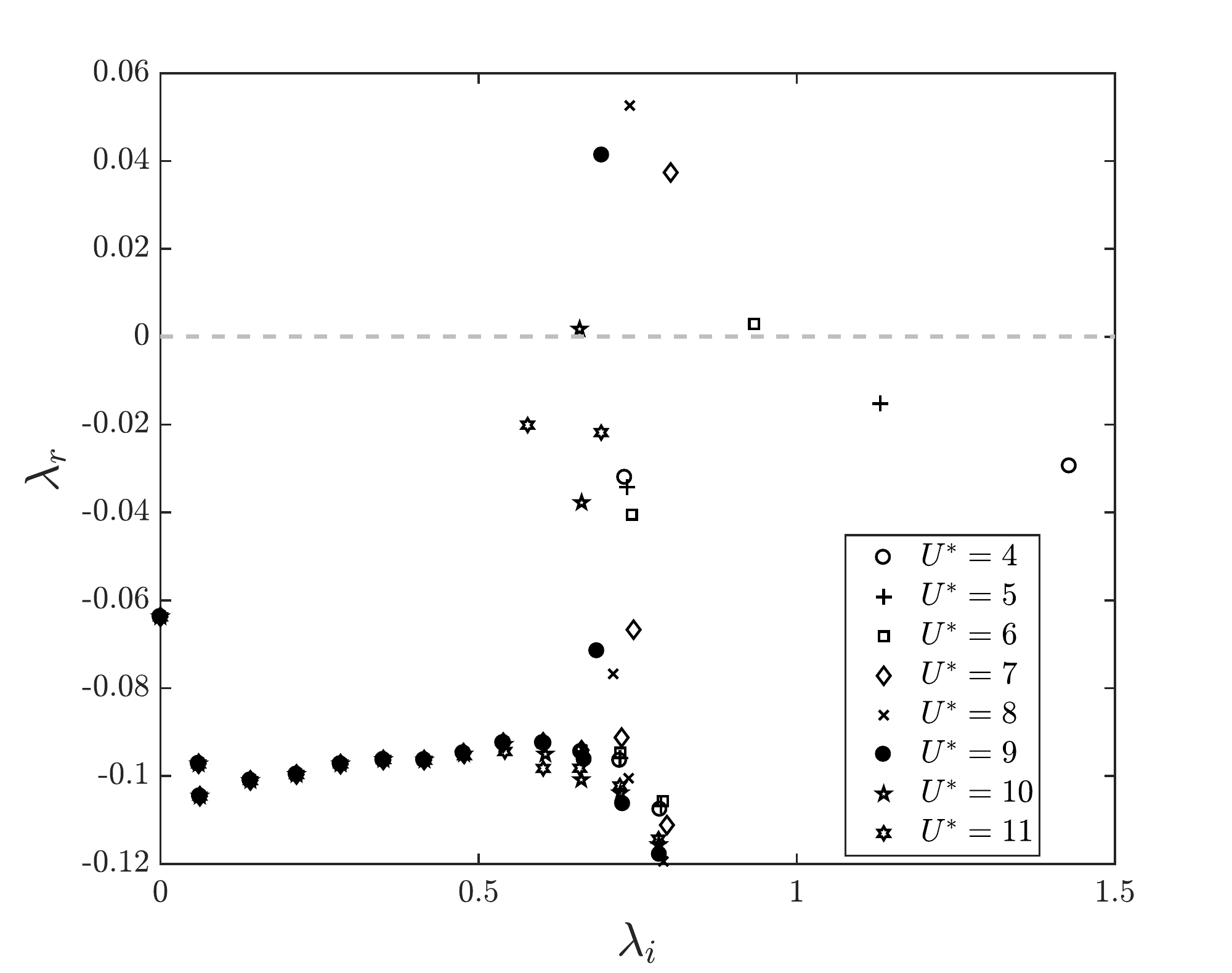}
		\caption{}
		\label{fig:navrose40_spectra}		
	\end{subfigure}
	\begin{subfigure}{0.48\textwidth}
		\includegraphics[width=1.0\textwidth]{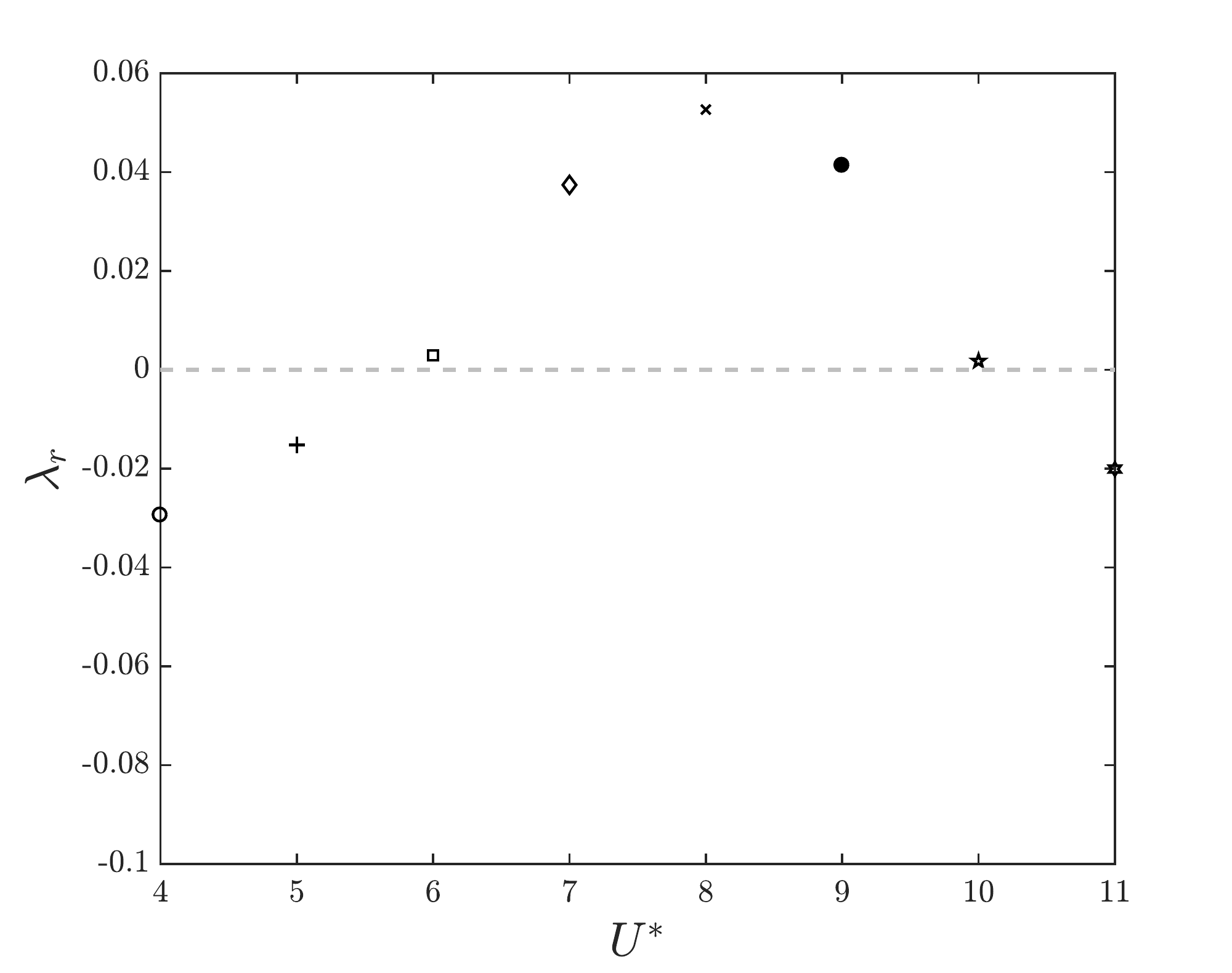}
		\caption{}
		\label{fig:navrose40_growth}
	\end{subfigure}
	\caption{(a) One-sided spectra for an oscillating cylinder at $Re=40$ for varying reduced velocities $U^{*}$. (b) Variation of the growth rate with reduced velocity $U^{*}$.}
\end{figure}

The variation of stability characteristics with varying reduced velocities is also investigated through the evaluation of the eigenvalue spectra for several different cases. The variation of spectra for different parameters is shown in figure~\ref{fig:navrose40_spectra}. Similar to the trend seen for the $Re=23.512$ case, the flow exhibits unstable eigenvalues within a narrow band of frequencies. The variation of  growth rate with the reduced velocities is shown in figure~\ref{fig:navrose40_growth}. The growth rates for $U^{*}=6$ and $U^{*}=10$ lie just above the stability threshold. This compares very well with the results of \cite{navrose16} who report the unstable range as $5.9<U^{*}<10.1$ as the parameter range of instability for the oscillating cylinder at $Re=40$. The peak growth rates occur at $U^{*}\approx8.0$ in \cite{navrose16}, which is also the case in the current work. 

\subsection{Asymmetric flow case of a rotated ellipse}
\begin{figure}
	\centering
	\includegraphics[width=0.70\textwidth]{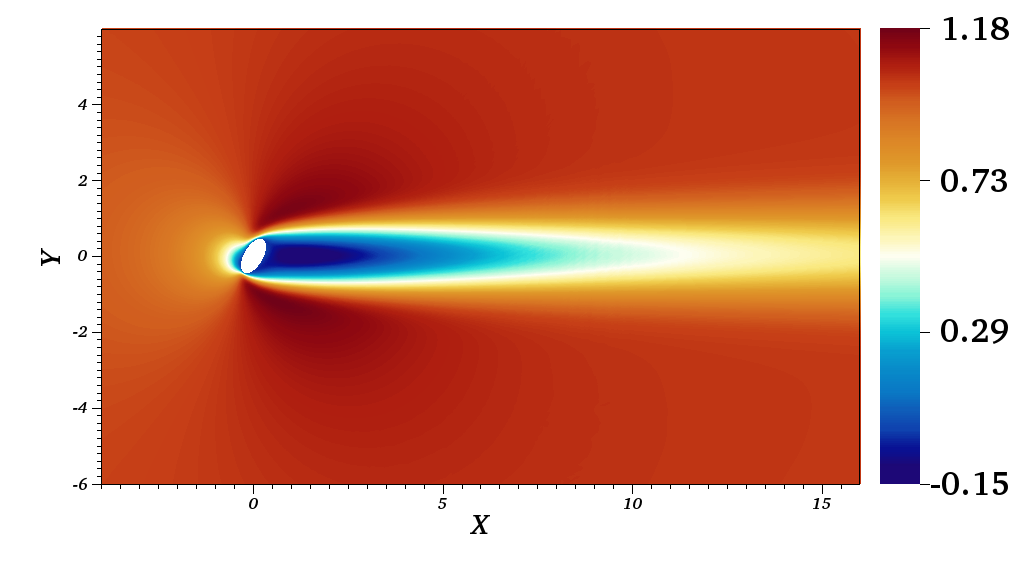}
	\caption{Streamwise velocity of the base flow state for a rotated ellipse at $Re=50$.}
	\label{fig:ellipse_base}
\end{figure}

The previous subsection investigated the flow around an oscillating circular cylinder through the linear stability analysis. The base flow for these cases exhibits symmetry about the horizontal axis passing through the origin. In order to test the linear formulation a case where no such symmetries arise we consider the case of a rotated ellipse in an open flow. The ellipse geometry is generated  with the minor axis length of $a=0.25$ aligned with the streamwise direction, the major axis length of $b=0.5$  aligned in the crossflow direction and the centre of the ellipse located at the origin of the coordinate system $(0,0)$. The ellipse is then rotated by an angle of $30^{\circ}$ clockwise. The stabilized base flow around the rotated ellipse is calculated at $Re=50$, where the diameter along the major axis is used as the length scale for the Reynolds number. The streamwise velocity for the stabilized base flow is shown in figure~\ref{fig:ellipse_base} which clearly shows the lack of symmetry of the base flow close to the ellipse.

\begin{figure}
	\centering
	\begin{subfigure}{0.48\textwidth}
		\includegraphics[width=1.0\textwidth]{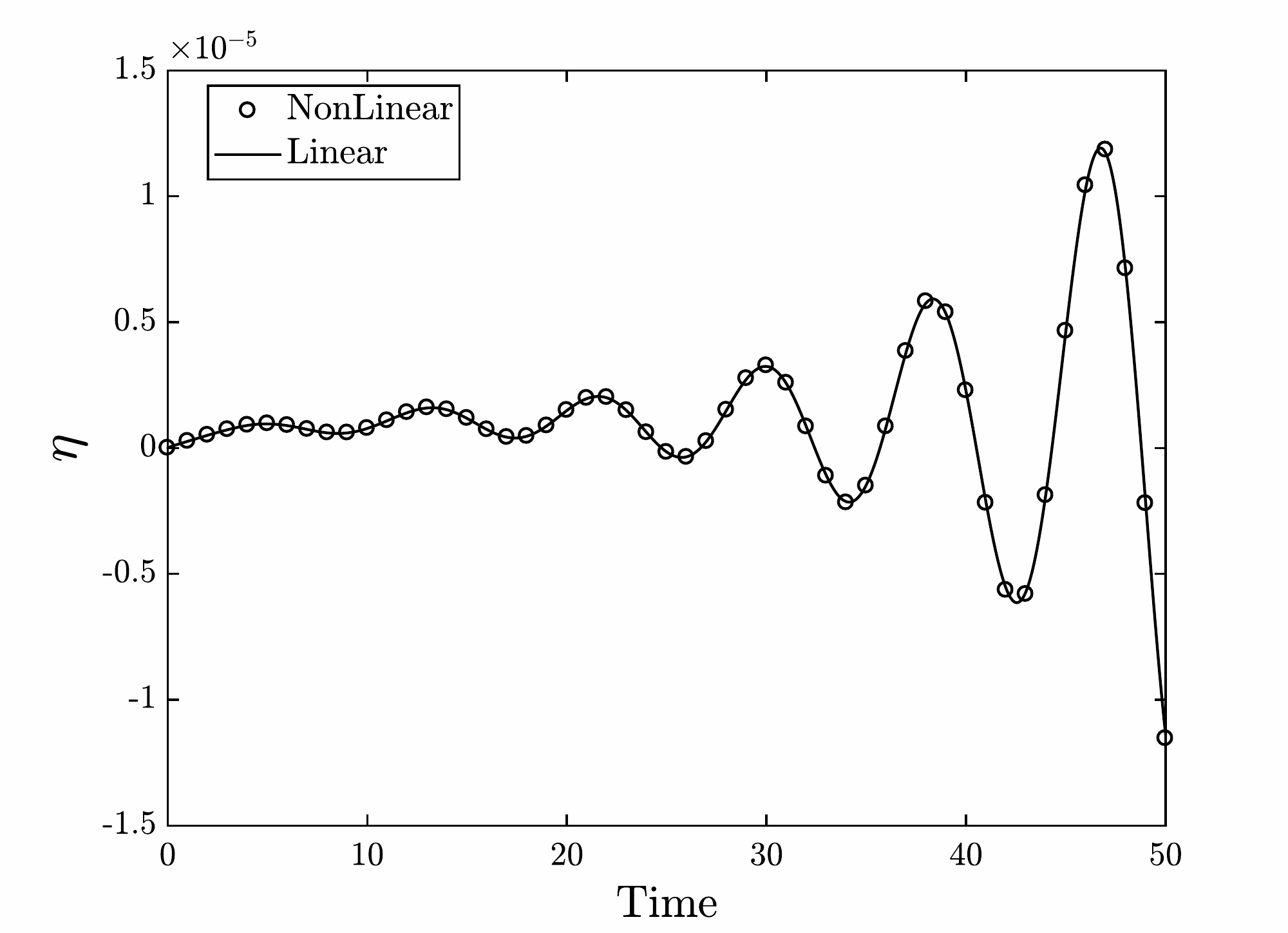}
		\caption{}
		\label{fig:ellipse_vert_evol}		
	\end{subfigure}
	\begin{subfigure}{0.48\textwidth}
		\includegraphics[width=1.0\textwidth]{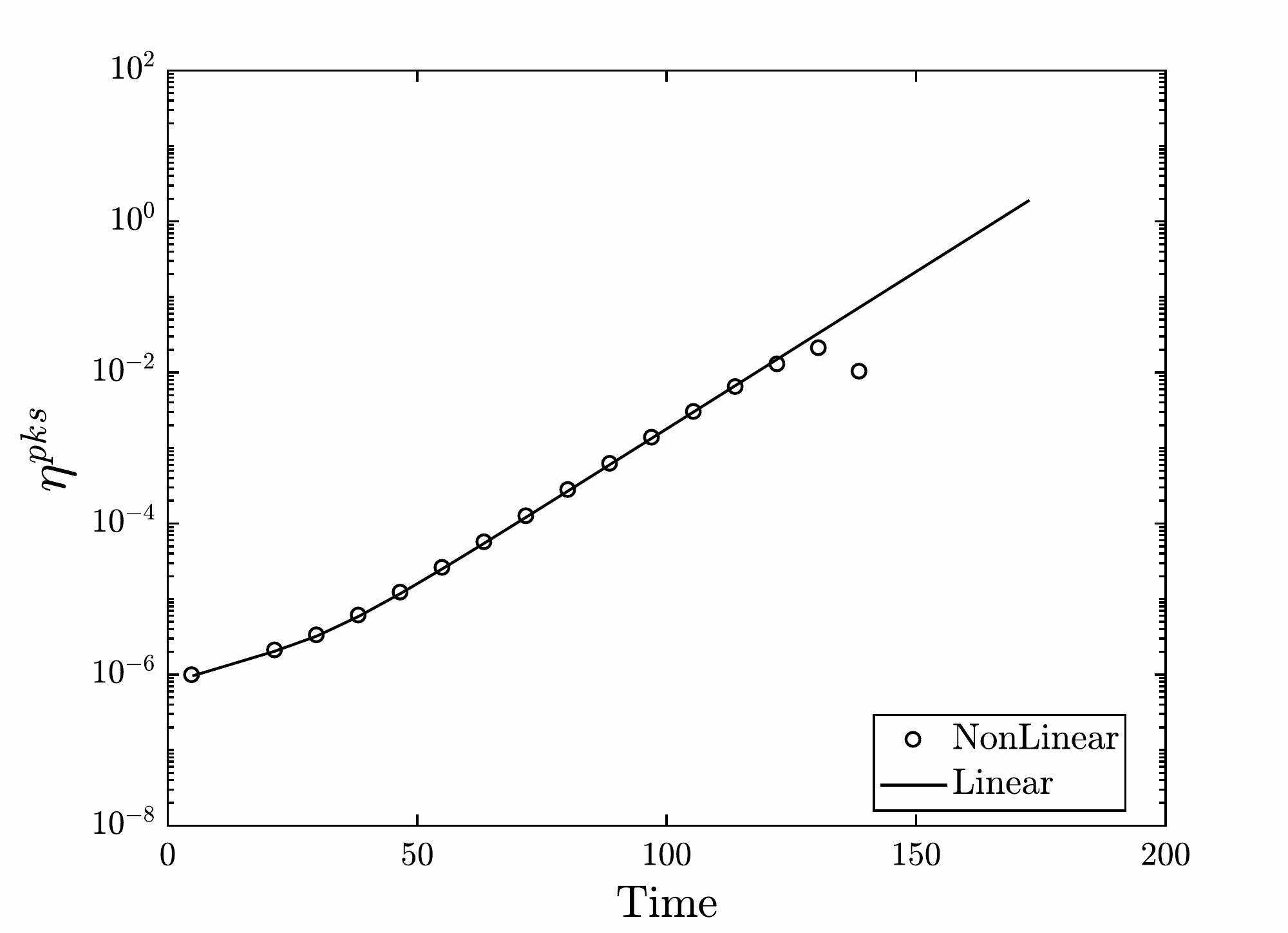}
		\caption{}
		\label{fig:ellipse_vert_pks}
	\end{subfigure}
	\begin{subfigure}{0.48\textwidth}
		\includegraphics[width=1.0\textwidth]{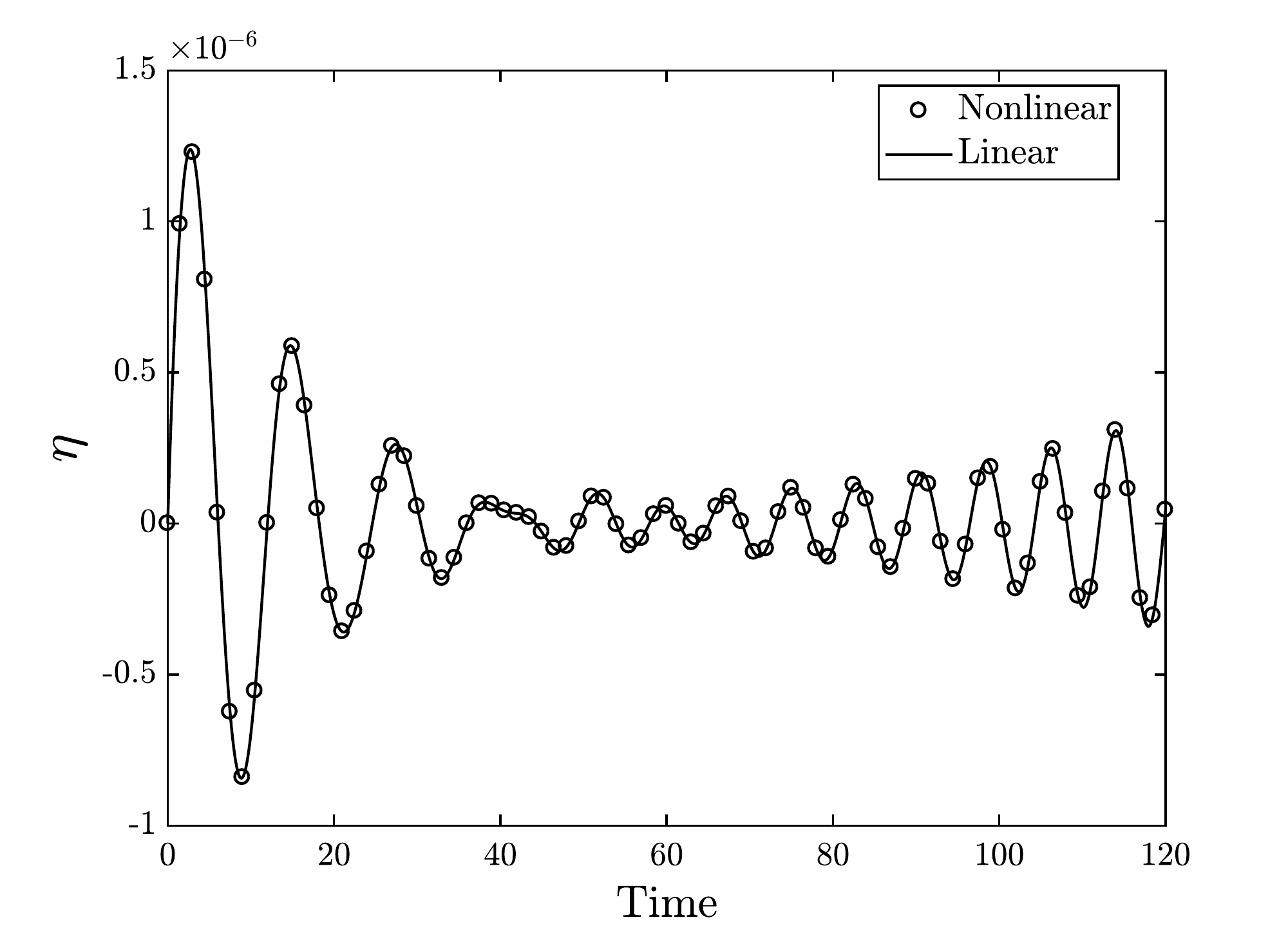}
		\caption{}
		\label{fig:ellipse30_evol}		
	\end{subfigure}
	\begin{subfigure}{0.48\textwidth}
		\includegraphics[width=1.0\textwidth]{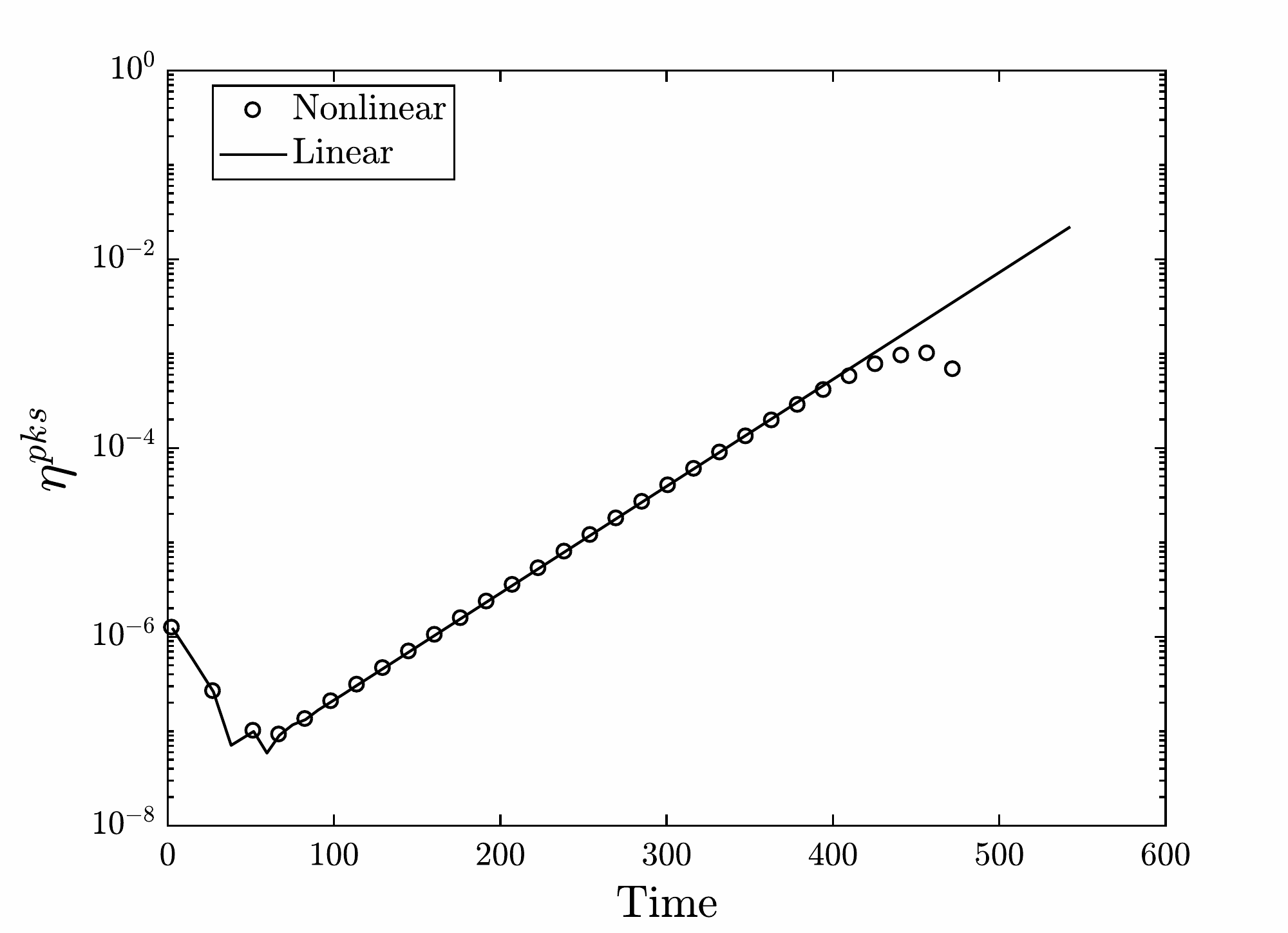}
		\caption{}
		\label{fig:ellipse30_pks}
	\end{subfigure}
	\caption{Comparison of linear and non-linear evolution of the position of a rotated ellipse at $Re=50$. Top two panels represent the vertical translation cases while the bottom two represent the rotational cases.	(a) Time evolution of vertical position $\eta$ in the first few cycles of the oscillation. (b) Evolution of peak amplitudes in a semi-log scale for vertical oscillations. (c) Time evolution of rotational angle $\eta$. (d) Evolution of peak amplitudes of the rotational angle in a semi-log scale.}
\end{figure}

\begin{figure}
	\centering
	\begin{subfigure}{0.48\textwidth}
		\includegraphics[width=1.0\textwidth]{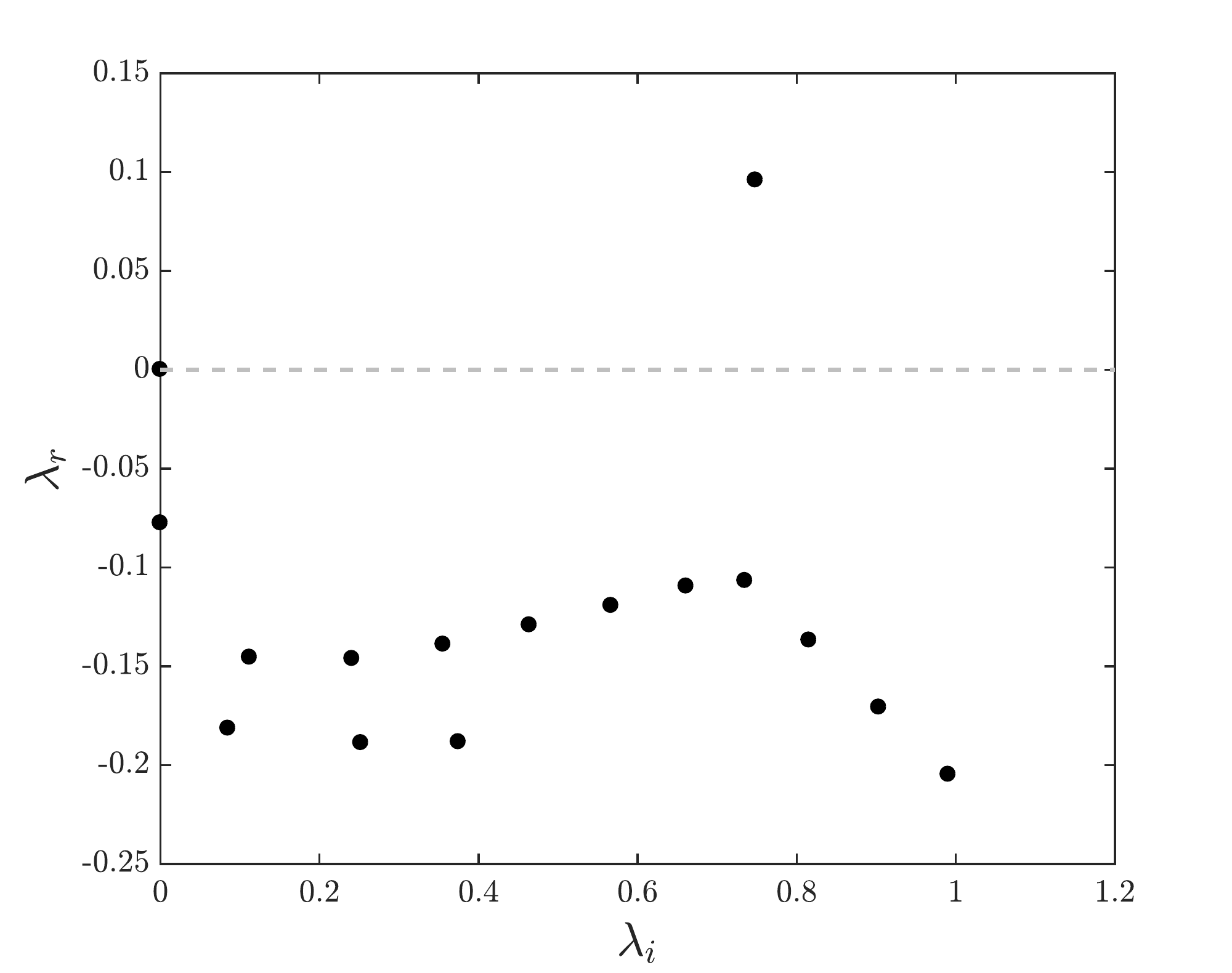}
		\caption{}
		\label{fig:ellipse30_spectra_vert}		
	\end{subfigure}
	\begin{subfigure}{0.48\textwidth}
		\includegraphics[width=1.0\textwidth]{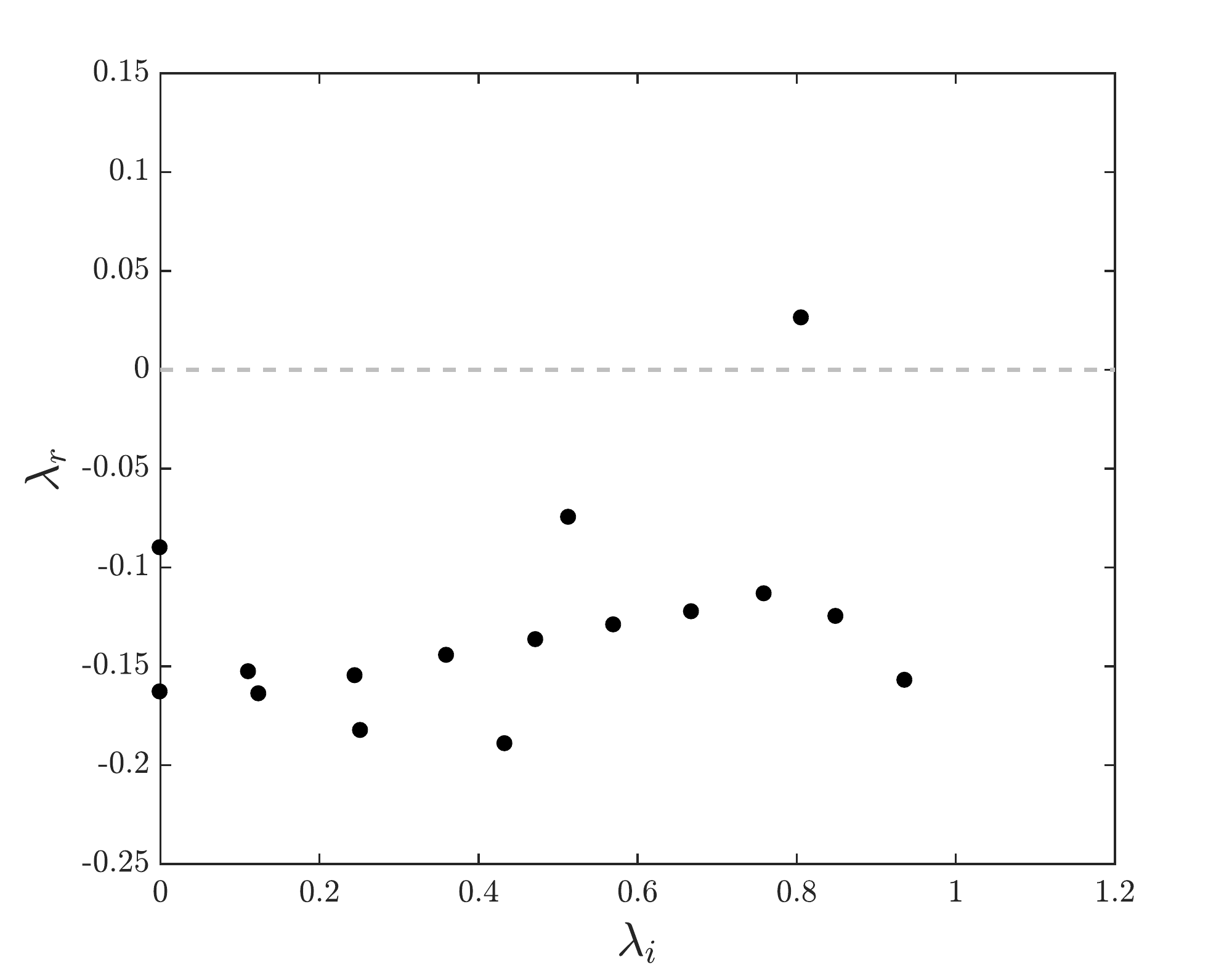}
		\caption{}
		\label{fig:ellipse30_spectra-rot}
	\end{subfigure}
	\caption{One-sided spectra for a rotated ellipse at $Re=50$.  (a) Spectra for an ellipse free to oscillate in the vertical direction and (b) spectra for the ellipse free to rotate about its center.}
	\label{fig:ellipse30_spectra}
\end{figure}

\begin{figure}
	\centering
	\includegraphics[width=0.48\textwidth]{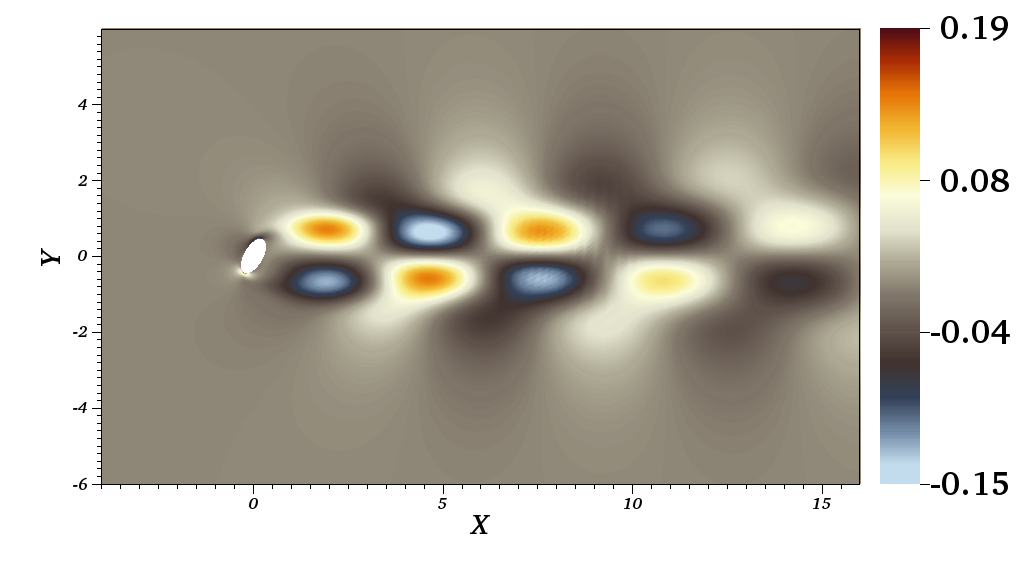}\hfill
	\includegraphics[width=0.48\textwidth]{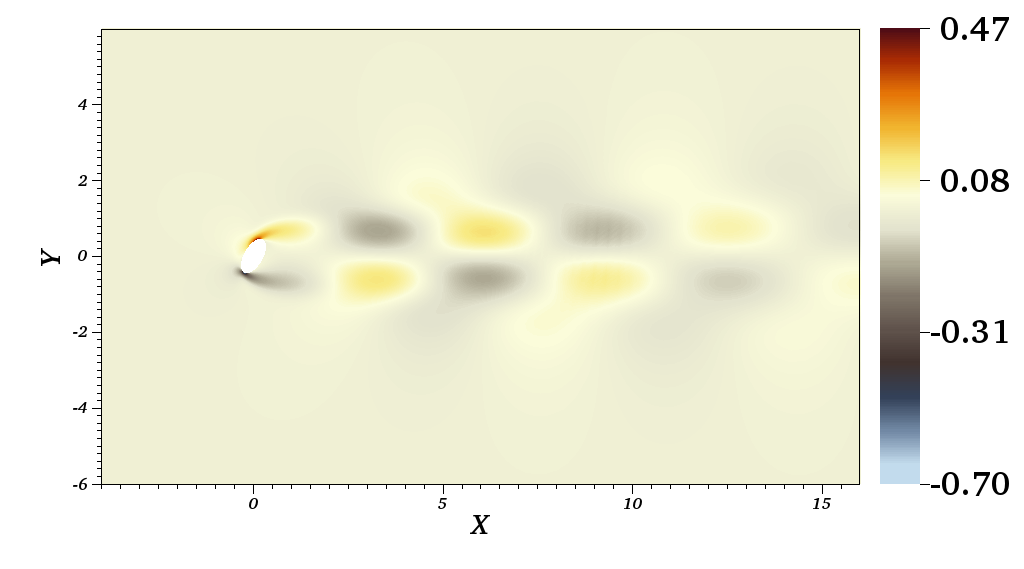} \vspace{2pt}\vfill
	\includegraphics[width=0.48\textwidth]{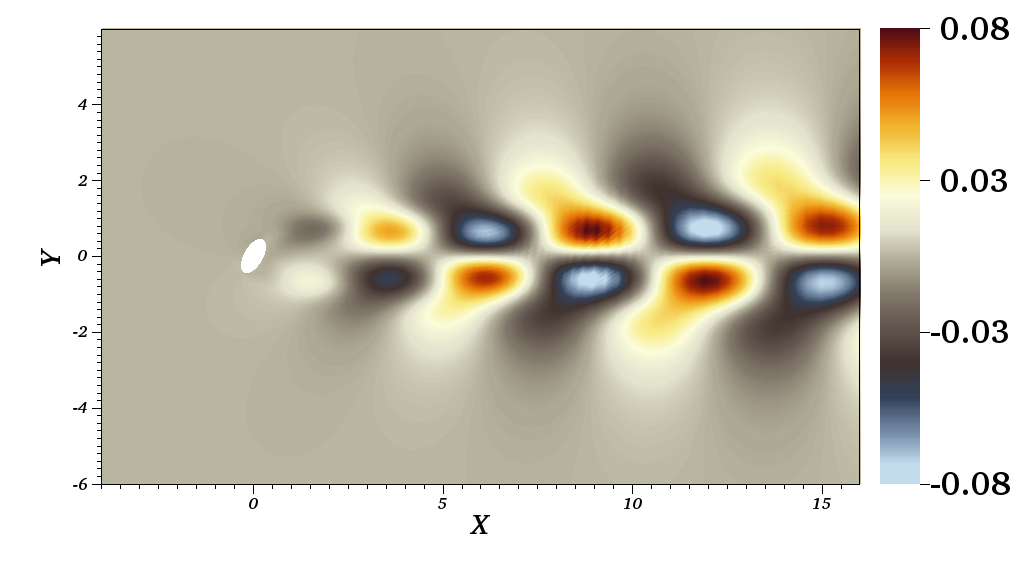}\hfill
	\includegraphics[width=0.48\textwidth]{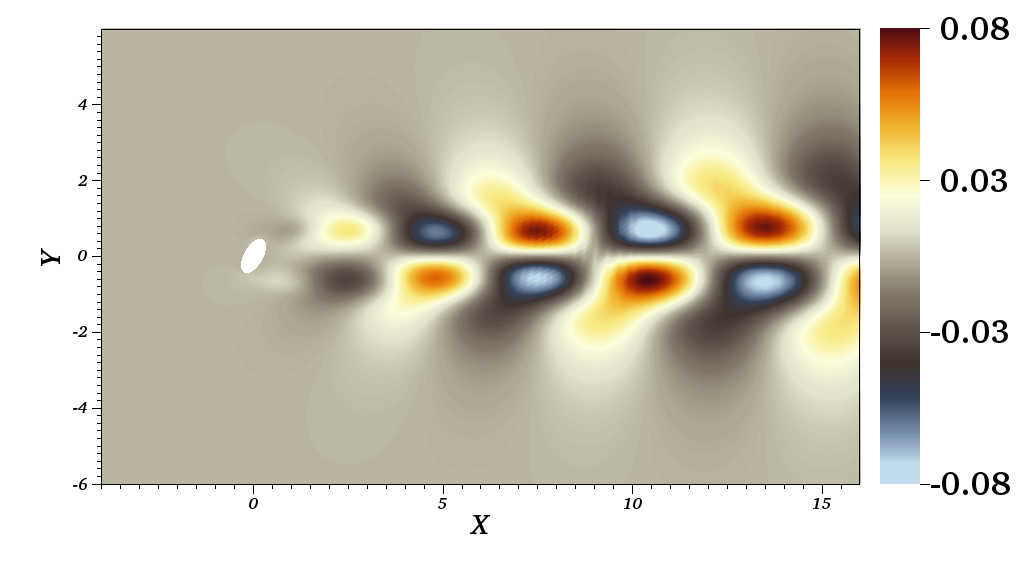}
	\caption{Streamwise velocity component of the eigenvector for (Top) vertical oscillation and (Bottom) rotational FSI cases. The left panels show the real part of the eigenvector and the right panels show the imaginary part of the eigenvector corresponding to the unstable eigenvalue. }
	\label{fig:ellipse_eigenvector}
\end{figure}

We consider two cases of FSI - an ellipse free to oscillate in the cross-flow direction and a case with the ellipse free to rotate about the out of plane axis passing through its center. In both cases the ellipse is considered to be held stationary due to a constant external force which balances the fluid forces at equilibrium. The density ratio is set to 10 for the ellipse and no spring or damping forces are considered. Thus the ellipse motion is governed entirely due to fluid forces and the inertia of the ellipse. We expect any modeling errors in fluid forces to show up strongly for such a case. Proceeding in the usual manner, the comparison of small-amplitude linear and non-linear evolution for the oscillating case is shown in figures~\ref{fig:ellipse_vert_evol} and \ref{fig:ellipse_vert_pks}, while the comparison for the rotational case is shown in figures~\ref{fig:ellipse30_evol} and \ref{fig:ellipse30_pks}. The eigenvalue estimates for both cases are shown in table~\ref{tab:ellipse_vert_unstable}. A good agreement is found for all the cases considered. We note that for the case with vertical oscillations, a zero frequency unstable mode also exists. This was deduced from the time evolution of the simulations since the positive and negative peaks showed a marginally different growth rate. The Arnoldi procedure also showed the existence of both an oscillatory and zero frequency unstable mode. A non-linear least squares procedure was then used to estimate the growth rate and the unstable frequencies. The one-sided spectra for the two cases is shown in figure~\ref{fig:ellipse30_spectra} and the streamwise components of the eigenvector associated with the (oscillatory) unstable eigenvalues are shown in figure~\ref{fig:ellipse_eigenvector}.

\begin{table}[bh]
	\centering
	\begin{tabular}{x{2.0cm}x{2.8cm}x{2.8cm}x{2.8cm}x{2.8cm} }		
		Case  						& Non-linear &  Linear & Arnoldi \\
		\toprule
		Oscillation   	    &  $9.56\times10^{-2}\pm0.749i$ & $9.57\times10^{-2}\pm0.748i$ & $9.58\times10^{-2} \pm 0.748i$ \\
		Rotation       	      &  $2.60\times10^{-2}\pm0.806i$ & $2.60\times10^{-2}\pm0.806i$ & $2.60\times10^{-2} \pm 0.806i$ \\
		\bottomrule
	\end{tabular}
	\capspace
	\caption{Unstable eigenvalue estimates for vertical oscillation and rotational cases for a rotated ellipse at $Re=50$, obtained with three different methods.} \label{tab:ellipse_vert_unstable}
\end{table}

We mention that we have evaluated the ``added stiffness" terms of \cite{fanion00,fernandez03i} and \cite{fernandez03ii} for all of the tested cases and we find that these terms always remain several orders of magnitude smaller than the forces arising due to the fluid perturbations. This further confirms the earlier mathematical result (equation~\ref{basep_zero_force}) that the ``added stiffness" terms represent a higher-order correction and do not play a role in the linearized dynamics.

\subsection{Spontaneous symmetry breaking}
\begin{figure}
	\centering
	\begin{subfigure}{0.60\textwidth}
	\includegraphics[width=1.00\textwidth]{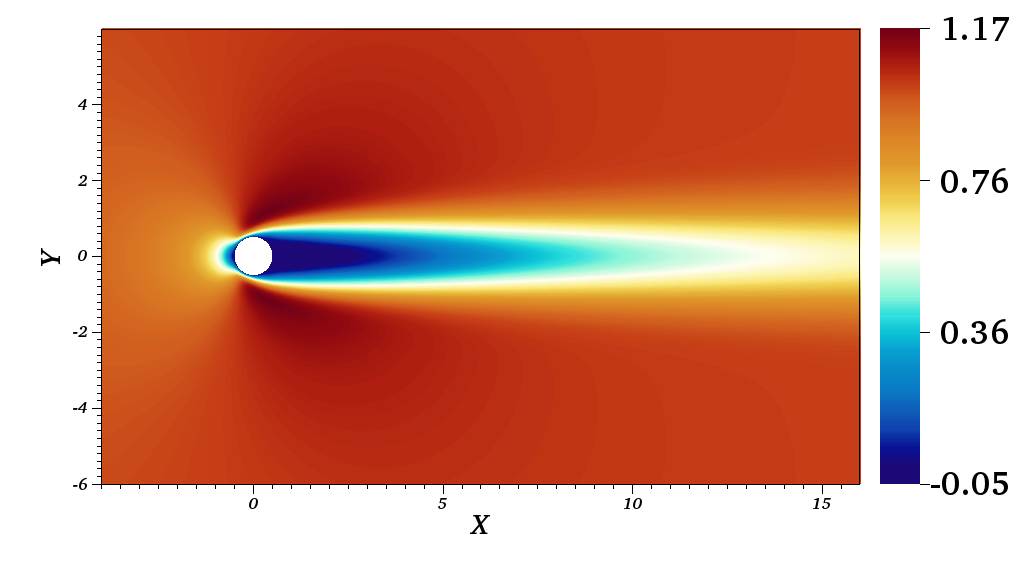}
	\caption{}
	\label{fig:ugis_base45}
	\end{subfigure}
	\begin{subfigure}{0.60\textwidth}
	\includegraphics[width=1.00\textwidth]{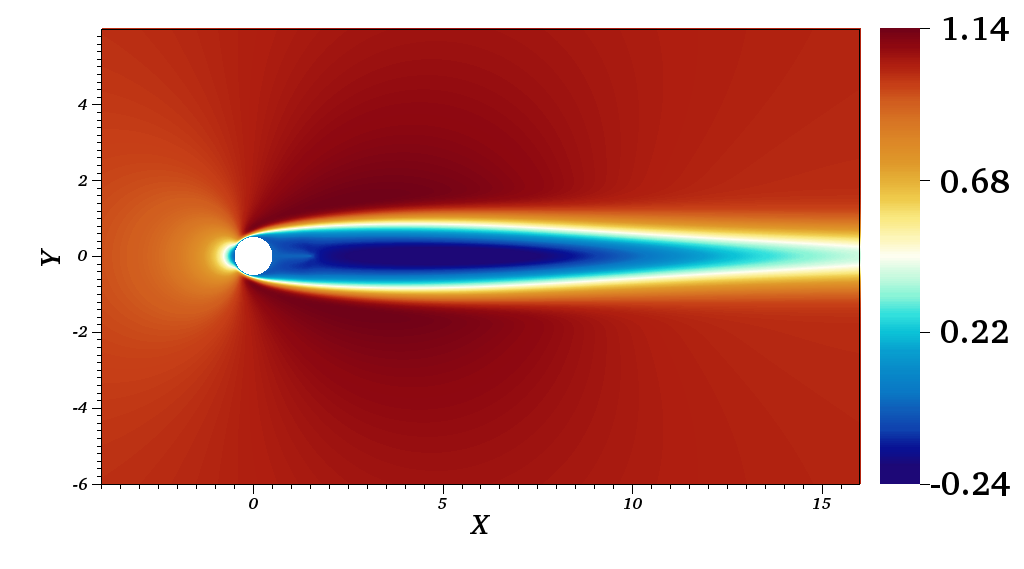}
	\caption{}
	\label{fig:ugis_base156}
	\end{subfigure}
	\caption{Base flow state for a cylinder with splitter plate at (a) $Re=45.0$ and (b) $Re=156$}
\end{figure}

We use the derived linear formulation to investigate the case of a circular cylinder with an attached splitter-plate set at zero incidence to the oncoming flow. The structural equations take the same form as in equation~\ref{struc_cossu}, with $\eta$ now representing the deviation of the rotational angle from the equilibrium position. The terms $\mathcal{M}, \mathcal{D}, \mathcal{K}$ and $\mathcal{F}$ represent the moment of inertia, rotational damping, rotational stiffness and the out of plane moment acting on the body respectively. The cylinder is free to rotate about its center and the rotational stiffness $\mathcal{K}$ and damping $\mathcal{D}$ are both set to zero. Thus the cylinder rotates only due to the action of the fluid forces. The particular setup, along with a variant where the splitter plate is flexible, has been a subject of investigation by several previous authors \citep{xu90,xu93,bagheri12,lacis14}. For certain lengths of the cylinder-splitter body, the system exhibits an interesting dynamic where the body spontaneously breaks symmetry and splitter-plate settles at a non-zero angle to the oncoming flow. The breaking of symmetry leads to the generation of lift force which could play a role in locomotion through passive mechanisms \citep{bagheri12}. The symmetry breaking phenomenon is known to occur for both sub-critical and super critical Reynolds numbers \citep{lacis14} and in both two and three dimensional configurations \citep{lacis17}. We investigate the phenomenon through our linear FSI framework. According to the results of \cite{lacis14}, symmetry breaking is expected to occur for splitter-plate lengths of less than $2D$. Accordingly we set the splitter-plate length to be $1D$ and a thickness of $0.02D$. Following \cite{lacis14} we use a solid to fluid density ratio of $1.001$ for $Re=45$ and $1.01$ for $Re=156$. Figures~\ref{fig:ugis_base45} and \ref{fig:ugis_base156} show the baseflow states for the diameter based Reynolds numbers of $Re=45$ (sub-critical) and $Re=156$ (super-critical).

\begin{figure}
	\centering
	\includegraphics[width=0.60\textwidth]{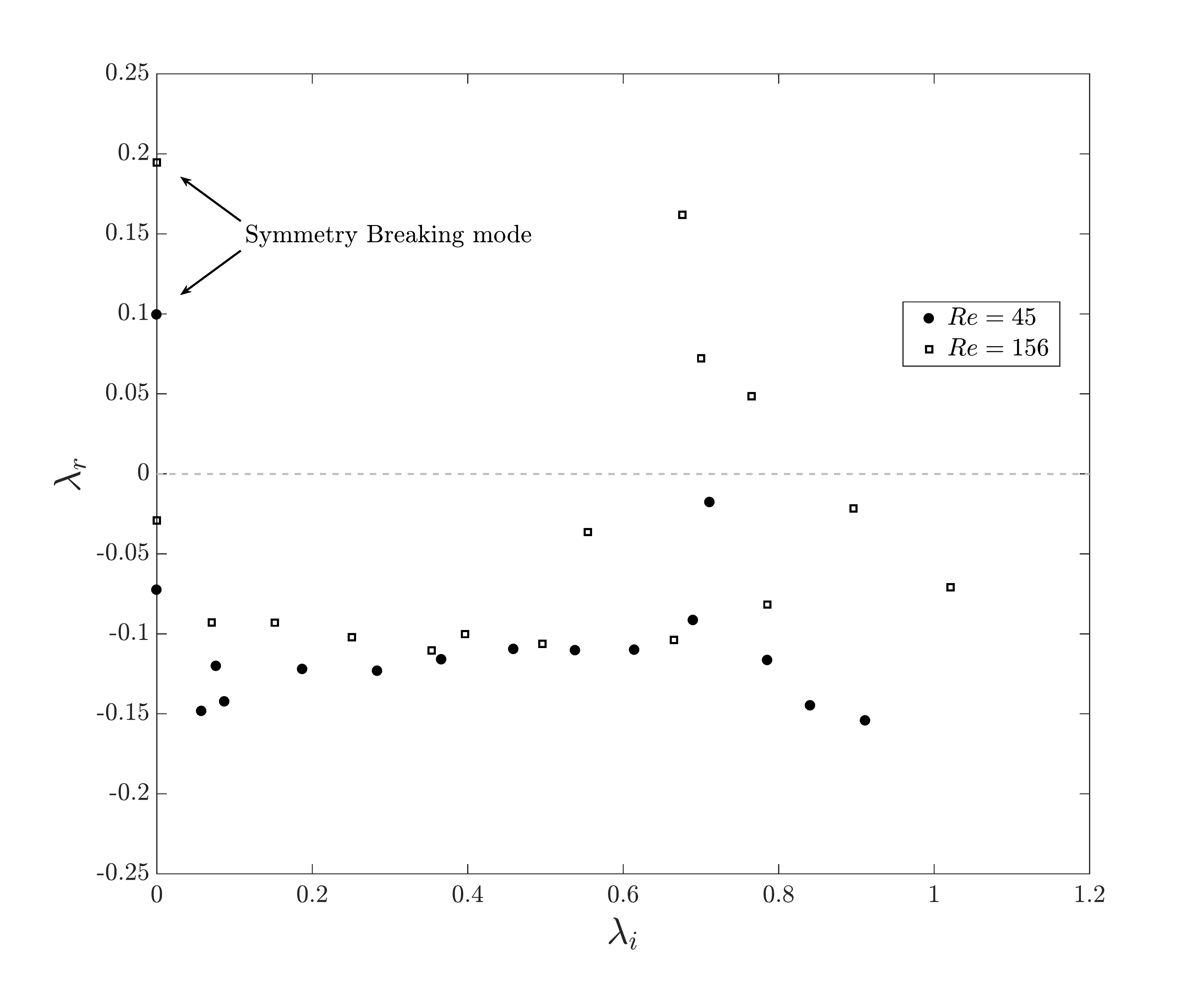}
	\caption{One-sided spectra for the cases exhibiting spontaneous symmetry breaking at $Re=45$ and $Re=156$ for a rotatable cylinder with splitter-plate.}
	\label{fig:ugis_spectra}
\end{figure}

\begin{figure}
	\centering
	\includegraphics[width=0.60\textwidth]{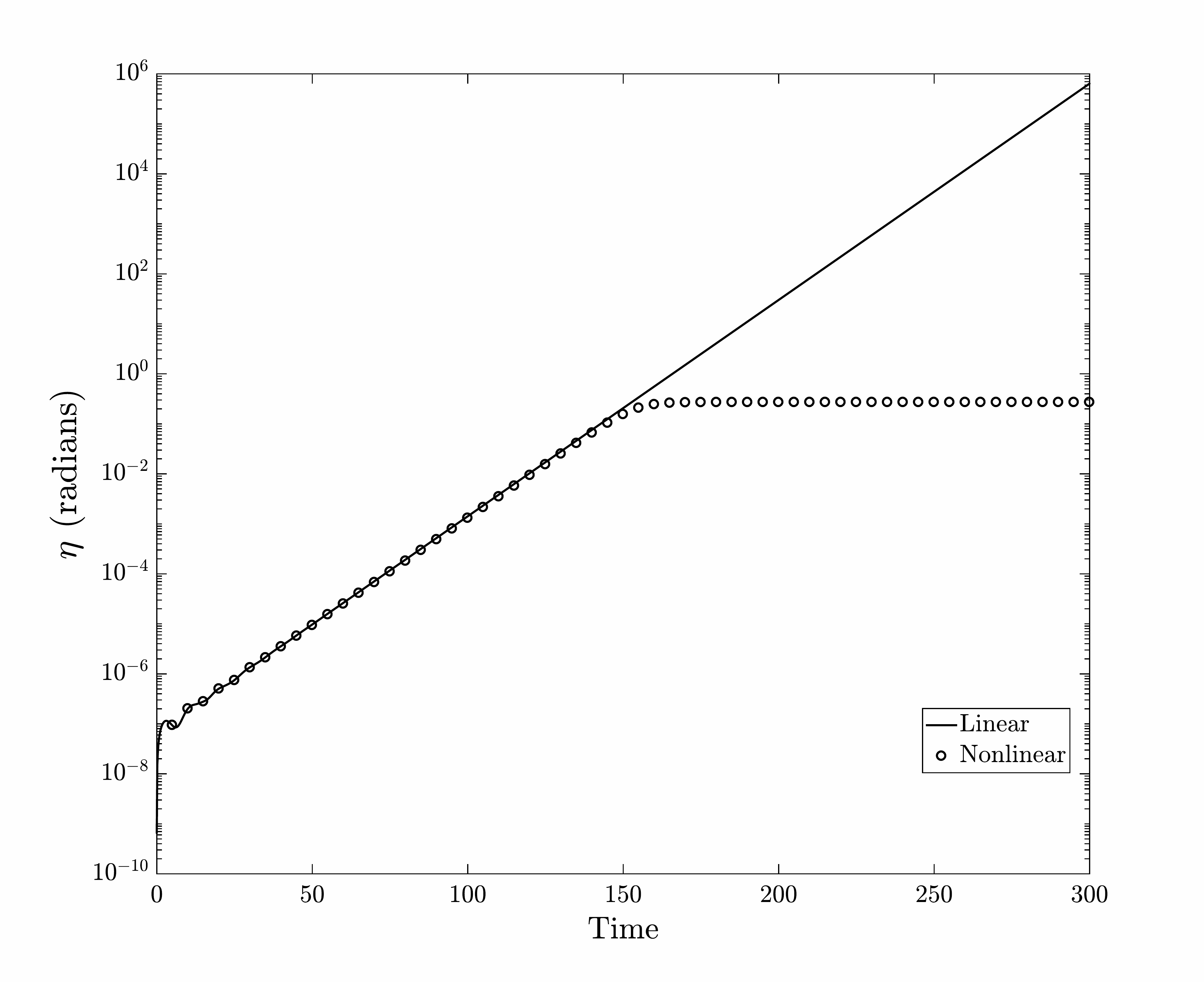}
	\caption{Time evolution of the angle for a rotatable cylinder with splitter-plate at $Re=45$.}
	\label{fig:ugis_log}
\end{figure}

Figure~\ref{fig:ugis_spectra} shows the spectra for the two different Reynolds numbers. Both cases have an eigenvalue lying on the positive $y$-axis ($\lambda=0.10+0i$ for $Re=45$ and $\lambda=0.19+0i$ for $Re=156$). This represents the symmetry-breaking eigenmode of the system since it does not oscillate about a zero mean but rather leads to a monotonic growth in the rotational angle. For the higher Reynolds number,  several other modes of instability exist in the flow which represent the von-K\'{a}rm\'{a}n modes of the flow. From the perspective of linear analysis, these modes simply oscillate about the mean angle with growing oscillation amplitudes. However the zero frequency mode leads to monotonic rise in the angle and thus causes the symmetry breaking effect.
For $Re=45$, we also simulate the linear and non-linear cases after adding a random small-amplitude perturbations to both flow cases. Figure~\ref{fig:ugis_log} shows the time evolution of the angle $\eta$ of the cylinder-splitter body. Both the simulations undergo the same exponential growth of about five orders of magnitude before the non-linear case saturates, while the linear case continues its exponential growth. The saturation of the non-linear simulations occur at $\eta\approx15.5^{\circ}$, which is the same turn angle reported by \cite{lacis14} for a splitter-plate of length $1D$. Thus the linear FSI framework predicts the onset of symmetry breaking instability. Of course for the system to finally exhibit symmetry breaking a non-linear mechanism is required since the flow must equilibrate at the new position. However the onset can be traced to the zero frequency unstable mode.

\section{Structural sensitivity of the eigenvalue}
\label{fsi_struc_sensitivity}
As noted in figure~\ref{fig:re23_spectra_k} and \ref{fig:navrose40_spectra}, the unstable eigenvalue of the coupled FSI system changes as the structural parameters are varied, while the base flow is held constant. The variation occurs both for the growth rate as well as the frequency of the eigenvalue. One might expect the sensitivity of the eigenvalue to structural perturbations to change as well. We investigate the eigenvalue sensitivity to structural perturbations for different values of the structural parameters for a spring-mounted 2D circular cylinder, which is free to oscillate in the cross-stream direction. The linear FSI problem is defined by equations~\ref{NS_cossu_lin}--\ref{fluid_forces_cossu_lin}. In the following sections we assume all derivatives are evaluated in the reference configuration and we drop the superscript $^{0}$ from the derivative terms. In addition, uppercase letters denote base-flow quantities and lowercase letters denote perturbation quantities and the superscripts $^{0}$ and $'$ are dropped from the base flow and perturbation quantities.

The eigenvalue sensitivity is typically studied through the use of adjoint equations \citep{giannetti07,luchini14}. For any linear operator $\mathcal{L}$, the adjoint operator $\mathcal{L}^{\dagger}$ is defined such that it satisfies the Lagrange identity 

\begin{eqnarray}
	\langle \mathcal{L}^{\dagger}p,q\rangle = \langle p,\mathcal{L}q\rangle, \nonumber 
\end{eqnarray}
for any arbitrary vectors $q$ and $p$ in the domain of $\mathcal{L}$ and $\mathcal{L}^{\dagger}$ respectively. The symbol $\langle \cdot,\cdot\rangle$ denotes the inner product under which the above identity holds. In the current context the operator $\mathcal{L}$ represents the linearized Navier--Stokes equations for FSI, also referred to as the direct problem, $\mathcal{L}^{\dagger}$ is the corresponding adjoint operator with the definition of the inner product as the integral over the domain $\Omega$ and time horizon $T$
\begin{eqnarray}
	\langle p,q\rangle = \int_{T}\int_{\Omega} p^{H}q\ d\Omega\ dt \nonumber.
\end{eqnarray}
Defining the vector $\mathbf{q}=(\mathbf{u},p,\eta,\varphi)$, which lies in the domain of $\mathcal{L}$, and an adjoint vector $\mathbf{q}^{\dagger}=(\mathbf{u}^{\dagger},p^{\dagger},\eta^{\dagger},\varphi^{\dagger})$ which lies in the domain of $\mathcal{L^{\dagger}}$, the adjoint equations may be found by taking the inner product between the adjoint variables and the direct equations. The use of integration-by-parts, divergence theorem and a judicious manipulation of terms leads to a set of expressions for the adjoint equations. The full derivation of the equations along with the boundary conditions may be found in Appendix~\ref{appB}. The final form of the adjoint equations is expressed below.

\begin{subequations}
\begin{align}
-\frac{\partial u^{\dagger}_{i}}{\partial t} + u^{\dagger}_{j}\frac{\partial U_{j}}{\partial x_{i}}-U_{j}\frac{\partial u^{\dagger}_{i}}{\partial x_{j}}
+ \frac{\partial p^{\dagger}}{\partial x_{i}} - \frac{1}{Re}\frac{\partial^{2} u^{\dagger}_{i}}{\partial x_{j}\partial x_{j}}   =& 0, & \text{in } \Omega^{f} \label{NS_adj},\\
\frac{\partial u^{\dagger}_{i}}{\partial x_{i}} =  &0,  &\text{in } \Omega^{f}, \label{incompressibility_adj} \\
-\mathcal{M}\frac{d \varphi^{\dagger}}{d t}  +\mathcal{D}\varphi^{\dagger}  -\eta^{\dagger}  + \oint_{\partial\Omega} \sigma^{\dagger}_{2j}n_{j} d\Omega =& 0, &\text{in } \Omega^{s},  \label{struc_adj} \\
-\frac{d\eta^{\dagger}}{dt} + \mathcal{K}\varphi^{\dagger} - \oint_{\partial\Omega} \left(\frac{\partial U_{k}}{\partial x_{2}}\right) \sigma^{\dagger}_{kj}n_{j} d\Omega = &0&\text{in } \Omega^{s},  \\
u^{\dagger}_{1} = & 0, & \text{on } \Gamma, \label{adj_bc1} \\
u^{\dagger}_{2}  - \varphi^{\dagger}_{2} = & 0, & \text{on } \Gamma, \label{adj_bc2} \\
\sigma_{ij}^{\dagger} - \left(-p^{\dagger}\delta_{ij} + \frac{1}{Re}\frac{\partial u^{\dagger}_{i}}{\partial x_{j}}   \right) = &0, & \text{adjoint forces on } \Gamma. \label{adj_fluid_forces} \\
	u^{\dagger}_{i} = & 0, & \text{on } \partial\Omega_{v},  \label{adj_dirichlet}\\
(\sigma^{\dagger}_{ij}+u_{i}^{\dagger}U_{j})n_{j} =& 0& \text{on } \partial\Omega_{o}. \label{adj_O}
\end{align}
\end{subequations}
Here $\sigma_{ij}^{\dagger}$ is referred to as the adjoint fluid forces acting on the cylinder. Unlike in the direct equations, the adjoint fluid velocities are homogeneous on the FSI interface $\Gamma$. The sign of the diffusive term indicates the equations evolve an adjoint field backward in time. A variable substitution $\tau = - t$ can be made to return the equations to forward (in $\tau$) propagation.

\begin{figure}
	\centering
	\includegraphics[width=0.60\textwidth]{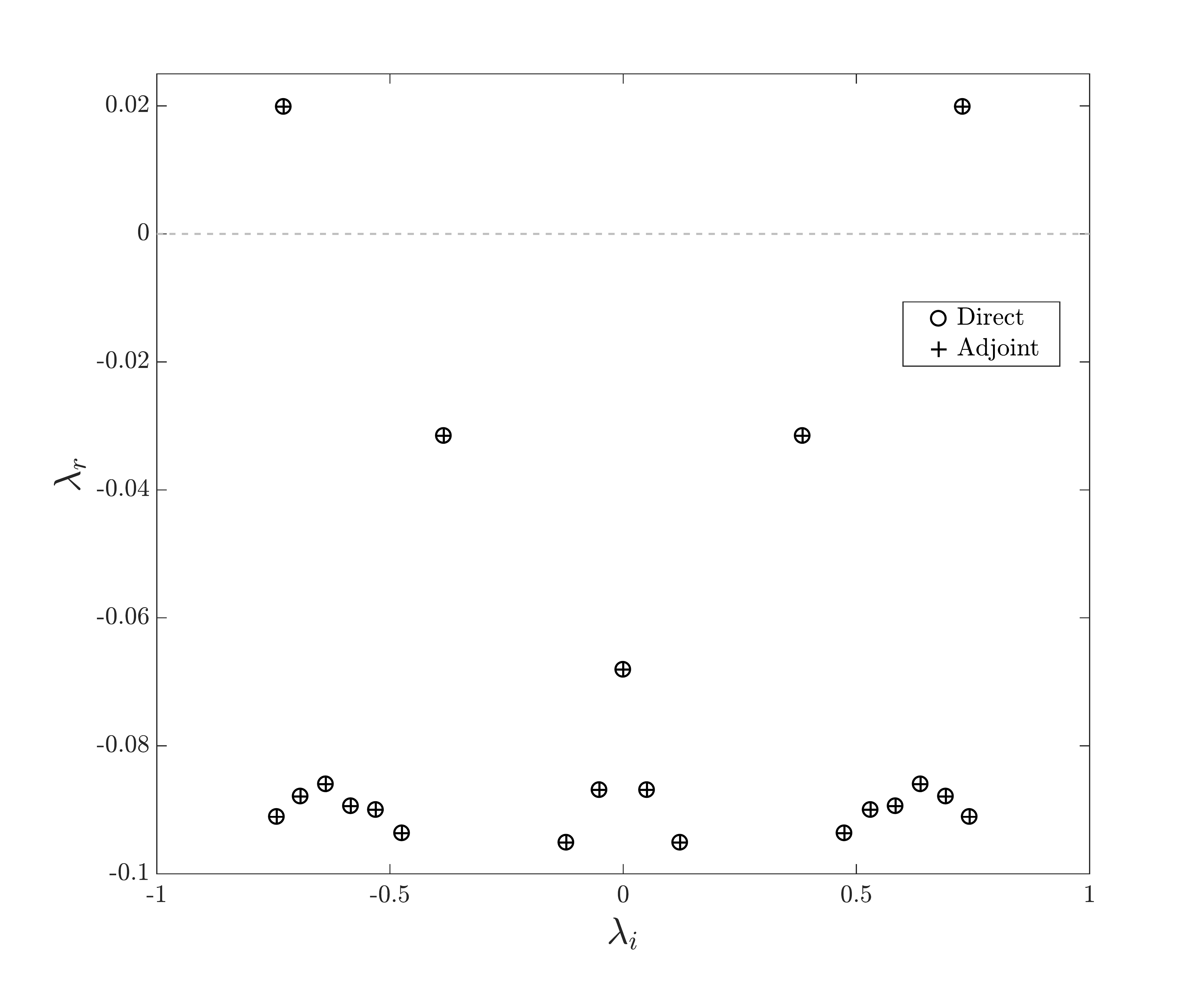}
	\caption{Comparison of the eigenspectra for the direct and adjoint problems for density ratio of $10$ and natural frequency $\omega_{n}=0.3833$}
	\label{fig:dir_adj}
\end{figure}

A validation of the adjoint equations can be found by comparing the spectra of the direct and adjoint problems. The eigenvalues of the adjoint problem are the complex conjugate of the eigenvalues of the direct problem. For real matrices, the spectra for both problems is the same. Figure~\ref{fig:dir_adj} shows the direct and adjoint spectra for a spring-mounted 2D cylinder at a diameter based Reynolds number of $Re=50$, mass ratio of $10$, corresponding to $\mathcal{M}=7.854$, $\mathcal{K}=1.1537$ which corresponds to the natural frequency of the spring-mass system of $\omega_{n}=0.3833$. The damping is set to zero ($\mathcal{D}=0$). As observed in the figure, the spectra for both the direct and adjoint problems have a very good agreement with each other.

For a general linear eigenvalue problem $\mathcal{L}q = \lambda q$, a first-order approximation to the perturbed eigenvalue problem $(\mathcal{L} +\Delta\mathcal{L})(q + \Delta q) = (\lambda + \Delta\lambda)(q +\Delta q)$ can be obtained by making use of the corresponding adjoint eigenvector  vector field $q^{\dagger}$ \citep{giannetti07}, resulting in an expression for the eigenvalue perturbation as
\begin{eqnarray}
\Delta\lambda = \frac{\langle q^{\dagger},\Delta\mathcal{L}q\rangle}{\langle q^{\dagger},q\rangle}, \nonumber \\
(\mathcal{L}^{\dagger} - \lambda^{*}\mathcal{I})q^{\dagger} = 0, \nonumber
\end{eqnarray}
where $^{*}$ implies complex conjugation and $q^{\dagger}$ represents the right eigenvector of the adjoint operator $\mathcal{L}^{\dagger}$ with the eigenvalue of $\lambda^{*}$. To estimate the drift in eigenvalue, knowledge of the specific operator perturbation $\Delta\mathcal{L}$ is required. However, as shown by \cite{giannetti07}, a spatial sensitivity map of the eigenvalue perturbation can be found by assuming the perturbation operator to be of the form of a spatially localized feedback, with the feedback being proportional to the local values of the field variables (velocities). For a such a case we may estimate the eigenvalue drift along with its upper bound for each spatially localized operator perturbation as
\begin{eqnarray}
\Delta\lambda &=& \frac{\langle q^{\dagger},\delta(x-x_{0},y-y_{0})C_{0} q\rangle}{\langle q^{\dagger},q\rangle}, \nonumber \\
|\Delta\lambda|  &\le& ||C_{0}||\boldsymbol{\Theta}(x_{0},y_{0}), \nonumber \\
\boldsymbol{\Theta}(x,y) &=& \frac{||q^{\dagger}(x,y)||\cdot ||q(x,y)||}{||\langle q^{\dagger},q\rangle||}, \label{eig_sensitivity}
\end{eqnarray}
where $C_{0}$ is a matrix that defines the feedback due to the localized operator perturbation and the local coupling between different velocity components. The quantity $\boldsymbol{\Theta}(x,y)$ gives an indication of regions where the localized feedback will produce a large drift in the eigenvalue, thus representing the sensitivity of the eigenvalue to structural perturbations.

\begin{figure}
\begin{subfigure}{1\textwidth}
	\centering
	\includegraphics[width=0.49\textwidth]{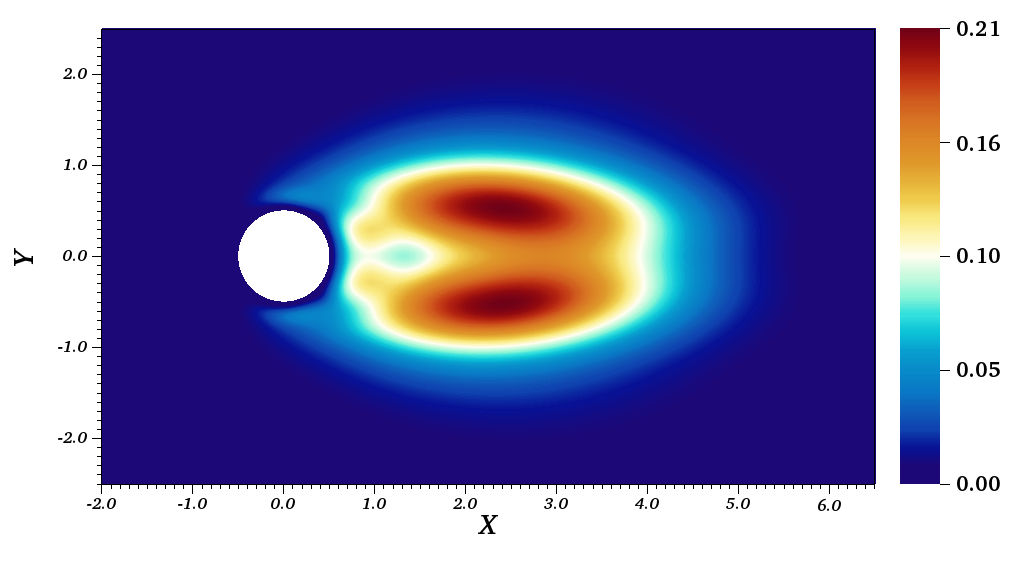}
	\caption{Stationary Cylinder}	\label{fig:st0}
\end{subfigure}	
	\begin{subfigure}{0.49\textwidth}
		\centering
		\includegraphics[width=1\textwidth]{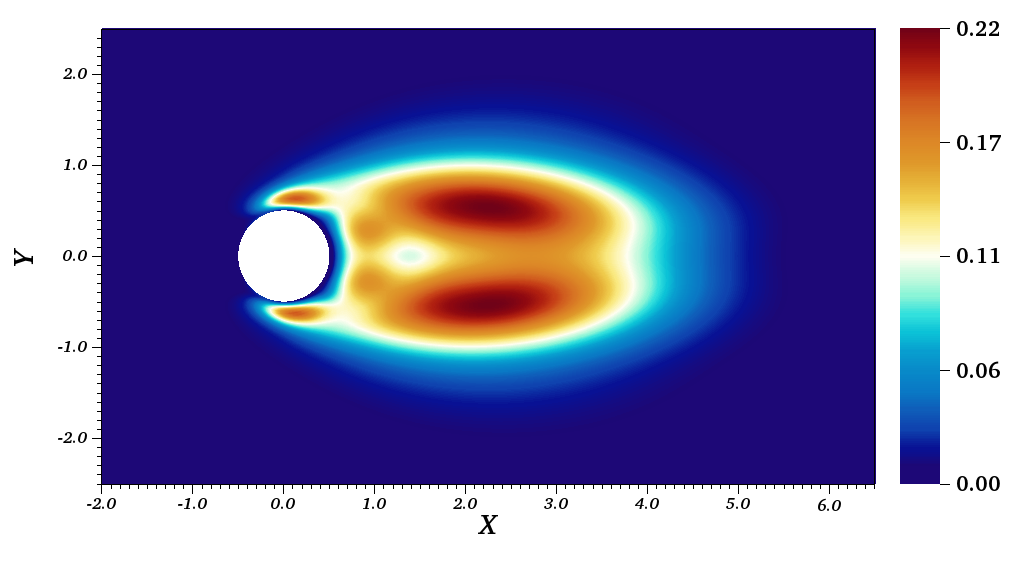}
		\caption{$\omega_{n}=0.0767$} \label{fig:st1}
	\end{subfigure}
	\begin{subfigure}{0.49\textwidth}
	\centering
	\includegraphics[width=1\textwidth]{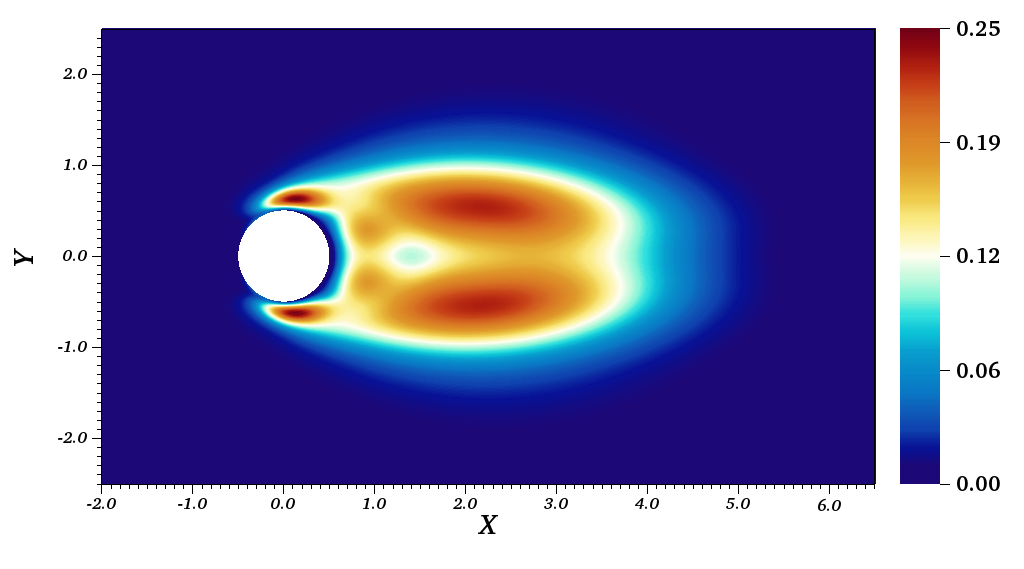}
	\caption{$\omega_{n}=0.3833$} \label{fig:st2}	
\end{subfigure}
	\begin{subfigure}{0.49\textwidth}
	\centering
	\includegraphics[width=1\textwidth]{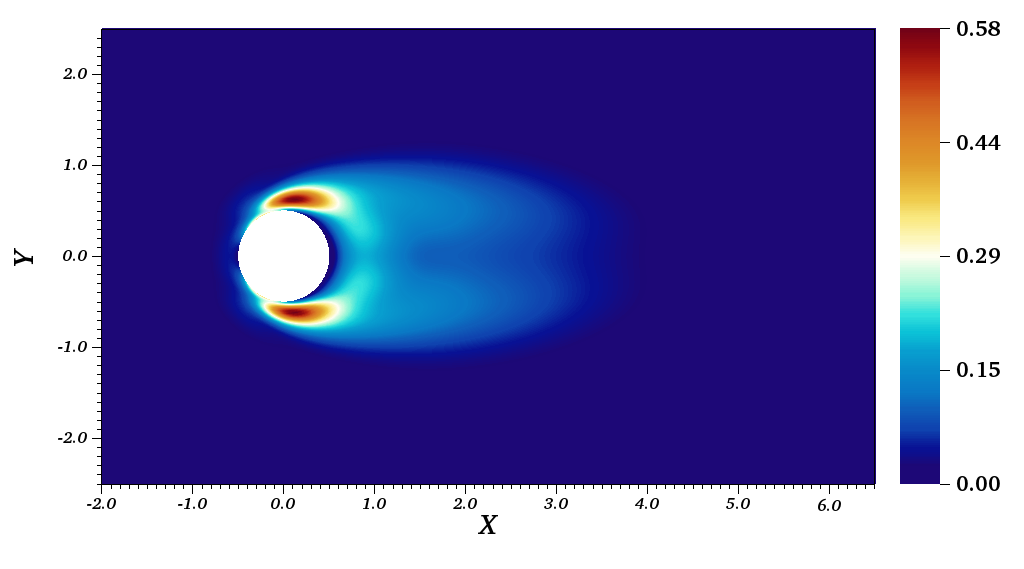}
	\caption{$\omega_{n}=0.7665$} \label{fig:st3}
\end{subfigure}
	\begin{subfigure}{0.49\textwidth}
	\centering
	\includegraphics[width=1\textwidth]{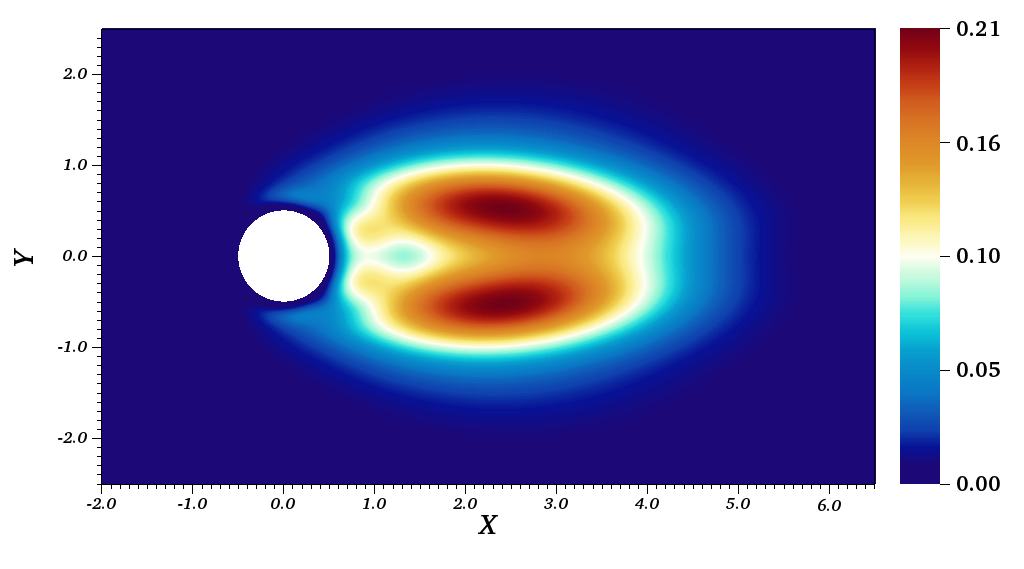}
	\caption{$\omega_{n}=3.8327$} \label{fig:st4}	
\end{subfigure}
	\caption{Comparison of the structural sensitivity map for the least stable eigenvalue for different values of the natural frequency $\omega_{n}$. The top panel (a) is the structural sensitivity for the stationary cylinder case. Bottom four panels (b)-(e) represent the cases with FSI.}
	\label{fig:struc_sensitivity}
\end{figure}

\begin{figure}
	\centering
	\includegraphics[width=0.49\textwidth]{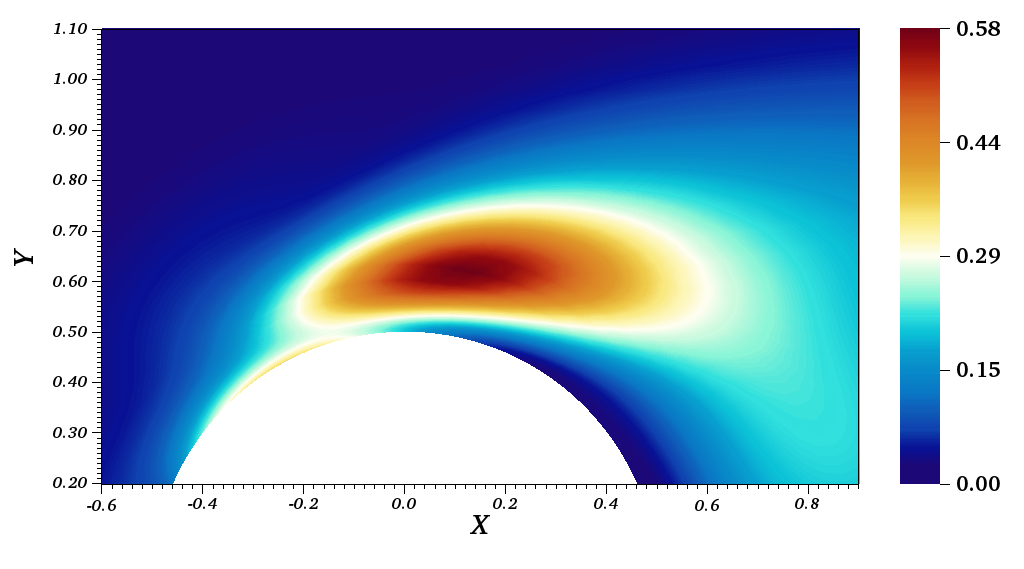}
	\caption{Closeup of the sensitivity map for $\omega_{n}=0.7665$}
	\label{fig:wavemaker_closeup}
\end{figure}

The structural sensitivity of spring-mounted cylinder is investigated for varying natural frequencies $\omega_{n}$ of the spring-mass system. In all subsequent cases the structural damping is set to zero and the density ratio is set to $10$. The natural frequency of the spring-mass system is varied and the sensitivity map $\boldsymbol{\Theta}$ of the least stable eigenvalue is evaluated. Table~\ref{tab:struc_param2} shows the variation of the natural frequency $\omega_{n}$ and the corresponding least stable eigenvalue. The table also shows the unstable eigenvalue for a stationary cylinder at $Re=50$. Figure~\ref{fig:struc_sensitivity} shows the corresponding changes in structural sensitivity of the least stable eigenvalue. When the structural frequency is much lower than the unstable frequency (figure~\ref{fig:st1}), the sensitivity map resembles the sensitivity map of the stationary cylinder with small changes close to the cylinder. A small region of sensitivity is generated close to the cylinder, which is located symmetrically with respect to the horizontal axis. This region grows in intensity as $\omega_{n}$ approaches closer to the unstable frequency (\ref{fig:st2}). Close to resonance (\ref{fig:st3}), the dominant region shifts to the cylinder, lying symmetrically on the top and bottom. The near wake region is still sensitive to structural perturbations however it is no longer the dominant region. As $\omega_{n}$ is increased to be much larger than the unstable frequency, the sensitivity map resembles the stationary case again. In all cases with FSI, the sensitivity is non-zero at the cylinder surface. However, the values remain small if $\omega_{n}$ is far from resonance. Close to resonance, small regions on the cylinder surface have high sensitivities. A close up of the sensitivity map for $\omega_{n}=0.7665$ is shown in figure~\ref{fig:wavemaker_closeup} which shows non-zero values of $\boldsymbol{\Theta}$ at the cylinder surface. The non-zero sensitivity on the cylinder surface has implications for control of vortex-induced vibrations. In particular it is found that control techniques for elimination of vortex streets which apply to stationary cylinders  do not always work for vibrating cylinders \citep{dong08}. 

\begin{table}
	\centering
	\begin{tabular}{x{4cm}x{2cm}}
		$\lambda$ & $\omega_{n}$  \\ 
		\toprule
		$1.3325\times10^{-2}  \pm 0.74189i$ 	     & Stationary 	  \\
		$1.8128\times10^{-2}  \pm 0.73109i$	   	     & $0.0767$ 	 \\
		$1.9773\times10^{-2} \pm  0.72781i$	   	     & $0.3833$ 	 \\
		$7.2151\times10^{-2}  \pm 0.74010i$	   	     & $0.7665$ 	  \\
		$1.3119\times10^{-2}  \pm 0.74237i$	   	     & $3.8327$ 	  \\		
		\bottomrule
	\end{tabular}
	\capspace
	\caption{Unstable eigenvalue corresponding to the structural natural frequency $\omega_{n}$ for the oscillating cylinder cases.} \label{tab:struc_param2}
\end{table}

\section{Conclusions}
\label{conclusion}
An Eulerian formulation for the linear stability analysis of rigid-body-motion fluid-structure-interaction problems is rigorously derived and validated. The final form of the linear equations is evaluated on the equilibrium grid on which the base state is defined and the first-order effects of moving interfaces are captured by the modification of the interface boundary conditions. The ``added-stiffness" terms that arise in the formulation of \cite{fanion00,fernandez03i,fernandez03ii} are shown to vanish to a first order-order approximation and play no role in the linearized perturbation dynamics. The formulation is validated by comparing the evolution of the linear equations with the non-linear system when both systems are perturbed with identical small amplitude disturbances. This has been performed for several different cases for both rotation and translation motion in symmetric as well as asymmetric flow cases. The linear and nonlinear equations evolve in a near identical manner and the extracted frequency and growth rates for the two cases match to within $0.1\%$ relative difference. 

The FSI linear framework is used to analyze the case of symmetry breaking for a rotating cylinder with a rigid splitter-plate. The linear stability analysis predicts the existence of a zero frequency unstable mode. This mode is identified as the cause of symmetry breaking since it causes the system to monotonically deviate from the equilibrium position and thus causing the onset of the symmetry breaking effect. The zero frequency unstable mode can be found for both the sub-critical and super-critical Reynolds numbers.

Finally, the eigenvalue sensitivity to structural perturbations is investigated using the adjoint equations for FSI, for a 2D cylinder oscillating in the cross-stream direction at $Re=50$, for varying natural frequencies of the spring-mass system.  When the structural frequencies are far from the unstable frequency, the sensitivity is hardly affected. However, close to resonance, the sensitivity map changes completely and the dominant region of sensitivity lies close to the cylinder, located symmetrically above and below. In all cases of FSI, the sensitivity at the surface is non-zero, however it attains large values only close to the resonance condition between the fluid and the structure.

\section*{Acknowledgements}
Financial support for this work was provided by the European Research Council under grant agreement 694452-TRANSEP-ERC-2015-AdG.\ The computations were performed on resources provided by the Swedish National Infrastructure for Computing (SNIC) at the PDC Center for High Performance Computing at the Royal Institute of Technology (KTH). The authors would like to thank Dr. L{\=a}cis for his helpful comments on the manuscript.

\appendix
\section{Navier--Stokes of Taylor expanded velocity}
\label{appA_fsi}

We evaluate the Navier--Stokes at the perturbed locations using the Taylor expansion of the velocity and pressure fields and then use the geometric linearization to consistently evaluate derivatives at the perturbed configuration. Non-linear terms arising due to interaction of perturbations terms $\mathbf{u}', p', \mathcal{R}'$ and $\mathbf{\Delta x}'$ are dropped.
\begin{eqnarray}
\begin{split}
	\left[U^{0}_{j}\frac{\partial U^{0}_{i}}{\partial x_{j}} + 
	U^{0}_{j}\frac{\partial u^{\xi}_{i}}{\partial x_{j}} + 
	u^{\xi}_{j}\frac{\partial U^{0}_{i}}{\partial x_{j}}  +
	\frac{\partial P^{0}}{\partial x_{i}} +
	\frac{\partial p^{\xi}}{\partial x_{i}}	-
	\frac{1}{Re}\frac{\partial^{2}U^{0}_{i}}{\partial x_{j}\partial x_{j}} -
	\frac{1}{Re}\frac{\partial^{2}u^{\xi}_{i}}{\partial x_{j}\partial x_{j}}
	\right]& 
	\\
	+\left[ \left.\frac{\partial U_{i}}{\partial t}\right\vert_{\mathbf{w}}
	-w_{j}\frac{\partial U^{0}_{i}}{\partial x_{j}} +
	U^{0}_{j}\frac{\partial u'_{i}}{\partial x_{j}} + 
	u'_{j}\frac{\partial U^{0}_{i}}{\partial x_{j}} +
	\frac{\partial p'}{\partial x_{i}} 
	-\frac{1}{Re}\frac{\partial^{2}u'_{i}}{\partial x_{j}\partial x_{j}}
	\right] &= 0.
\end{split} \nonumber \\
\begin{split}
\implies \overbrace{\left[U^{0}_{j}\frac{\partial U^{0}_{i}}{\partial x^{0}_{k}}(\mathcal{I}_{kj} - \mathcal{R'}_{kj})\right]}^{I}
+ \overbrace{\left[U^{0}_{j}\frac{\partial }{\partial x^{0}_{k}}\left(\frac{\partial U^{0}_{i}}{\partial x^{0}_{l}}\Delta x_{l}\right)(\mathcal{I}_{kj} - \mathcal{R'}_{kj})\right]}^{II} &\\
+ \overbrace{\left[\left(\frac{\partial U^{0}_{j}}{\partial x^{0}_{l}}\Delta x_{l}\right)\frac{\partial U^{0}_{i}}{\partial x^{0}_{k}}(\mathcal{I}_{kj} - \mathcal{R'}_{kj})\right]}^{III}
+ \overbrace{\left[\frac{\partial P^{0}}{\partial x^{0}_{k}}(\mathcal{I}_{ki} - \mathcal{R'}_{ki})\right]}^{IV}
+\overbrace{\left[u'_{j}\frac{\partial U^{0}_{i}}{\partial x^{0}_{k}}(\mathcal{I}_{kj} - \mathcal{R'}_{kj})\right]}^{V} \\
+\overbrace{\left[U^{0}_{j}\frac{\partial u'_{i}}{\partial x^{0}_{k}}(\mathcal{I}_{kj} - \mathcal{R'}_{kj})\right]}^{VI}
+ \overbrace{\left[\frac{\partial p'}{\partial x^{0}_{k}}(\mathcal{I}_{ki} - \mathcal{R'}_{ki})\right]}^{VII}
+ \overbrace{\left[\frac{\partial }{\partial x^{0}_{k}}\left(\frac{\partial P^{0}}{\partial x_{l}}\Delta x_{l}  \right)(\mathcal{I}_{ki} - \mathcal{R'}_{ki})\right]}^{VIII} &\\
- \overbrace{\left[\frac{1}{Re}\frac{\partial^{2}U^{0}_{i}}{\partial x^{0}_{k}\partial x^{0}_{n}}(\mathcal{I}_{kj}\mathcal{I}_{nj} - \mathcal{I}_{kj}\mathcal{R'}_{nj}  -\mathcal{R'}_{kj}\mathcal{I}_{nj} + \mathcal{R'}_{kj}\mathcal{R'}_{nj})\right]}^{IX} \\
-\overbrace{\left[\frac{1}{Re}\left\lbrace\frac{\partial }{\partial x^{0}_{k}}\left\lbrace \frac{\partial }{\partial x^{0}_{n}}\left(\frac{\partial U^{0}_{i}}{\partial x^{0}_{l}}\Delta x_{l}\right) \right\rbrace\right\rbrace(\mathcal{I}_{kj}\mathcal{I}_{nj} - \mathcal{I}_{kj}\mathcal{R'}_{nj}  -\mathcal{R'}_{kj}\mathcal{I}_{nj} + \mathcal{R'}_{kj}\mathcal{R'}_{nj})\right]}^{X} &\\
-\overbrace{\left[\frac{1}{Re}\frac{\partial^{2}u'_{i}}{\partial x^{0}_{k}\partial x^{0}_{n}}(\mathcal{I}_{kj}\mathcal{I}_{nj} - \mathcal{I}_{kj}\mathcal{R'}_{nj}  -\mathcal{R'}_{kj}\mathcal{I}_{nj} + \mathcal{R'}_{kj}\mathcal{R'}_{nj})\right]}^{XI} & \\
-\overbrace{\left[w_{j}\frac{\partial U^{0}_{i}}{\partial x^{0}_{k}}(\mathcal{I}_{kj} - \mathcal{R'}_{kj})\right]}^{XII}
+ \left.\frac{\partial U_{i}}{\partial t}\right\vert_{\mathbf{w}} &=0
\end{split} \nonumber
\end{eqnarray}
For clarity the terms are numbered and the expansion of the individual terms after dropping the terms higher than first order is given below.

\begin{align}
	I &=U^{0}_{j}\frac{\partial U^{0}_{i}}{\partial x^{0}_{j}} - \mathcal{R'}_{kj}U^{0}_{j}\frac{\partial U^{0}_{i}}{\partial x^{0}_{k}}, & 	VII &= \frac{\partial p'}{\partial x^{0}_{i}}, \nonumber \\
	II &= \Delta x_{l}U^{0}_{j}\frac{\partial^{2} U^{0}_{i}}{\partial x^{0}_{j}\partial x^{0}_{l}} + U^{0}_{j}\frac{\partial U^{0}_{i}}{\partial x^{0}_{l}}\mathcal{R'}_{lj}, & 	VIII &= \frac{\partial^{2} P^{0}}{\partial x^{0}_{i}\partial x^{0}_{l}}\Delta x_{l} + \frac{\partial P^{0}}{\partial x^{0}_{l}}\mathcal{R'}_{li}, \nonumber \\
	III &=\Delta x_{l}\frac{\partial U^{0}_{j}}{\partial x^{0}_{l}}\frac{\partial U^{0}_{i}}{\partial x^{0}_{j}}, & IX &=-\frac{1}{Re}\left[\frac{\partial^{2}U^{0}_{i}}{\partial x^{0}_{j}\partial x^{0}_{j}} -2\frac{\partial^{2}U^{0}_{i}}{\partial x^{0}_{j}\partial x^{0}_{k}}\mathcal{R'}_{kj}\right], \nonumber \\
	IV &=\frac{\partial P^{0}}{\partial x^{0}_{i}} - \frac{\partial P^{0}}{\partial x^{0}_{k}}\mathcal{R'}_{ki}, &	X &= -\frac{1}{Re}\left[\frac{\partial^{3} U^{0}_{i}}{\partial x^{0}_{j}\partial x^{0}_{j}\partial x^{0}_{l}}\Delta x_{l} + 2\frac{\partial^{2} U^{0}_{i}}{\partial x^{0}_{j}\partial x^{0}_{l}}\mathcal{R'}_{lj}\right], \nonumber \\
	V &= u'_{j}\frac{\partial U^{0}_{i}}{\partial x^{0}_{j}}, & 	XI &= -\frac{1}{Re}\left[\frac{\partial^{2}u'_{i}}{\partial x^{0}_{j}\partial x^{0}_{j}}\right], \nonumber \\
	VI &= U^{0}_{j}\frac{\partial u'_{i}}{\partial x^{0}_{j}}, & 	XII &= -w_{j}\frac{\partial U^{0}_{i}}{\partial x^{0}_{j}}. \nonumber
\end{align}
On collecting all terms of the expansions, several terms cancel and we obtain the following simplified expression
\begin{eqnarray}
\begin{split}
\left[U^{0}_{j}\frac{\partial U^{0}_{i}}{\partial x^{0}_{j}}   +
\frac{1}{\rho}\frac{\partial P^{0}}{\partial x^{0}_{i}}	-
\frac{1}{Re}\frac{\partial^{2} U^{0}_{i}}{\partial x^{0}_{j}\partial x^{0}_{j}}
\right] +
\Delta x_{l}\frac{\partial }{\partial x^{0}_{l}}\left[U^{0}_{j}\frac{\partial U^{0}_{i}}{\partial x^{0}_{j}}   +
\frac{1}{\rho}\frac{\partial P^{0}}{\partial x^{0}_{i}}	-
\frac{1}{Re}\frac{\partial^{2} U^{0}_{i}}{\partial x^{0}_{j}\partial x^{0}_{j}}
\right] & &\\
+\left[ \left.\frac{\partial U_{i}}{\partial t}\right\vert_{\mathbf{w}}
-w_{j}\frac{\partial U^{0}_{i}}{\partial x^{0}_{j}} +
U^{0}_{j}\frac{\partial u'_{i}}{\partial x^{0}_{j}} + 
u'_{j}\frac{\partial U^{0}_{i}}{\partial x^{0}_{j}} +
\frac{\partial p'}{\partial x^{0}_{i}}
-\frac{1}{Re}\frac{\partial^{2}u'_{i}}{\partial x^{0}_{j}\partial x^{0}_{j}}
\right]&=0 &
\end{split} \label{NS_taylor_expansion}
\end{eqnarray}
The first two bracketed terms together in equation~\ref{NS_taylor_expansion} represent the first-order Taylor expansion of the steady state Navier--Stokes evaluated at the perturbed location $\mathbf{\Delta x}$. Since the steady state equation for the base flow is identically zero everywhere, the Taylor expansion also vanishes.



\section{Derivation of Adjoint}
\label{appB}
The adjoint equations are derived for the linearized FSI equations under the inner-product defined as the integral over the domain $\Omega$ and time horizon $T$
\begin{eqnarray}
	\langle \mathbf{q}^{\dagger},\mathbf{q} \rangle = \int_{T}\int_{\Omega} \mathbf{q}^{\dagger H}\mathbf{q}\ d\Omega\ dt \label{appB_inner_prod}.
\end{eqnarray}
In the context of FSI $\mathcal{L}$ represents the linearized FSI equations defined by equations~\ref{NS_cossu_lin}--\ref{fluid_forces_cossu_lin} and $\mathcal{L}^{\dagger}$ represents the adjoint operator satisfying the Lagrange identity under the inner-product defined in equation~\ref{appB_inner_prod}
\begin{eqnarray}
	\langle \mathcal{L}^{\dagger}\mathbf{q}^{\dagger},\mathbf{q} \rangle = \langle \mathbf{q}^{\dagger},\mathcal{L}\mathbf{q} \rangle .
\end{eqnarray}
For linear FSI, we consider the case of a 2D circular cylinder free to oscillate in the vertical direction subject to the action of a spring-mass-damper system and the fluid forces. The structural system is then defined by the position $\eta$ and velocity $\varphi$ of its center of mass in the vertical direction. In the following $i=2$ represents the vertical direction. The vectors are defined as $\mathbf{q}=(\mathbf{u},p,\eta,\varphi)$, which lies in the domain of $\mathcal{L}$, and  $\mathbf{q}=(\mathbf{u}^{\dagger},p^{\dagger},\eta^{\dagger},\varphi^{\dagger})$, which lies in the domain of $\mathcal{L}^{\dagger}$. The Lagrange identity may be wriiten as 

\begin{eqnarray}
\langle \mathcal{L}^{\dagger}\mathbf{q}^{\dagger},\mathbf{q}\rangle &=& \langle \mathbf{q}^{\dagger},\mathcal{L}\mathbf{q}\rangle, \nonumber \\
\implies	\langle \mathcal{L}^{\dagger}\mathbf{q}^{\dagger},\mathbf{q}\rangle & = &\left\lbrace
\begin{split}
\int_{T}\int_{\Omega} u^{\dagger}_{i}\left[\frac{\partial u_{i}}{\partial t} + u_{j}\frac{\partial U_{i}}{\partial x_{j}} + U_{j}\frac{\partial u_{i}}{\partial x_{j}}\right] d\Omega dt \\
+\int_{T}\int_{\Omega}u^{\dagger}_{i}\left[\frac{\partial p}{\partial x_{i}} - \frac{1}{Re}\frac{\partial^{2} u_{i}}{\partial x_{j} \partial x_{j}}\right] d\Omega dt\\
- \int_{T}\int_{\Omega} p^{\dagger}\frac{\partial u_{i}}{\partial x_{i}}d\Omega dt + \int_{T}\eta^{\dagger}\left(\frac{d\eta}{dt} - \varphi\right)dt\\
+ \int_{T}\varphi^{\dagger}\left(\mathcal{M}\frac{d \varphi}{d t}  +\mathcal{D}\varphi + \mathcal{K}\eta + \oint_{\partial\Omega} \sigma_{2j}n_{j}d\Omega\right)dt
\end{split}\right. \nonumber
\end{eqnarray}
Integrating by parts and using the divergence theorem leads to 

\begin{eqnarray}
\langle \mathcal{L}^{\dagger}\mathbf{q}^{\dagger},\mathbf{q}\rangle &=\left\lbrace\begin{split}
\int_{T}\int_{\Omega} u_{i}\left[-\frac{\partial u^{\dagger}_{i}}{\partial t} +
u^{\dagger}_{j}\frac{\partial U_{j}}{\partial x_{i}}
- U_{j}\frac{\partial u^{\dagger}_{i}}{\partial x_{j}}
- \frac{1}{Re} \frac{\partial^{2} u^{\dagger}_{i}}{\partial x_{j}\partial x_{j}}
+ \frac{\partial p^{\dagger}}{\partial x_{i}}
\right] d\Omega dt \\
- \int_{T}\int_{\Omega}p\left[\frac{\partial u^{\dagger}_{i}}{\partial x_{i}}\right]d\Omega dt
	+ \int_{T}\oint_{\partial\Omega}  \left[(u^{\dagger}_{i}u_{i}U_{j})n_{j} 
	+ (u^{\dagger}_{i} p)n_{i} 
	 \right]d\Omega dt \\
	+\int_{T}\oint_{\partial\Omega}  \left[+\frac{1}{Re}\left(u_{i}\frac{\partial u^{\dagger}_{i}}{\partial x_{j}}
	-u^{\dagger}_{i}\frac{\partial u_{i}}{\partial x_{j}} \right)n_{j} 
	- (u_{i}p^{\dagger})n_{i}
	 \right]d\Omega dt \\
     +\int_{T}\left[\varphi\left(-\mathcal{M}\frac{d \varphi^{\dagger}}{d t}  +\mathcal{D}\varphi^{\dagger}  -\eta^{\dagger}  \right) 
     + \eta\left(-\frac{d\eta^{\dagger}}{dt} + \mathcal{K}\varphi^{\dagger}\right) \right] dt \\
      + \int_{T}\varphi^{\dagger}\left[\oint_{\partial\Omega} \sigma_{2j}n_{j}d\Omega\right]dt
     + \left[\int_{\Omega}  (u^{\dagger}_{i}u_{i}) d\Omega\right]_{0}^{T}
     +\left[\mathcal{M}\frac{d (\varphi^{\dagger}\varphi)}{d t} + \frac{d (\eta^{\dagger}\eta)}{d t} \right]_{0}^{T}
	 \end{split}\right. \nonumber
\end{eqnarray}
The temporal boundary terms for appropriately defined direct and adjoint operators vanish \citep{bagheri09}.
For the remaining boundary terms, we denote the direct and adjoint stresses in compact notation as
\begin{eqnarray}
	\sigma_{ij} = -p\delta_{ij} +\frac{1}{Re}\frac{\partial u_{i}}{\partial x_{j}}, \nonumber \\
	\sigma_{ij}^{\dagger} = -p^{\dagger}\delta_{ij}  + \frac{1}{Re}\frac{\partial u^{\dagger}_{i}}{\partial x_{j}}  \nonumber
\end{eqnarray}
For the outer boundaries of the domain the structural variables play no role and and the boundary terms may be written compactly as
\begin{eqnarray}
	\int_{T}\oint_{\partial\Omega}  \left[u_{i}(u_{i}^{\dagger}U_{j} +\sigma^{\dagger}_{ij})n_{j} - u^{\dagger}_{i}\sigma_{ij}n_{j}\right]\ d\Omega dt = 0.
\end{eqnarray}
For homogeneous Dirichlet and stress-free conditions for the direct problem, these lead to
\begin{align}
	u^{\dagger}_{i}	&=0	&	\text{on } \partial\Omega_{v}, \nonumber \\
	(\sigma^{\dagger}_{ij}+u_{i}^{\dagger}U_{j})n_{j} &=0& \text{on } \partial\Omega_{o}. \nonumber
\end{align}
On the interface boundary $\Gamma_{0}$, the structural equations are part of the boundary terms. The baseflow $\mathbf{U}$ is identically zero and using the velocity boundary condition at the interface for the direct variables, one obtains
\begin{eqnarray}
	\begin{split}
	-\left[\int_{T}\oint_{\partial\Omega}  u^{\dagger}_{i}\sigma_{ij}n_{j}d\Omega dt\right]
	+ \left[\int_{T}\eta\left(-\frac{d\eta^{\dagger}}{dt} + \mathcal{K}\varphi^{\dagger}  - \oint_{\partial\Omega}\frac{\partial U_{k}}{\partial x_{2}}\sigma^{\dagger}_{kj}n_{j}d\Omega\right)dt\right]\\
+\left[\int_{T}\varphi\left(-\mathcal{M}\frac{d \varphi^{\dagger}}{d t}  +\mathcal{D}\varphi^{\dagger}  -\eta^{\dagger}  + \oint_{\partial\Omega}\sigma^{\dagger}_{ij}n_{j}d\Omega \right)  dt  \right]
 +\left[\int_{T}  \oint_{\partial\Omega} \varphi^{\dagger}\sigma_{2j}n_{j}d\Omega dt\right] = 0
	\end{split} \nonumber
\end{eqnarray}

The second and the third brackets form the two adjoint structural equations, while the first and the fourth bracket together forms the boundary conditions for the coupling between the fluid and the structure in the adjoint equations.
The final set of adjoint equations for FSI of the 2D oscillating cylinder can be written as

\begin{subequations}
	\begin{align}
	-\frac{\partial u^{\dagger}_{i}}{\partial t} + u^{\dagger}_{j}\frac{\partial U_{j}}{\partial x_{i}}-U_{j}\frac{\partial u^{\dagger}_{i}}{\partial x_{j}}
	+ \frac{\partial p^{\dagger}}{\partial x_{i}} - \frac{1}{Re}\frac{\partial^{2} u^{\dagger}_{i}}{\partial x_{j}\partial x_{j}}   =& 0, & \text{in } \Omega^{f} ,\\
	\frac{\partial u^{\dagger}_{i}}{\partial x_{i}} =  &0,  &\text{in } \Omega^{f}, \\
	-\mathcal{M}\frac{d \varphi^{\dagger}}{d t}  +\mathcal{D}\varphi^{\dagger}  -\eta^{\dagger}  + \oint_{\Gamma} \sigma^{\dagger}_{2j}n_{j} d\Omega =& 0, &\text{in } \Omega^{s},  \\
	-\frac{d\eta^{\dagger}}{dt} + \mathcal{K}\varphi^{\dagger} - \oint_{\partial\Omega} \left(\frac{\partial U_{k}}{\partial x_{2}}\right) \sigma^{\dagger}_{kj}n_{j} d\Omega = &0&\text{in } \Omega^{s},  \\
	u^{\dagger}_{1} = & 0, & \text{on } \Gamma,  \\
	u^{\dagger}_{2}  - \varphi^{\dagger}_{2} = & 0, & \text{on } \Gamma,  \\
	\sigma_{ij}^{\dagger} - \left(-p^{\dagger}\delta_{ij} + \frac{1}{Re}\frac{\partial u^{\dagger}_{i}}{\partial x_{j}}   \right) = &0, & \text{adjoint forces on } \Gamma. \\
	u^{\dagger}_{i} = & 0, & \text{on } \partial\Omega_{v},  \\
	(\sigma^{\dagger}_{ij}+u_{i}^{\dagger}U_{j})n_{j} =& 0& \text{on } \partial\Omega_{o}.  
	\end{align}
\end{subequations}


\section{Convergence tests}
\label{fsi_convergence}
\begin{figure}
		\centering
		\includegraphics[width=0.48\textwidth]{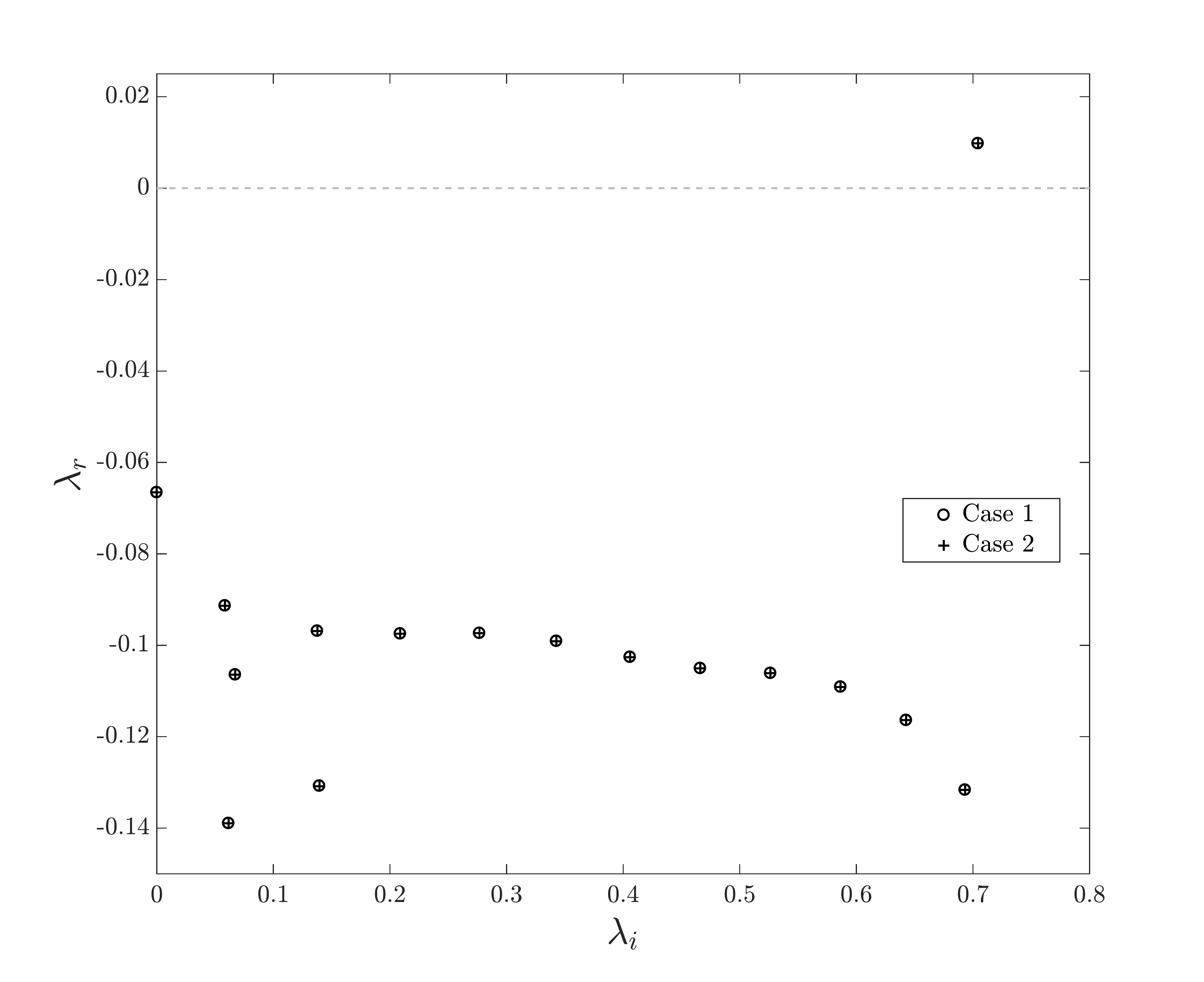}
		\includegraphics[width=0.48\textwidth]{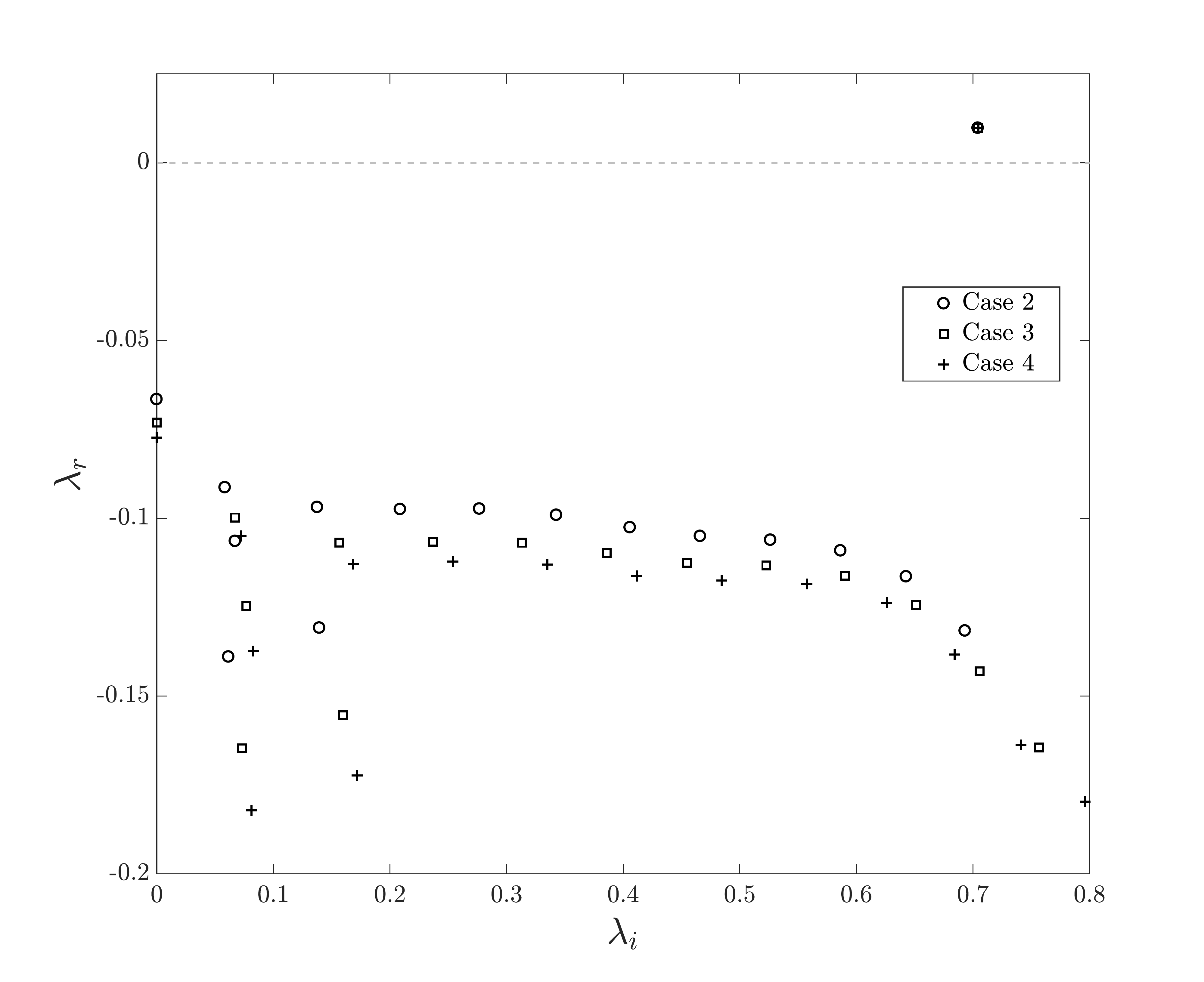}
		\caption{A comparison of spectra for different domain studies. (Left) shows the spectra for two different polynomial orders of $N=7$ (case 1) and $N=9$ (case 2). (Right) shows the comparison of spectra for 3 different domain sizes.}
		\label{fig:conv_tests}
\end{figure}

The convergence tests for an oscillating cylinder are reported in this section. The convergence test is performed for the case of a spring-mounted 2D cylinder at a diameter based Reynolds number of $Re=23.512$. The structural parameters are the same as the ones reported in table~\ref{tab:struc_param}. Table~\ref{tab:conv_tests} reports the domain size and polynomial order of the various tests performed to ensure converged results. A comparison of the spectra for case $1$ and case $2$ is shown in figure~\ref{fig:conv_tests}. The panel on the left shows the spectra comparison for two different polynomial orders. The two cases have nearly identical spectra implying the convergence of results with grid resolution. The panel on the right shows the spectra for three different domain sizes. For all three cases the physical eigenvalue of the system is always converged. The highly damped modes represent the box modes of the system which vary slightly depending on the size of the computational box. A similar behavior of the damped eigenmodes can be found in \citealp{assemat12}. Overall, the results show the convergence of the physical eigenvalue both in terms of grid resolution and numerical domain. 
\begin{table}
	\centering
	\begin{tabular}{*{4}{x{2.5cm}}}
		\toprule
		$Case$ & Polynomial order (N) & Streamwise domain & Crossflow domain \\
		\toprule
		1	   	     & 7 	   &  $x\in[-25,60]$ 	& $y\in[-50,50]$\\
		2	   	     & 9 	   &  $x\in[-25,60]$ 	& $y\in[-50,50]$\\
		3	   	     & 9 	   &  $x\in[-25,50]$ 	& $y\in[-40,40]$\\
		4	   	     & 9 	   &  $x\in[-25,45]$ 	& $y\in[-35,35]$\\				
	\end{tabular}
	\caption{Structural parameters for a 2D cylinder oscillating in the cross-stream direction.} \label{tab:conv_tests}
\end{table}

%
\bibliographystyle{apa}

\begin{thebibliography}{}

\bibitem[\protect\astroncite{Assemat et~al.}{2012}]{assemat12}
Assemat, P., Fabre, D., and Magnaudet, J. (2012).
\newblock The onset of unsteadiness of two-dimensional bodies falling or rising
  freely in a viscous fluid: a linear study.
\newblock {\em Journal of Fluid Mechanics}, 690:173–202.

\bibitem[\protect\astroncite{Bagheri et~al.}{2009}]{bagheri09}
Bagheri, S.and~Schlatter, P., Schmid, P.~J., and Henningson, D.~S. (2009).
\newblock Global stability of a jet in crossflow.
\newblock {\em Journal of Fluid Mechanics}, 624:33–44.

\bibitem[\protect\astroncite{Bagheri et~al.}{2012}]{bagheri12}
Bagheri, S., Mazzino, A., and Bottaro, A. (2012).
\newblock Spontaneous symmetry breaking of a hinged flapping filament generates
  lift.
\newblock {\em Physical Review Letters}, 109(15):154502.

\bibitem[\protect\astroncite{Barkley et~al.}{2002}]{barkley02}
Barkley, D., Gomes, M. G.~M., and HENDERSON, R.~D. (2002).
\newblock Three-dimensional instability in flow over a backward-facing step.
\newblock {\em Journal of Fluid Mechanics}, 473:167–190.

\bibitem[\protect\astroncite{Benjamin}{1959}]{benjamin59}
Benjamin, T.~B. (1959).
\newblock Shearing flow over a wavy boundary.
\newblock {\em Journal of Fluid Mechanics}, 6(2):161–205.

\bibitem[\protect\astroncite{Benjamin}{1960}]{benjamin60}
Benjamin, T.~B. (1960).
\newblock Effects of a flexible boundary on hydrodynamic stability.
\newblock {\em Journal of Fluid Mechanics}, 9(4):513–532.

\bibitem[\protect\astroncite{Brynjell-Rahkola et~al.}{2017}]{rahkola17}
Brynjell-Rahkola, M., Shahriari, N., Schlatter, P., Hanifi, A., and Henningson,
  D.~S. (2017).
\newblock Stability and sensitivity of a cross-flow-dominated
  {F}alkner–{S}kan–{C}ooke boundary layer with discrete surface roughness.
\newblock {\em Journal of Fluid Mechanics}, 826:830–850.

\bibitem[\protect\astroncite{Cano-Lozano et~al.}{2016}]{lozano16}
Cano-Lozano, J.~C., Tchoufag, J., Magnaudet, J., and Martínez-Bazán, C.
  (2016).
\newblock A global stability approach to wake and path instabilities of nearly
  oblate spheroidal rising bubbles.
\newblock {\em Physics of Fluids}, 28(1):014102.

\bibitem[\protect\astroncite{Carpenter and Garrad}{1985}]{carpenter85}
Carpenter, P.~W. and Garrad, A.~D. (1985).
\newblock The hydrodynamic stability of flow over kramer-type compliant
  surfaces. part 1. tollmien-schlichting instabilities.
\newblock {\em Journal of Fluid Mechanics}, 155:465–510.

\bibitem[\protect\astroncite{Carpenter and Garrad}{1986}]{carpenter86}
Carpenter, P.~W. and Garrad, A.~D. (1986).
\newblock The hydrodynamic stability of flow over kramer-type compliant
  surfaces. part 2. flow-induced surface instabilities.
\newblock {\em Journal of Fluid Mechanics}, 170:199–232.

\bibitem[\protect\astroncite{Carpenter and Morris}{1990}]{carpenter90}
Carpenter, P.~W. and Morris, P.~J. (1990).
\newblock The effect of anisotropic wall compliance on boundary-layer stability
  and transition.
\newblock {\em Journal of Fluid Mechanics}, 218:171–223.

\bibitem[\protect\astroncite{Citro et~al.}{2017}]{citro17}
Citro, V., Luchini, P., Giannetti, F., and Auteri, F. (2017).
\newblock Efficient stabilization and acceleration of numerical simulation of
  fluid flows by residual recombination.
\newblock {\em Journal of Computational Physics}, 344:234 -- 246.

\bibitem[\protect\astroncite{Cossu and Morino}{2000}]{cossu00}
Cossu, C. and Morino, L. (2000).
\newblock On the instability of a spring-mounted circular cylinder in a viscous
  flow at low reynolds numbers.
\newblock {\em Journal of Fluids and Structures}, 14(2):183 -- 196.

\bibitem[\protect\astroncite{Davies and Carpenter}{1997}]{davies97}
Davies, C. and Carpenter, P.~W. (1997).
\newblock Instabilities in a plane channel flow between compliant walls.
\newblock {\em Journal of Fluid Mechanics}, 352:205–243.

\bibitem[\protect\astroncite{Dong et~al.}{2008}]{dong08}
Dong, S., Triantafyllou, G.~S., and Karniadakis, G.~E. (2008).
\newblock Elimination of vortex streets in bluff-body flows.
\newblock {\em Phys. Rev. Lett.}, 100:204501.

\bibitem[\protect\astroncite{Dowell and Hall}{2001}]{dowell01}
Dowell, E.~H. and Hall, K.~C. (2001).
\newblock Modeling of fluid-structure interaction.
\newblock {\em Annual Review of Fluid Mechanics}, 33(1):445--490.

\bibitem[\protect\astroncite{Eriksson and Rizzi}{1985}]{eriksson85}
Eriksson, L.~E. and Rizzi, A. (1985).
\newblock Computer-aided analysis of the convergence to steady state of
  discrete approximations to the euler equations.
\newblock {\em Journal of Computational Physics}, 57(1):90 -- 128.

\bibitem[\protect\astroncite{Ern et~al.}{2012}]{ern12}
Ern, P., Frédéric, R., Fabre, D., and Magnaudet, J. (2012).
\newblock Wake-induced oscillatory paths of bodies freely rising or falling in
  fluids.
\newblock {\em Annual Review of Fluid Mechanics}, 44(1):97--121.

\bibitem[\protect\astroncite{Fabre et~al.}{2011}]{fabre11}
Fabre, D., Assemat, P., and Magnaudet, J. (2011).
\newblock A quasi-static approach to the stability of the path of heavy bodies
  falling within a viscous fluid.
\newblock {\em Journal of Fluids and Structures}, 27(5):758 -- 767.
\newblock IUTAM Symposium on Bluff Body Wakes and Vortex-Induced Vibrations
  (BBVIV-6).

\bibitem[\protect\astroncite{Fanion et~al.}{2000}]{fanion00}
Fanion, T., Fernández, M., A., and {Le Tallec}, P. (2000).
\newblock Deriving adequate formulations for fluid-structure interaction
  problems: from {ALE} to transpiration.
\newblock {\em Revue Européenne des Éléments Finis}, 9(6-7):681--708.

\bibitem[\protect\astroncite{Fern\'{a}ndez and {Le
  Tallec}}{2003a}]{fernandez03i}
Fern\'{a}ndez, M.~A. and {Le Tallec}, P. (2003a).
\newblock Linear stability analysis in fluid–structure interaction with
  transpiration. {P}art i: {F}ormulation and mathematical analysis.
\newblock {\em Computer Methods in Applied Mechanics and Engineering},
  192(43):4805 -- 4835.

\bibitem[\protect\astroncite{Fern\'{a}ndez and {Le
  Tallec}}{2003b}]{fernandez03ii}
Fern\'{a}ndez, M.~A. and {Le Tallec}, P. (2003b).
\newblock Linear stability analysis in fluid–structure interaction with
  transpiration. {P}art ii: {N}umerical analysis and applications.
\newblock {\em Computer Methods in Applied Mechanics and Engineering},
  192(43):4837 -- 4873.

\bibitem[\protect\astroncite{Fischer et~al.}{2017}]{fischer17}
Fischer, P., Schmitt, M., and Tomboulides, A. (2017).
\newblock {\em Recent Developments in Spectral Element Simulations of
  Moving-Domain Problems}, pages 213--244.
\newblock Springer New York, New York, NY.

\bibitem[\protect\astroncite{Fischer et~al.}{2008}]{nek5000}
Fischer, P.~F., Lottes, J.~W., and Kerkemeier, S.~G. (2008).
\newblock Nek5000 web page.
\newblock \url{http://nek5000.mcs.anl.gov}.

\bibitem[\protect\astroncite{Freund}{2014}]{freund14}
Freund, J.~B. (2014).
\newblock Numerical simulation of flowing blood cells.
\newblock {\em Annual Review of Fluid Mechanics}, 46(1):67--95.

\bibitem[\protect\astroncite{Gianetti and Luchini}{2007}]{giannetti07}
Gianetti, F. and Luchini, P. (2007).
\newblock Structural sensitivity of the first instability of the cylinder wake.
\newblock {\em Journal of Fluid Mechanics}, 581:167–197.

\bibitem[\protect\astroncite{Goza et~al.}{2018}]{goza18}
Goza, A., Colonius, T., and Sader, J.~E. (2018).
\newblock Global modes and nonlinear analysis of inverted-flag flapping.
\newblock {\em Journal of Fluid Mechanics}, 857:312–344.

\bibitem[\protect\astroncite{Heil and Hazel}{2011}]{heil11}
Heil, M. and Hazel, A.~L. (2011).
\newblock Fluid-structure interaction in internal physiological flows.
\newblock {\em Annual Review of Fluid Mechanics}, 43(1):141--162.

\bibitem[\protect\astroncite{Ho and Patera}{1990}]{ho90}
Ho, L.-W. and Patera, A.~T. (1990).
\newblock {A Legendre spectral element method for simulation of unsteady
  incompressible viscous free-surface flows}.
\newblock {\em Computer Methods in Applied Mechanics and Engineering},
  80(1):355 -- 366.

\bibitem[\protect\astroncite{Ho and Patera}{1991}]{ho91}
Ho, L.-W. and Patera, A.~T. (1991).
\newblock Variational formulation of three-dimensional viscous free-surface
  flows: Natural imposition of surface tension boundary conditions.
\newblock {\em International Journal for Numerical Methods in Fluids},
  13(6):691--698.

\bibitem[\protect\astroncite{Kumaran}{2003}]{kumaran03}
Kumaran, V. (2003).
\newblock Hydrodynamic stability of flow through compliant channels and tubes.
\newblock In Carpenter, P.~W. and Pedley, T.~J., editors, {\em Flow Past Highly
  Compliant Boundaries and in Collapsible Tubes}, pages 95--118, Dordrecht.
  Springer Netherlands.

\bibitem[\protect\astroncite{K{\"u}ttler and Wall}{2008}]{kuttler08}
K{\"u}ttler, U. and Wall, W.~A. (2008).
\newblock Fixed-point fluid--structure interaction solvers with dynamic
  relaxation.
\newblock {\em Computational Mechanics}, 43(1):61--72.

\bibitem[\protect\astroncite{L{\=a}cis et~al.}{2014}]{lacis14}
L{\=a}cis, U., Brosse, N., Ingremeau, F., Mazzino, A., Lundell, F., Kellay, H.,
  and Bagheri, S. (2014).
\newblock Passive appendages generate drift through symmetry breaking.
\newblock {\em Nature communications}, 5:5310.

\bibitem[\protect\astroncite{L{\=a}cis et~al.}{2017}]{lacis17}
L{\=a}cis, U., Olivieri, S., Mazzino, A., and Bagheri, S. (2017).
\newblock Passive control of a falling sphere by elliptic-shaped appendages.
\newblock {\em Phys. Rev. Fluids}, 2:033901.

\bibitem[\protect\astroncite{Landahl}{1962}]{landahl62}
Landahl, M.~T. (1962).
\newblock On the stability of a laminar incompressible boundary layer over a
  flexible surface.
\newblock {\em Journal of Fluid Mechanics}, 13(4):609–632.

\bibitem[\protect\astroncite{Lehoucq et~al.}{1998}]{lehoucq98}
Lehoucq, R., Sorensen, D., and Yang, C. (1998).
\newblock {\em ARPACK Users' Guide}.
\newblock Society for Industrial and Applied Mathematics.

\bibitem[\protect\astroncite{Lesoinne et~al.}{2001}]{lesoinne00}
Lesoinne, M., Sarkis, M., Hetmaniuk, U., and Farhat, C. (2001).
\newblock A linearized method for the frequency analysis of three-dimensional
  fluid/structure interaction problems in all flow regimes.
\newblock {\em Computer Methods in Applied Mechanics and Engineering},
  190(24):3121 -- 3146.
\newblock Advances in Computational Methods for Fluid-Structure Interaction.

\bibitem[\protect\astroncite{Luchini and Bottaro}{2014}]{luchini14}
Luchini, P. and Bottaro, A. (2014).
\newblock Adjoint equations in stability analysis.
\newblock {\em Annual Review of Fluid Mechanics}, 46(1):493--517.

\bibitem[\protect\astroncite{{Maday} and {Patera}}{1989}]{maday89}
{Maday}, Y. and {Patera}, A.~T. (1989).
\newblock {Spectral element methods for the incompressible Navier-Stokes
  equations}.
\newblock In {\em State-of-the-art surveys on computational mechanics
  (A90-47176 21-64). New York, American Society of Mechanical Engineers, 1989,
  p. 71-143. Research supported by DARPA.}, pages 71--143.

\bibitem[\protect\astroncite{Mittal et~al.}{2004}]{mittal04}
Mittal, R., Seshadri, V., and Udaykumar, H.~S. (2004).
\newblock Flutter, tumble and vortex induced autorotation.
\newblock {\em Theoretical and Computational Fluid Dynamics}, 17(3):165--170.

\bibitem[\protect\astroncite{Mittal and Singh}{2005}]{mittal05}
Mittal, S. and Singh, S. (2005).
\newblock Vortex-induced vibrations at subcritical {Re}.
\newblock {\em Journal of Fluid Mechanics}, 534:185–194.

\bibitem[\protect\astroncite{Mougin and Magnaudet}{2001}]{guillaume01}
Mougin, G. and Magnaudet, J. (2001).
\newblock Path instability of a rising bubble.
\newblock {\em Phys. Rev. Lett.}, 88:014502.

\bibitem[\protect\astroncite{Navrose and Mittal}{2016}]{navrose16}
Navrose and Mittal, S. (2016).
\newblock Lock-in in vortex-induced vibration.
\newblock {\em Journal of Fluid Mechanics}, 794:565–594.

\bibitem[\protect\astroncite{Negi et~al.}{2018}]{negi18}
Negi, P., Vinuesa, R., Hanifi, A., Schlatter, P., and Henningson, D. (2018).
\newblock Unsteady aerodynamic effects in small-amplitude pitch oscillations of
  an airfoil.
\newblock {\em International Journal of Heat and Fluid Flow}, 71:378 -- 391.

\bibitem[\protect\astroncite{Negi}{2017}]{negi17}
Negi, P.~S. (2017).
\newblock {\em Boundary Layers over wing sections}.
\newblock Licentiate thesis, KTH Royal Institute of Technology.

\bibitem[\protect\astroncite{Pfister et~al.}{2019}]{pfister19}
Pfister, J.-L., Marquet, O., and Carini, M. (2019).
\newblock Linear stability analysis of strongly coupled fluid–structure
  problems with the {A}rbitrary-{L}agrangian–{E}ulerian method.
\newblock {\em Computer Methods in Applied Mechanics and Engineering}, 355:663
  -- 689.

\bibitem[\protect\astroncite{Riley et~al.}{1988}]{riley88}
Riley, J.~J., Gad-el Hak, M., and Metcalfe, R.~W. (1988).
\newblock Complaint coatings.
\newblock {\em Annual Review of Fluid Mechanics}, 20(1):393--420.

\bibitem[\protect\astroncite{Schlatter et~al.}{2004}]{schlatter04}
Schlatter, P., Stolz, S., and Kleiser, L. (2004).
\newblock \text{LES} of transitional flows using the approximate deconvolution
  model.
\newblock {\em International Journal of Heat and Fluid Flow}, 25(3):549 -- 558.

\bibitem[\protect\astroncite{Schlatter et~al.}{2006}]{schlatter06}
Schlatter, P., Stolz, S., and Kleiser, L. (2006).
\newblock Large-eddy simulation of spatial transition in plane channel flow.
\newblock {\em Journal of Turbulence}, 7:N33.

\bibitem[\protect\astroncite{Shelley and Zhang}{2011}]{shelley11}
Shelley, M.~J. and Zhang, J. (2011).
\newblock Flapping and bending bodies interacting with fluid flows.
\newblock {\em Annual Review of Fluid Mechanics}, 43(1):449--465.

\bibitem[\protect\astroncite{Sorensen}{1992}]{sorensen92}
Sorensen, D. (1992).
\newblock Implicit application of polynomial filters in a k-step arnoldi
  method.
\newblock {\em SIAM Journal on Matrix Analysis and Applications},
  13(1):357--385.

\bibitem[\protect\astroncite{Tchoufag et~al.}{2014a}]{tchoufag14}
Tchoufag, J., Fabre, D., and Magnaudet, J. (2014a).
\newblock Global linear stability analysis of the wake and path of
  buoyancy-driven disks and thin cylinders.
\newblock {\em Journal of Fluid Mechanics}, 740:278–311.

\bibitem[\protect\astroncite{Tchoufag et~al.}{2014b}]{tchoufag14a}
Tchoufag, J., Magnaudet, J., and Fabre, D. (2014b).
\newblock Linear instability of the path of a freely rising spheroidal bubble.
\newblock {\em Journal of Fluid Mechanics}, 751:R4.

\bibitem[\protect\astroncite{Williamson and Govardhan}{2004}]{williamson04}
Williamson, C. and Govardhan, R. (2004).
\newblock Vortex-induced vibrations.
\newblock {\em Annual Review of Fluid Mechanics}, 36(1):413--455.

\bibitem[\protect\astroncite{Xu et~al.}{1990}]{xu90}
Xu, J.~C., Sen, M., and Gad-el Hak, M. (1990).
\newblock Low-{R}eynolds number flow over a rotatable cylinder--splitter plate
  body.
\newblock {\em Physics of Fluids A: Fluid Dynamics}, 2(11):1925--1927.

\bibitem[\protect\astroncite{Xu et~al.}{1993}]{xu93}
Xu, J.~C., Sen, M., and Gad-el Hak, M. (1993).
\newblock Dynamics of a rotatable cylinder with splitter plate in uniform flow.
\newblock {\em Journal of fluids and structures}, 7(4):401--416.

\end{thebibliography}

\end{document}